\documentclass[camera,letterpaper,nomarginnotes,nonarrowgutter]{jpaper}
\usepackage{cite}
\usepackage{amsmath,amssymb,amsfonts}
\usepackage{algorithmic}
\usepackage{graphicx}
\usepackage{textcomp}
\usepackage{fancyhdr}
\usepackage{glossaries} 

\usepackage{duckuments}
\usepackage{atbegshi} 
\usepackage{filecontents}
\usepackage{multirow}
\usepackage{booktabs}
\usepackage{balance}
\usepackage{setspace}
\usepackage{listings}
\usepackage[linesnumbered]{algorithm2e}
\usepackage{siunitx} 
\usepackage{subcaption}
\usepackage{tikz}
\usepackage{multicol}
\usepackage{threeparttable}
\usepackage{titlesec}
\usepackage[colorinlistoftodos,prependcaption,textsize=tiny]{todonotes}
\usepackage{marginnote}
\usepackage{enumitem}
\usepackage[bookmarks=true,breaklinks=true,letterpaper=true,colorlinks,citecolor=blue,linkcolor=blue,urlcolor=blue]{hyperref}
\usepackage{float}
\usepackage{xcolor}
\usepackage[calc,useregional]{datetime2}
\usepackage{dblfloatfix}
\usepackage{bm}
\usepackage{makecell}

\marginparwidth=\dimexpr \marginparwidth + 1.2cm\relax

\newif\ifdraft
\draftfalse

\newif\ifcameraready
\camerareadyfalse

\newif\ifrev
\revfalse

\newcounter{version}
\ifcameraready
\else
\fi

\newcommand{\affilETH}[0]{\textsuperscript{\S}}
\newcommand{\affilETU}[0]{\textsuperscript{$\dagger$}}
\newcommand{\affilUOS}[0]{\textsuperscript{$\ddagger$}}
\author{
{O\u{g}uzhan Canpolat\affilETH\affilETU}\qquad
{A. Giray Ya\u{g}l{\i}kç{\i}\affilETH}\qquad
{Ataberk Olgun\affilETH}\qquad
{Ismail Emir Yuksel\affilETH}\\
{Yahya Can Tu\u{g}rul\affilETH\affilETU}\qquad
{Konstantinos Kanellopoulos\affilETU}\qquad
{O\u{g}uz Ergin\affilUOS\affilETH\affilETU}\qquad
{Onur Mutlu\affilETH}\\
{\affilETH\emph{ETH Z{\"u}rich}}\qquad \affilETU\emph{TOBB University of Economics and Technology}\qquad \affilUOS\emph{University of Sharjah}\vspace{-0.5em}
}

\definecolor{gfored}{rgb}{0.580, 0.050, 0.211}
\definecolor{ao}{rgb}{0.007, 0.520, 0.867}
\definecolor{moegi}{rgb}{0.357, 0.537, 0.188}
\definecolor{jl}{rgb}{1.0, 0.2, 0.8}
\definecolor{brown(web)}{rgb}{0.65, 0.16, 0.16}
\definecolor{bisque}{rgb}{1.0, 0.89, 0.77}
\definecolor{nbs}{rgb}{0.88, 0.07, 0.37}
\definecolor{yt}{rgb}{0.58, 0.44, 0.86}
\definecolor{iy}{rgb}{0.0, 0.36, 0.05}
\definecolor{burntorange}{rgb}{0.8, 0.33, 0.0}
\definecolor{lightmauve}{rgb}{0.86, 0.82, 1.0}
\definecolor{frenchblue}{rgb}{0.19, 0.55, 0.91}
\definecolor{amber}{rgb}{1.0, 0.49, 0.0}
\definecolor{awesome}{rgb}{1.0, 0.13, 0.32}
\definecolor{dollarbill}{rgb}{0.52,0.73,0.4}
\definecolor{moegi}{rgb}{0.357, 0.537, 0.188}
\definecolor{burgundy}{rgb}{0.5, 0.0, 0.13}
\definecolor{ballblue}{rgb}{0.13, 0.67, 0.8}
\definecolor{ups-truck}{rgb}{0.53, 0.28, 0.21}
\definecolor{airforceblue}{rgb}{0.36, 0.54, 0.66}
\definecolor{cadmiumgreen}{rgb}{0.0, 0.42, 0.24}
\definecolor{darkcyan}{rgb}{0.0, 0.55, 0.55}
\definecolor{caribbeangreen}{rgb}{0.0, 0.8, 0.6}
\definecolor{flamingopink}{rgb}{0.99, 0.56, 0.67}
\definecolor{jazzberryjam}{rgb}{0.65, 0.04, 0.37}
\definecolor{mediumpersianblue}{rgb}{0.0, 0.4, 0.65}
\definecolor{coolblack}{rgb}{0.0, 0.18, 0.39}
\definecolor{bleudefrance}{rgb}{0.19, 0.55, 0.91}
\definecolor{ao}{rgb}{0.0, 0.0, 1.0}
\definecolor{babyblueeyes}{rgb}{0.63, 0.79, 0.95}
\definecolor{darkwarmgray}{rgb}{0.2,0,0}
\definecolor{brightpink}{rgb}{1.0, 0.0, 0.5}
\definecolor{darkblue}{rgb}{0.0, 0.0, 0.55}

\newcommand*\circled[1]{\tikz[baseline=(char.base)]{\node[shape=circle,fill,inner sep=0.5pt] (char) {\textcolor{white}{#1}};}}

\newcommand*\circledblue[1]{\tikz[baseline=(char.base)]{\node[shape=circle,fill={rgb:red,44;green,46;blue,136},inner sep=0.5pt] (char) {\textcolor{white}{#1}};}}

\definecolor{pastelpink}{RGB}{250,176,228}
\newcommand*\circledpastelpink[1]{\tikz[baseline=(char.base)]{\node[shape=circle,fill=pastelpink,inner sep=0.5pt] (char) {\textcolor{white}{#1}};}}

\definecolor{pastelblue}{RGB}{161,201,244}
\newcommand*\circledpastelblue[1]{\tikz[baseline=(char.base)]{\node[shape=circle,fill=pastelblue,inner sep=0.5pt] (char) {\textcolor{white}{#1}};}}

\newcommand{\ignore}[1]{}

\ifrev
    \let\marginpar\marginnote
    \newcommand{\cql}[2]{#2\todo[size=\small,color=jazzberryjam]{\textbf{\textrm{\textcolor{white}{#1}}}}}
    \newcommand{\iql}[2]{#2\todo[size=\small,color=babyblueeyes]{\textbf{\textrm{\textcolor{black}{#1}}}}}

\else

    \newcommand{\cql}[1]{}
    \newcommand{\iql}[1]{}
\fi

\ifdraft
    
    \newcommand{\param}[1]{\textcolor{red}{#1}} 

\else

\fi

\definecolor{frenchblue}{rgb}{0.19, 0.55, 0.91}

\usepackage[most]{tcolorbox} 
\tcbset{before skip=1.5pt, after skip=4pt}

\newtcolorbox[auto counter]{obsx}[3][]{%
    colframe = #2!45,
    colback  = #2!10,
    coltitle = #2!20!black, 
    colbacktitle=#2!20,
    coltitle=black,
    fonttitle=\bfseries, 
    title=#3~\thetcbcounter.\ ,
    enhanced,
    attach boxed title to top left={yshift=-2.8mm, xshift=0.15cm},
    bottom=-2.2pt,
    #1%
}

\usepackage[most]{tcolorbox} 
\newtcolorbox[auto counter]{tkx}[2][]{%
    enhanced, breakable, center title,
    colframe = #2!45,
    colback  = #2!10,
    colbacktitle=#2!20,
    left=-0.5pt,
    right=-0.5pt,
    bottom=-2pt,
    top=-0.25pt,
    #1%
}
\newcounter{obs}
\newcounter{tkw}
\definecolor{amber}{rgb}{1.0, 0.49, 0.0}
\definecolor{awesome}{rgb}{1.0, 0.13, 0.32}
\definecolor{dollarbill}{rgb}{0.52,0.73,0.4}
\definecolor{moegi}{rgb}{0.357, 0.537, 0.188}
\definecolor{burgundy}{rgb}{0.5, 0.0, 0.13}
\definecolor{ballblue}{rgb}{0.13, 0.67, 0.8}
\definecolor{ups-truck}{rgb}{0.53, 0.28, 0.21}
\definecolor{airforceblue}{rgb}{0.36, 0.54, 0.66}
\definecolor{cadmiumgreen}{rgb}{0.0, 0.42, 0.24}
\definecolor{darkcyan}{rgb}{0.0, 0.55, 0.55}
\definecolor{caribbeangreen}{rgb}{0.0, 0.8, 0.6}
\definecolor{flamingopink}{rgb}{0.99, 0.56, 0.67}
\definecolor{jazzberryjam}{rgb}{0.65, 0.04, 0.37}
\definecolor{mediumpersianblue}{rgb}{0.0, 0.4, 0.65}
\definecolor{coolblack}{rgb}{0.0, 0.18, 0.39}
\definecolor{bleudefrance}{rgb}{0.19, 0.55, 0.91}
\definecolor{ao}{rgb}{0.0, 0.0, 1.0}
\definecolor{babyblueeyes}{rgb}{0.63, 0.79, 0.95}
\definecolor{darkwarmgray}{rgb}{0.2,0,0}
\definecolor{brightpink}{rgb}{1.0, 0.0, 0.5}
\definecolor{iy}{rgb}{0.0, 0.36, 0.05}

\newcommand{\squishlist}{
 \begin{list}{$\circ$}
  { \setlength{\itemsep}{0pt}
     \setlength{\parsep}{0pt}
     \setlength{\topsep}{0pt}
     \setlength{\partopsep}{0pt}
     \setlength{\leftmargin}{1em}
     \setlength{\labelwidth}{1em}
     \setlength{\labelsep}{0.5em} } }

\newcommand{\squishsublist}{
\begin{list}{$\rightarrow$}
 { \setlength{\itemsep}{0pt}
    \setlength{\parsep}{0pt}
    \setlength{\topsep}{-10em}
    \setlength{\partopsep}{-3pt}
    \setlength{\leftmargin}{1em}
    \setlength{\labelwidth}{1em}
    \setlength{\labelsep}{0.5em} } }

\newcommand{\squishend}{
  \end{list}  }

\newcommand{\head}[1]{\noindent\textbf{#1.}}

\newcounter{take}
\ifdraft
\paperwidth=\dimexpr \paperwidth + 4cm\relax
\oddsidemargin=\dimexpr\oddsidemargin + 2cm\relax
\evensidemargin=\dimexpr\evensidemargin + 2cm\relax
\marginparwidth=\dimexpr \marginparwidth + 2cm\relax
\fi
\newcommand{\gf}[2]{\ifnum#1=\value{version}\textcolor{red}{#2}\else{#2}\fi}
\newcommand{\agy}[2]{\ifnum#1=\value{version}\textcolor{blue}{#2}\else{#2}\fi}
\newcommand{\atb}[2]{\ifnum#1=\value{version}\textcolor{orange}{#2}\else{#2}\fi}
\newcommand{\yct}[2]{\ifnum#1=\value{version}\textcolor{yt}{#2}\else{#2}\fi}
\newcommand{\ous}[2]{\ifnum#1=\value{version}\textcolor{RoyalBlue3}{#2}\else{#2}\fi}
\newcommand{\iey}[2]{\ifnum#1=\value{version}\textcolor{iy}{#2}\else{#2}\fi}
\newcommand{\om}[2]{\ifnum#1=\value{version}\textcolor{gfored}{#2}\else#2\fi}

\newcommand{\agytodo}[2]{\ifnum#1=\value{version}\todo[size=\scriptsize, linecolor=orange, bordercolor=orange, backgroundcolor=white]{\textcolor{blue}{TODO: #2}}\else{}\fi}
\newcommand{\ycttodo}[2]{\ifnum#1=\value{version}\todo[size=\scriptsize, linecolor=orange, bordercolor=orange, backgroundcolor=white]{\textcolor{yt}{TODO: #2}}\else{}\fi}
\newcommand{\ieytodo}[2]{\ifnum#1=\value{version}\todo[size=\scriptsize, linecolor=orange, bordercolor=orange, backgroundcolor=white]{\textcolor{iey}{TODO: #2}}\else{}\fi}

\newcommand{\agycomment}[2]{\ifnum#1=\value{version}\todo[size=\scriptsize, linecolor=orange, bordercolor=orange, backgroundcolor=white]{\textcolor{blue}{Giray: #2}}\else{}\fi}
\newcommand{\atbcomment}[2]{\ifnum#1=\value{version}\todo[size=\scriptsize, linecolor=orange, bordercolor=orange, backgroundcolor=white]{\textcolor{orange}{Atb: #2}}\else{}\fi}
\newcommand{\yctcomment}[2]{\ifnum#1=\value{version}\todo[size=\scriptsize, linecolor=orange, bordercolor=orange, backgroundcolor=white]{\textcolor{yt}{Yahya: #2}}\else{}\fi}
\newcommand{\ouscomment}[2]{\ifnum#1=\value{version}\todo[size=\scriptsize, linecolor=orange, bordercolor=orange, backgroundcolor=white]{\textcolor{RoyalBlue3}{Oguzhan: #2}}\else{}\fi}
\newcommand{\ieycomment}[2]{\ifnum#1=\value{version}\todo[size=\scriptsize, linecolor=orange, bordercolor=orange, backgroundcolor=white]{\textcolor{iy}{Ismail: #2}}\else{}\fi}
\newcommand{\omcomment}[2]{\ifnum#1=\value{version}\todo[size=\scriptsize, linecolor=orange, bordercolor=orange, backgroundcolor=white]{\textcolor{gfored}{Onur: #2}}\else{}\fi}
\newcommand{\versionedparam}[2]{\ifnum#1=\value{version}{#2}\else{#2}\fi}
\providecommand{\param}[1]{\versionedparam{\value{version}}{#1}}

\newcommand{\X}[0]{BreakHammer}
\newcommand{\Xshort}[0]{BH}

\newcommand{\secref}[1]{§\ref{#1}}

\newcommand{\tabref}[1]{Table~\ref{#1}}
\newcommand{\algref}[1]{Alg.~\ref{#1}}
\newcommand{\figref}[1]{Fig.~\ref{#1}}
\newcommand{\figsref}[1]{Figs.~\ref{#1}}
\newcommand{\expref}[1]{Expression~\ref{#1}}

\newcommand{\revtag}[1]{}

\newcommand{\rhmemisolationrefs}[0]{\cite{fournaris2017exploiting, poddebniak2018attacking, tatar2018throwhammer, carre2018openssl, barenghi2018software, zhang2018triggering, bhattacharya2018advanced, google-project-zero, kim2014flipping, rowhammergithub, seaborn2015exploiting, van2016drammer, gruss2016rowhammer, razavi2016flip, pessl2016drama, xiao2016one, bosman2016dedup, bhattacharya2016curious, burleson2016invited, qiao2016new, brasser2017can, jang2017sgx, aga2017good, mutlu2017rowhammer, tatar2018defeating, gruss2018another, lipp2018nethammer, van2018guardion, frigo2018grand, cojocar2019eccploit,  ji2019pinpoint, mutlu2019rowhammer, hong2019terminal, kwong2020rambleed, frigo2020trrespass, cojocar2020rowhammer, weissman2020jackhammer, zhang2020pthammer, yao2020deephammer, deridder2021smash, hassan2021utrr, jattke2022blacksmith, tol2022toward, kogler2022half, orosa2022spyhammer, zhang2022implicit, liu2022generating, cohen2022hammerscope, zheng2022trojvit, fahr2022frodo, tobah2022spechammer, rakin2022deepsteal, park2016statistical, park2016experiments,lim2017active, ryu2017overcoming, yun2018study, yang2019trap, walker2021ondramrowhammer, kim2020revisiting, orosa2021deeper, yaglikci2022understanding, khan2018analysis, agarwal2018rowhammer, li2014write, ni2018write, genssler2022reliability, mutlu2023fundamentally, tol2023dont}}

\newcommand{\exploitingRowHammerAllCitations}[0]{\cite{fournaris2017exploiting, poddebniak2018attacking, tatar2018throwhammer, carre2018openssl, barenghi2018software, zhang2018triggering, bhattacharya2018advanced, google-project-zero, kim2014flipping, rowhammergithub, seaborn2015exploiting, van2016drammer, gruss2016rowhammer, razavi2016flip, pessl2016drama, xiao2016one, bosman2016dedup, bhattacharya2016curious, burleson2016invited, qiao2016new, brasser2017can, jang2017sgx, aga2017good, mutlu2017rowhammer, tatar2018defeating, gruss2018another, lipp2018nethammer, van2018guardion, frigo2018grand, cojocar2019eccploit,  ji2019pinpoint, mutlu2019rowhammer, hong2019terminal, kwong2020rambleed, frigo2020trrespass, cojocar2020rowhammer, weissman2020jackhammer, zhang2020pthammer, yao2020deephammer, deridder2021smash, hassan2021utrr, jattke2022blacksmith, tol2022toward, kogler2022half, orosa2022spyhammer, zhang2022implicit, liu2022generating, cohen2022hammerscope, zheng2022trojvit, fahr2022frodo, tobah2022spechammer, rakin2022deepsteal, aydin2022cyber, mus2022jolt, wang2022research, lefforge2023reverse,fahr2022effects, kaur2022work, cai2022feasibility, li2022cyberradar, roohi2022efficient, staudigl2022neurohammer, yang2022socially, islam2022signature, france2022modeling, kurmus2017from, li2023fphammer, tol2023dont, tomita2022extracting}}

\newcommand{\understandingRowHammerAllCitations}[0]{\cite{redeker2002investigation, kim2014flipping, park2014active, park2016statistical, yang2016suppression, park2016experiments,lim2017active, ryu2017overcoming, yang2017scanning, lim2018study, yun2018study, yang2019trap, gautam2019rowhammering, walker2021ondramrowhammer, kim2020revisiting, orosa2021deeper, jiang2021quantifying, orosa2022spyhammer, cohen2022hammerscope, yaglikci2022understanding, khan2018analysis, agarwal2018rowhammer, li2014write, ni2018write, genssler2022reliability, mutlu2023fundamentally, he2023whistleblower, baeg2022estimation, frigo2020trrespass, mutlu2017rowhammer, mutlu2018rowhammer, mutlu2019rowhammer, olgun2023hbm, olgun2023drambender, zhou2023double, luo2023rowpress, lang2023blaster, li2024understanding, zhou2024understanding, zhou2024unveiling}}

\newcommand{\mitigatingRowHammerAllCitations}[0]{\cite{apple2015about, hp2015rowhammer, lenovo2015rowhammer, greenfield2012throttling, kim2014flipping, kim2014architectural, jedec2017ddr4, aichinger2015ddr, aweke2016anvil, bains-merged, bains2015row, bains2016distributed, bains2016row, gomez2016dummy, yang2016suppression, son2017making, seyedzadeh2018mitigating, irazoqui2016mascat, ryu2017overcoming, yang2017scanning, you2019mrloc, lee2019twice, park2020graphene, yaglikci2021security, yaglikci2021blockhammer, frigo2020trrespass, kang2020cattwo, hassan2021utrr, qureshi2022hydra, saileshwar2022randomized, brasser2017can, konoth2018zebram, vig2018rapid, hassan2019crow, gautam2019rowhammering, kim2022mithril, lee2021cryoguard, marazzi2022protrr, zhang2022softtrr, joardar2022learning, juffinger2023csi, yaglikci2022hira, saxena2022aqua, manzhosov2022revisiting, ajorpaz2022evax, naseredini2022alarm, joardar2022machine, hassan2022case, zhang2020leveraging,loughlin2021stop, devaux2021method, han2021surround, fakhrzadehgan2022safeguard, saroiu2022price, saroiu2022configure, loughlin2022moesiprime, zhou2022lt, hong2023dsac, mutlu2023fundamentally, marazzi2023rega, di2023copy, sharma2022review, woo2023scalable, park2022row, wi2023shadow, kim2023a11v, gude2023defending, guha2022criticality, france2022modeling, france2022reducing, bennett2021panopticon, enomoto2022efficient, arikan2022processor, tomita2022extracting, saxena2023pt, zhou2023dnndefender, bostanci2024comet, didio2023copyonflip, jedec2024jesd795c, seyedzadeh2017counterbased, seyedzadeh2017mitigating, van2018guardion, van2024dram, van2024dynamic, verma2023defense, olgun2024abacus}}

\newcommand{\refreshBasedRowHammerDefenseCitations}[0]{\cite{lee2019twice, seyedzadeh2017cbt, seyedzadeh2018mitigating, kang2020cattwo, park2020graphene, kim2022mithril, kim2014architectural, bains2015row, bains2016distributed, bains2016row, aweke2016anvil, apple2015about, kim2014flipping, son2017making, you2019mrloc, yaglikci2021security, frigo2020trrespass, hassan2021utrr, qureshi2022hydra, devaux2021method, lee2021cryoguard, marazzi2022protrr, zhang2022softtrr, joardar2022learning}}

\makeatletter
\def\bstctlcite{\@ifnextchar[{\@bstctlcite}{\@bstctlcite[@auxout]}}
\def\@bstctlcite[#1]#2{\@bsphack
 \@for\@citeb:=#2\do{%
   \edef\@citeb{\expandafter\@firstofone\@citeb}%
   \if@filesw\immediate\write\csname #1\endcsname{\string\citation{\@citeb}}\fi}%
 \@esphack}
\makeatother

\newcommand{\artifactbadges}{
  \AtBeginShipoutNext{\AtBeginShipoutUpperLeft{%
    \raisebox{-2cm}[0pt][0pt]{%
      \hspace*{\dimexpr\paperwidth-8.5cm\relax}
      \href{https://www.acm.org/publications/policies/artifact-review-and-badging-current}{\includegraphics[width=1.75cm]{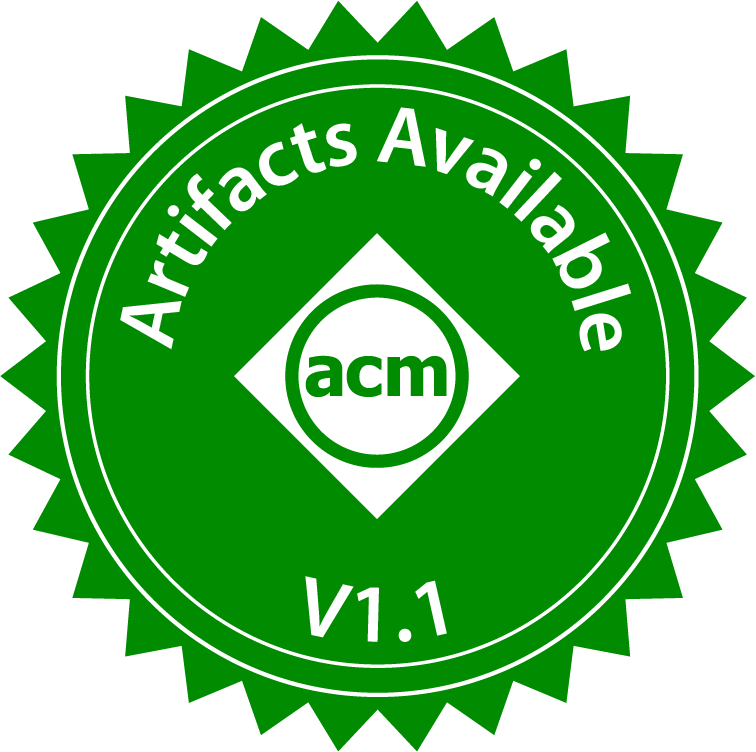}}%
      \href{https://www.acm.org/publications/policies/artifact-review-and-badging-current}{\includegraphics[width=1.75cm]{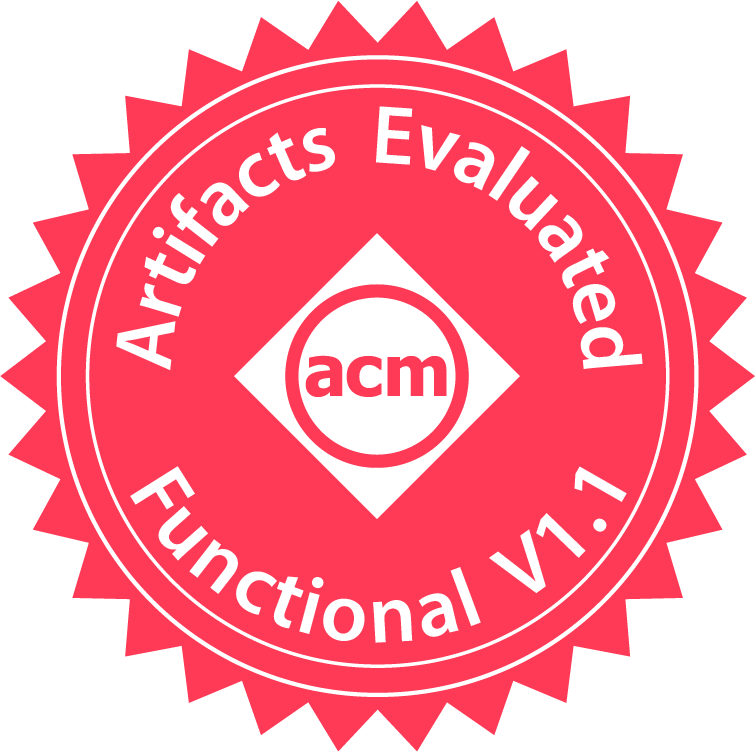}}%
      \href{https://www.acm.org/publications/policies/artifact-review-and-badging-current}{\includegraphics[width=1.75cm]{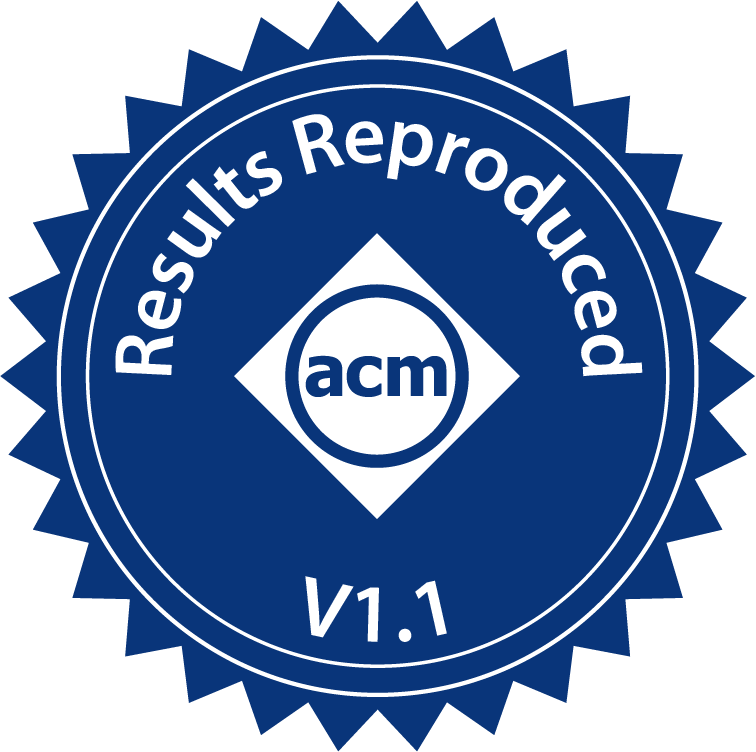}}%
      \href{https://www.acm.org/publications/policies/artifact-review-and-badging-current}{\includegraphics[width=1.75cm]{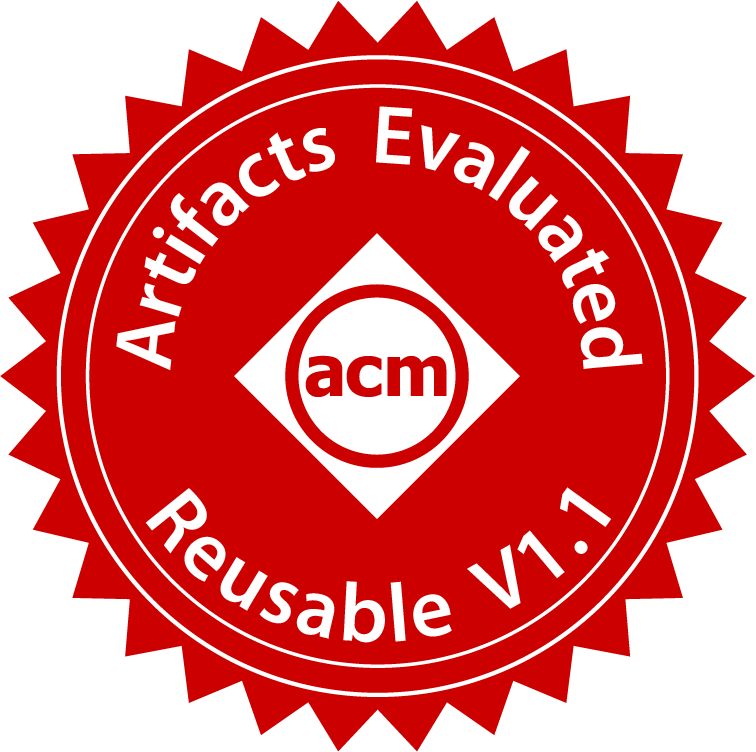}}%
    }}}%
}

\newcommand{\nrh}[0]{N_{RH}}
\newacronym{nrh}{$\nrh{}$}{\emph{RowHammer threshold}}

\newcommand{\rsatki}[0]{RS_{\text{atk}i}}
\newcommand{\rsatkmax}[0]{RS_{\text{atk}}^{\text{max}}}
\newcommand{\rsbeni}[0]{RS_{\text{ben}i}}
\newcommand{\rsbenavg}[0]{RS_{\text{ben}}^{\text{avg}}}

\newcommand{\trcd}[0]{t_{RCD}}
\newacronym{trcd}{$\trcd{}$}{\emph{row activation latency}}
\newcommand{\tras}[0]{t_{RAS}}
\newacronym{tras}{$\tras{}$}{\emph{charge restoration latency}}
\newcommand{\trp}[0]{t_{RP}}
\newacronym{trp}{$\trp{}$}{\emph{precharge latency}}
\newcommand{\trc}[0]{t_{RC}}
\newacronym{trc}{$\trc{}$}{\emph{the minimum time window between two row activation commands targeting different banks}}
\newcommand{\trefi}[0]{t_{REFI}}
\newacronym{trefi}{$\trefi$}{\emph{refresh interval}}
\newcommand{\trefw}[0]{t_{REFW}}
\newacronym{trefw}{$\trefw$}{\emph{refresh window}}
\newcommand{\trfc}[0]{t_{RFC}}
\newacronym{trfc}{$\trfc{}$}{\emph{refresh latency}}
\newcommand{\trrd}[0]{t_{RRD}}
\newacronym{trrd}{$\trrd{}$}{\emph{the minimum time window between two row activation commands targeting different banks}}

\newcommand{\rfmth}[0]{RFM_{th}}
\newacronym{rfmth}{$\rfmth{}$}{\emph{bank activation threshold to issue an RFM command}}

\newcommand{\tfilter}[0]{T_{Filter}}
\newacronym{tfilter}{$\tfilter{}$}{\emph{the filtering threshold}}

\newcommand{\act}[0]{ACT}
\newcommand{\pre}[0]{PRE}
\newcommand{\refresh}[0]{REF}
\newcommand{\wri}[0]{\texttt{WR}}
\newcommand{\rd}[0]{\texttt{RD}}

\newcommand{\apa}[0]{\texttt{APA}}

\newcommand{\vddh}[0]{\texttt{VDD/2}}

\newcommand{\pum}[0]{\texttt{PuM}}

\usepackage[shortcuts]{extdash}

\newcounter{passversion}
\definecolor{midnightblue}{rgb}{0.1, 0.1, 0.44}
\DTMnewtimestyle{ampm}{%
  \renewcommand{\DTMdisplaytime}[3]{%
    \def\THEHOUR{##1}\def\THEAMPM{am}
    \ifnum##1>12
      \edef\THEHOUR{\number\numexpr##1-12}
      \edef\THEAMPM{pm}
    \fi
    \THEHOUR
    .%
    \DTMtwodigits{##2}
    \THEAMPM
  }%
}
\DTMsettimestyle{ampm}

\title{\huge \X{}: Enhancing RowHammer Mitigations\\by Carefully Throttling Suspect Threads}

\def\BibTeX{{\rm B\kern-.05em{\sc i\kern-.025em b}\kern-.08em
    T\kern-.1667em\lower.7ex\hbox{E}\kern-.125emX}}

\begin{document}
\artifactbadges
\maketitle
\pagestyle{plain}

\begin{abstract}

RowHammer is a \om{4}{major} read disturbance mechanism in DRAM where repeatedly accessing (hammering) a row of DRAM cells (DRAM row) induces bitflips in other physically nearby DRAM rows. 
RowHammer solutions perform preventive actions (e.g., refresh neighbor rows of the hammered row) that mitigate such bitflips to preserve \emph{memory isolation}, a fundamental building block of security and privacy in modern computing systems.
However, preventive actions induce non-negligible memory request latency and system performance overheads as they interfere with memory requests.
As shrinking technology node size over DRAM chip generations exacerbates RowHammer, the overheads of RowHammer solutions become prohibitively expensive.
As a result, a malicious program can effectively hog the memory system and deny service to benign applications by causing many RowHammer-preventive actions.

In this work, we tackle the performance overheads of RowHammer solutions by tracking and throttling the generators of memory accesses that trigger RowHammer solutions.
To this end, we propose \X{}.
\X{} 1) observes the time-consuming RowHammer-preventive actions of existing RowHammer mitigation mechanisms,
2) identifies \ous{4}{hardware threads that trigger} many of these actions, and
3) reduces the memory bandwidth usage of \ous{4}{each} identified thread.
As such, \X{} significantly reduces the number of RowHammer-preventive actions performed, thereby improving 1) system performance and DRAM energy, and 2) reducing the maximum slowdown induced on a benign application, with near-zero area overhead.
Our extensive evaluations demonstrate that \X{} effectively reduces the \om{5}{negative performance, energy, and fairness effects} of \param{eight}\omcomment{5}{eight? seven?}\ouscomment{5}{we combine with eight (<- excludes BlockHammer) and compare against BlockHammer} RowHammer mitigation mechanisms.
To foster further research we open-source our \X{} implementation and scripts at \url{https://github.com/CMU-SAFARI/BreakHammer}.

\end{abstract}
\vspace{-0.5em}
\section{Introduction}
\label{sec:introduction}

To ensure system \agy{0}{robustness (i.e., reliability, security, and safety)}, it is critical to maintain memory isolation: accessing a memory address should \emph{not} cause unintended side-effects on data stored on other addresses.
Unfortunately, with aggressive technology node scaling, dynamic random access memory (DRAM)~\cite{dennard1968dram}, the prevalent {main} memory technology, suffers from increased read disturbance:
accessing (reading) a DRAM cell degrades the data integrity of other physically close \om{3}{yet} \emph{unaccessed} DRAM cells by exacerbating the unaccessed cells' inherent charge leakage and \agy{0}{consequently} causing bitflips.
\emph{RowHammer}~\cite{kim2014flipping,mutlu2017rowhammer,mutlu2019rowhammer,mutlu2023fundamentally} is a prime example of such DRAM read disturbance \om{3}{phenomena} where a DRAM row (i.e., victim row) can experience bitflips when a nearby DRAM row (i.e., aggressor row) is repeatedly opened (i.e., hammered)~\rhmemisolationrefs{}.

Many prior works demonstrate attacks on a wide range of systems where they exploit read disturbance to escalate privilege, leak private data, and manipulate critical application outputs~\exploitingRowHammerAllCitations{}.
Recent experimental studies~\cite{mutlu2017rowhammer, mutlu2019rowhammer, frigo2020trrespass, cojocar2020rowhammer, kim2020revisiting, kim2014flipping, luo2023rowpress} \om{3}{find} that newer DRAM chip generations are more susceptible to read disturbance.
For example, chips manufactured in 2018-2020 can experience RowHammer bitflips after an order of magnitude fewer row activations compared to the chips manufactured in 2012-2013~\cite{kim2020revisiting}.
As RowHammer worsens, ensuring robust (i.e., reliable, secure, and safe) operation causes prohibitively large performance overheads~\cite{kim2020revisiting, yaglikci2021blockhammer, park2020graphene, yaglikci2022hira}.
Although RowHammer mitigation mechanisms can mitigate RowHammer bitflips, they aggressively and \om{3}{sometimes} excessively perform time- and energy-hungry operations to mitigate RowHammer (i.e., \agy{0}{RowHammer-preventive} actions) as fewer DRAM row activations can induce RowHammer bitflips in newer DRAM chips.
\om{3}{In fact}, adversarial memory access patterns can be crafted to trigger RowHammer mitigation mechanisms more frequently.
Thus, RowHammer mitigation mechanisms provide data integrity at the cost of \ous{3}{DRAM bandwidth availability} by incurring large performance overheads.
Therefore, reducing the performance overhead of such mechanisms is critical to provide data integrity without reducing \ous{3}{DRAM bandwidth availability}.

\head{Goal}
Our goal is to reduce the performance overhead of RowHammer mitigation mechanisms by \om{3}{carefully reducing} the number of performed RowHammer-preventive actions \emph{without} compromising system robustness.

\head{Key Idea}
\ous{3}{Our key idea is to limit the \ous{3}{dynamic} memory request count of a hardware thread based on how frequently the thread triggers RowHammer-preventive actions}.

\head{Key Mechanism}
We propose \agy{3}{a new \agy{3}{memory controller-based throttling} mechanism}, \X{}.\footnote{BreakHammer is analogous to breakwater, which \om{3}{is a structure that protects} docks against tides, waves, and storm surges; and consequently reduces their maintenance costs~\cite{de2022ancient}. Similarly, our mechanism, BreakHammer, protects the memory subsystem against hammering access patterns and, by doing so, reduces the overheads of RowHammer mitigation mechanisms.}
\om{3}{Cooperating with an existing memory controller-based or on-DRAM-die RowHammer mitigation mechanism (e.g., \cite{park2020graphene,kim2014flipping,saxena2022aqua,qureshi2022hydra,lee2018twice,marazzi2023rega,jedec2020ddr5,jedec2024jesd795c}) \om{4}{as we show}}, \X{}
1)~observes the triggered RowHammer-preventive actions,
2)~identifies \ous{4}{hardware threads that trigger} many RowHammer-preventive actions (we call these threads \emph{suspect threads}), and
3)~reduces the dynamic memory request count of \ous{4}{each} suspect thread.
Meanwhile, the existing RowHammer mitigation mechanism operates as usual and maintains security guarantees against RowHammer attacks.
\om{3}{As such}, \X{} significantly reduces the number of necessary RowHammer-preventive actions by throttling the malicious threads' memory requests \emph{without} compromising \om{3}{a} RowHammer mitigation mechanism's security guarantees.
To achieve this, \X{} divides execution \ous{5}{timeline} into equal periods that we call \emph{throttling windows} and in each throttling window performs \param{three} key operations when a RowHammer-preventive action is triggered.

First, \X{} attributes \ous{3}{a score} (i.e., \emph{RowHammer-preventive score}) to \ous{4}{hardware threads that cause} RowHammer-preventive actions that interfere with other memory accesses, e.g., blocking a DRAM bank due to refreshing victim rows.
For this purpose, \X{} maintains a \emph{RowHammer-preventive score counter} per hardware thread.
To showcase \X{}'s compatibility with various types of existing RowHammer mitigation mechanisms, we implement \X{} with \param{five} memory controller-based~\cite{park2020graphene,kim2014flipping,saxena2022aqua,qureshi2022hydra,lee2018twice} and \param{three} on-DRAM-die~\cite{marazzi2023rega,jedec2020ddr5,jedec2024jesd795c} state-of-the-art RowHammer mitigation mechanisms that we summarize under \param{five} categories.
1)~PARA~\cite{kim2014flipping}, Graphene~\cite{park2020graphene}, Hydra~\cite{qureshi2022hydra}, and TwiCe~\cite{lee2019twice} perform RowHammer-preventive refresh operations,
2)~AQUA~\cite{saxena2022aqua} migrates the contents of DRAM rows to a quarantine area within DRAM,
3)~REGA~\cite{marazzi2023rega} modifies the DRAM chip with a secondary row buffer to refresh multiple rows in parallel,
4)~\emph{Periodic Refresh Management} (RFM)~\cite{jedec2020jesd795} requires the memory controller to issue refresh management (RFM) commands when a \ous{0}{bank's} activation count exceeds a threshold, and
5)~\emph{Per Row Activation Counting} (PRAC)~\cite{jedec2024jesd795c} uses the \emph{alert\_n} signal to request a predetermined number of RFM commands when a row's activation count exceeds a threshold.

Second, \X{} identifies \ous{4}{suspect threads} using outlier analysis.
To do so, \X{} marks a hardware thread as suspect if the thread's RowHammer-preventive score both 1) exceeds a threshold and 2) \om{4}{largely} deviates from the average RowHammer-preventive score across all \om{4}{threads} in the system.
Third, \X{} reduces the memory bandwidth usage of \ous{4}{each} suspect thread until it is \emph{not} identified as a suspect in a future throttling window.
To \om{4}{this} end, \X{} limits the number of \emph{cache-miss buffers} (MSHRs~\cite{kroft1981lockup}) that track the last level cache (LLC) misses a suspect thread can allocate.

\head{\agy{3}{Security}}
\ous{3}{We mathematically evaluate \agy{3}{the impact of} a worst-case \agy{3}{memory performance attack for BreakHammer, where the} attacker operates \agy{3}{near} \X{}'s outlier detection threshold \agy{3}{\emph{without} being detected as an outlier by \X{}}.
Our security analysis shows that \X{} enforces a \agy{3}{theoretical} upper bound for the RowHammer-preventive \ous{3}{score} an attacker can \om{4}{gather} based on 1) the RowHammer-preventive \ous{3}{score} of benign threads and 2) the number of hardware threads \ous{3}{used} by the attacker.
For example, we show that an attacker \emph{cannot} trigger twice the RowHammer-preventive action count of concurrently running benign applications unless the attacker \ous{3}{uses 90\%} of all hardware threads.}

\head{\ous{3}{Hardware Implementation}}
\ous{3}{We model \X{}'s hardware design (RTL) in Chisel~\cite{bachrach2012chisel} and evaluate its latency and area overhead using Synopsys Design Compiler~\cite{synopsys} (65nm process technology) and CACTI~\cite{cacti}}.
Our \ous{3}{hardware complexity evaluation shows} that \X{} can be implemented off the critical path of memory \om{4}{request scheduling} because \X{} has a latency of \param{0.67ns}, which is lower than \gls{trrd} (e.g., 2.5ns in DDR4~\cite{jedec2017ddr4}) and incurs near-zero area overhead (i.e., 0.00042$mm^2$), which consumes 0.0002\% of a high-end Intel Xeon processor's chip area.

\head{\ous{3}{Key Results}}
We \ous{3}{rigorously} evaluate \ous{3}{\X{}'s impact on performance and DRAM energy using Ramulator 2.0~\cite{luo2023ramulator2,ramulator2github}, an open-source cycle-level simulator.
We execute
1)~\param{90} four-core workloads where one malicious application \om{4}{mounts} a memory performance attack by triggering many RowHammer-preventive actions and
2)~\param{90} four-core workloads where all applications are benign.
We select a diverse set of benign workloads from SPEC CPU2006~\cite{spec2006}, SPEC CPU2017~\cite{spec2017}, TPC~\cite{tpc}, MediaBench~\cite{fritts2009media}, and YCSB~\cite{ycsb} benchmark suites.}
Our evaluation shows that \X{} significantly improves system performance and \om{4}{reduces} DRAM energy (e.g., by \ous{3}{\param{90.1\%} and \param{55.7}\%}, respectively, on average across all tested workload mixes with a malicious application).

We make the following contributions:

\begin{itemize}
[noitemsep,topsep=0pt,parsep=0pt,partopsep=0pt,labelindent=0pt,itemindent=0pt,leftmargin=*]
    \item {We show that \om{3}{state-of-the-art} RowHammer mitigation mechanisms \agy{0}{significantly reduce \ous{3}{DRAM bandwidth availability}}.}
    
    \item {We \om{3}{introduce the idea that} it is possible to \ous{3}{1) mount memory performance attacks by exploiting the RowHammer-preventive actions of RowHammer mitigation mechanisms and 2)} reduce the performance overheads of RowHammer mitigation mechanisms by observing the time-consuming RowHammer-preventive actions and throttling the threads that trigger \agy{0}{such} actions.}
    
    \item {We propose \X{}, a memory controller-based low-cost mechanism that provides both memory controller-based and on-DRAM-die RowHammer mitigation mechanisms with throttling support \ous{0}{to 1)~reduce their \om{3}{performance and energy} overheads and 2)~prevent their actions from being \ous{1}{exploited} to reduce \ous{3}{DRAM bandwidth availability}}.}

    \item {We \om{3}{rigorously} evaluate \X{} and show that it significantly reduces the performance overhead of \ous{3}{eight state-of-the-art} RowHammer mitigation mechanisms without compromising \agy{0}{robustness} at near-zero area overhead.}

    \item {To foster further research \om{4}{in the design of more robust and high-performance} systems, we open-source our implementations at \url{https://github.com/CMU-SAFARI/BreakHammer}.}
\end{itemize}
\section{Background}
\label{sec:background}
\subsection{DRAM Organization and Operation}
\label{sec:dram_organization}

\begin{figure}[ht]
    \centering
    \includegraphics[width=\linewidth]{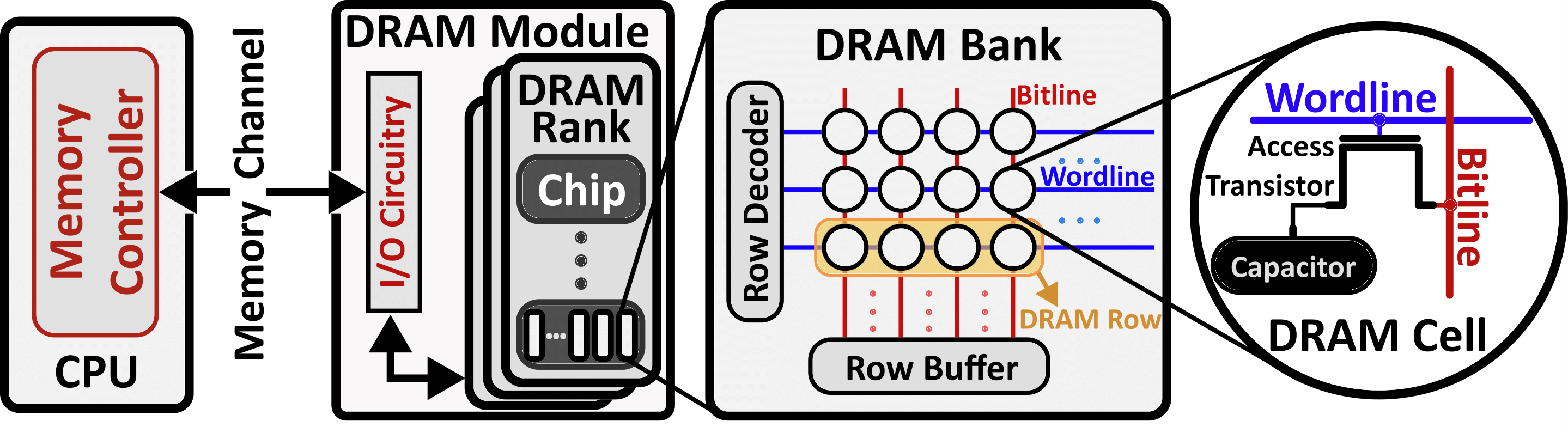}
    \caption{DRAM organization}
    \label{fig:dram_organization}
\end{figure}

\head{Organization}\ouscomment{1}{Copied from ABACUS}
\ous{1}{\figref{fig:dram_organization} shows the organization of \om{3}{a} DRAM-based memory system.
A memory channel connects the processor (CPU) to a set of DRAM chips.
This set of DRAM chips forms one or more DRAM ranks \om{3}{where each rank operates} in lockstep.
Each chip has multiple DRAM banks, where DRAM cells are organized as a two-dimensional array of DRAM rows and columns.
A DRAM cell is connected to the row buffer via a wire called bitline.
The DRAM cell stores one bit of data in the form of electrical charge in a capacitor.
The access transistor, controlled by the wordline, connects the cell to the bitline.}

\head{Operation}
\agy{0}{DRAM cells are internally accessed \om{3}{at} DRAM row granularity}.
To access a DRAM row, the memory controller issues a set of commands to DRAM over the memory channel.
The memory controller sends an \ous{3}{activate ($ACT$)} command to activate a DRAM row, which asserts the corresponding wordline and loads the row data into the row buffer.
Then, the memory controller can issue \ous{3}{read ($RD$)/write ($WR$)} commands to read from/write into the \agy{0}{row buffer}.
Subsequent accesses to the same \agy{0}{DRAM} row cause a row \agy{0}{buffer} hit.
\agy{1}{Accessing a different row causes a row buffer conflict.
Therefore}, to access a different \agy{0}{DRAM} row, the memory controller must first close the \agy{0}{open row} by issuing a \ous{3}{precharge ($PRE$)}  command \agy{1}{and open the row that contains the accessed data}.

\head{DRAM Refresh}
\om{3}{A} DRAM cell \om{3}{is} inherently leaky and \om{3}{loses its} charge over time due to charge leakage in the access transistor and the storage capacitor.
\om{3}{To} maintain data integrity, the memory controller periodically refreshes each row \ous{3}{every \gls{trefw}, which is} $32 ms$ for DDR5~\cite{jedec2020ddr5} and $64 ms$ for DDR4~\cite{jedec2017ddr4} \om{3}{(under normal operating temperature ranges)}.
To \ous{3}{timely refresh all cells}, the memory controller issues \ous{3}{a refresh ($REF$)} command \ous{3}{every \gls{trefi}, which is} $3.9 \mu s$ for DDR5~\cite{jedec2020ddr5} and $7.8 \mu s$ for DDR4~\cite{jedec2017ddr4}.


\subsection{DRAM Read Disturbance}
As DRAM manufacturing technology node size shrinks, interference \om{3}{between cells} increases, causing circuit-level read disturbance mechanisms.
Two prime examples of such read disturbance mechanisms are RowHammer~\cite{kim2014flipping,mutlu2023fundamentally,mutlu2019retrospective,mutlu2017rowhammer} and RowPress~\cite{luo2023rowpress,luo2024rowpress}, where repeatedly activating (i.e., opening) a DRAM row (i.e., aggressor row) or keeping the aggressor row active for a long time induces bitflips in physically nearby rows (i.e., victim rows), respectively.
To induce RowHammer bitflips, an aggressor row needs to be activated more \om{3}{times} than a threshold value called \gls{nrh}. 
\om{4}{Various} characterization studies~\understandingRowHammerAllCitations{} show that as DRAM \om{3}{scales} to smaller \om{3}{technology} nodes, DRAM chips \om{3}{become} more vulnerable to RowHammer (i.e., newer chips have lower \gls{nrh} values).
For example, \ous{3}{chips manufactured in 2018-2020 can experience bitflips at an order of magnitude fewer row activations than chips manufactured in 2012-2013~\cite{kim2020revisiting}}.
To make matters worse, RowPress~\cite{luo2023rowpress, luo2024experimental} reduces \gls{nrh} significantly (e.g., by orders of magnitude).

\head{DRAM Read Disturbance Mitigation Mechanisms}
Many prior works propose mitigation techniques~\mitigatingRowHammerAllCitations{} to protect DRAM chips against RowHammer bitflips.
These \om{3}{techniques usually} perform two tasks: 1)~execute a trigger algorithm and 2)~perform RowHammer-preventive actions.
The \om{3}{\emph{trigger algorithm}} observes memory access patterns and triggers a \om{3}{\emph{RowHammer-preventive action}} based on the result of a probabilistic or a deterministic process.
RowHammer-preventive actions include one of
1)~preventively refreshing victim row~\refreshBasedRowHammerDefenseCitations{},
2)~dynamically remapping aggressor rows~\cite{saileshwar2022randomized, saxena2022aqua, wi2023shadow, woo2023scalable}, and
3)~throttling unsafe accesses~\cite{greenfield2012throttling, yaglikci2021blockhammer}.
Existing RowHammer mitigation mechanisms can also prevent RowPress bitflips when their trigger algorithms are configured to be more aggressive \ous{3}{by configuring them against relatively lower} \gls{nrh} values~\cite{luo2023rowpress}.
\section{Motivation}\omcomment{4}{is the evaluation here benign-only?}\ouscomment{4}{yes}
\label{sec:motivation}

Prior works~\cite{kim2020revisiting, park2020graphene, yaglikci2021blockhammer, luo2023ramulator2} show that existing RowHammer mitigation \om{3}{mechanisms} incur prohibitively high performance overheads as \om{3}{the RowHammer threshold (\gls{nrh})} decreases because \om{3}{these mechanisms} perform RowHammer-preventive actions more aggressively.
Given that DRAM read disturbance worsens with shrinking technology node size, \gls{nrh} \om{3}{is} expected to reduce even more, e.g., $< 1K$~\cite{luo2023rowpress,luo2024rowpress}.
Therefore, reducing the performance overhead of existing RowHammer mitigation mechanisms is critical.

\begin{figure}[h]
\centering
\includegraphics[width=1\linewidth]{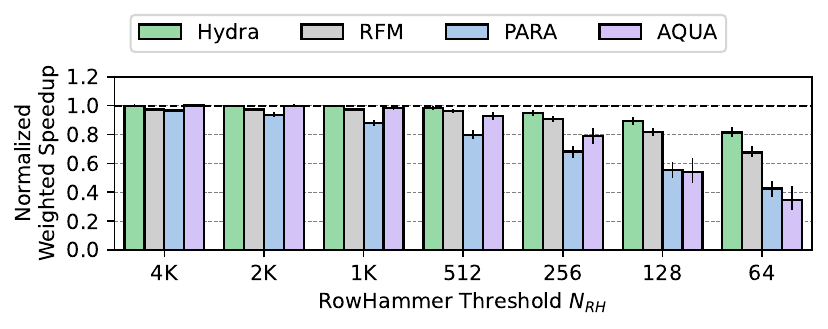}
\vspace{-16pt}\caption{System performance overheads of RowHammer mitigation mechanisms with worsening RowHammer vulnerability}
\label{fig:performance_motivation}
\end{figure}

We analyze the \om{3}{performance} impact of four state-of-the-art RowHammer mitigation mechanisms \om{3}{as \gls{nrh} decreases from \param{$4K$} to \param{$64$}}.\footnote{We conduct these simulations following the methodology described in \secref{sec:methodology} for \param{90} randomly chosen \ous{3}{four-core benign workloads from our benchmarks}.}
\figref{fig:performance_motivation}\omcomment{5}{Check figure placement and start final beautification} presents the system performance \ous{3}{(in terms of weighted speedup~\cite{eyerman2008systemlevel, snavely2000symbiotic})} of different RowHammer mitigation \om{3}{mechanisms}.\omcomment{5}{We motivate here with benign only but performance benefits are not high. Correct?}\ouscomment{5}{Correct.}
\ous{3}{The x- and y-axes show the \gls{nrh} values and system performance normalized to a baseline with no RowHammer mitigation mechanism, respectively}.
Different bars identify RowHammer mitigation mechanisms \ous{3}{and error bars mark the 100\% confidence interval across 90 workloads}.

We make \ous{1}{three} observations from \figref{fig:performance_motivation}.
First, \ous{2}{as \gls{nrh} decreases (i.e., DRAM chips get more vulnerable to read disturbance)}, system performance significantly reduces for all tested RowHammer mitigation mechanisms \om{3}{ranging from \param{18.7}\% \ous{3}{(Hydra)} to \param{65.9}\% \ous{3}{(AQUA)}}.
Second, at an \ous{3}{\gls{nrh} of $64$}, Hydra exhibits the least system performance overhead with a performance degradation of \param{18.7\%}.
Hydra achieves this at the expense of \ous{3}{1) storing a counter per row in DRAM and 2)} increasingly large \ous{3}{processor} chip area overhead that reaches \param{56.5KB} when configured for a dual rank system with 16 banks at each rank.
\ous{3}{Third, among the tested RowHammer mitigation mechanisms, PARA is the most lightweight and hardware-scalable RowHammer mitigation mechanism~\cite{yaglikci2022hira}, but it has a high system performance degradation of \param{57.6}\%.}
\ous{2}{We} conclude that RowHammer mitigation mechanisms become more expensive \ous{3}{in terms of area and performance} as \gls{nrh} decreases, and reducing their performance overheads at low cost is critical.\footnote{\ous{3}{This is \emph{not} a worst-case evaluation. There are worse workloads and system configurations that lead to higher overheads for \om{4}{\emph{all}} mechanisms (\secref{subsec:perfunderattack}).}}
\section{\X{}}
\label{sec:mechanism}

\X{} is the first mechanism that \om{4}{enhances} existing RowHammer mitigation mechanisms with throttling support to reduce their performance overheads.\footnote{BlockHammer~\cite{yaglikci2021blockhammer} proposes throttling as a mechanism to \emph{prevent} RowHammer bitflips.
On the other hand, \X{} uses throttling to reduce performance overheads of RowHammer mitigation mechanisms.}

\head{Overview}
\X{} is designed to cooperate with RowHammer mitigation mechanisms and reduce their performance overhead by reducing the number of performed RowHammer-preventive actions \emph{without} compromising system robustness.
\X{}'s key idea is to limit the \ous{3}{dynamic memory request count of a hardware thread that repeatedly triggers RowHammer-preventive actions.
\X{} splits the execution timeline into periods of equal length that we call \emph{throttling windows} (similar to a refresh window~\cite{jedec2017ddr4})}.\ouscomment{4}{multiple threads can be identified as suspect in a window, fixing everywhere}
\ous{3}{During each throttling window}, \X{} performs \param{three} key operations: \om{3}{it}
1)~observes the time-consuming RowHammer-preventive actions of existing RowHammer mitigation mechanisms,
2)~identifies hardware \om{4}{threads} that trigger many of these actions (we call such threads \emph{suspect threads}), and
3)~reduces the memory bandwidth usage of \om{4}{each} suspect thread until it is \emph{not} identified as a suspect in a future throttling window.
We implement \X{} near the memory controller without \om{4}{requiring} proprietary information \om{4}{from} or modifications to the DRAM chips.
Therefore, \X{} is compatible with commodity DRAM chips.

\begin{figure}[ht]
\centering
\includegraphics[width=1\linewidth]{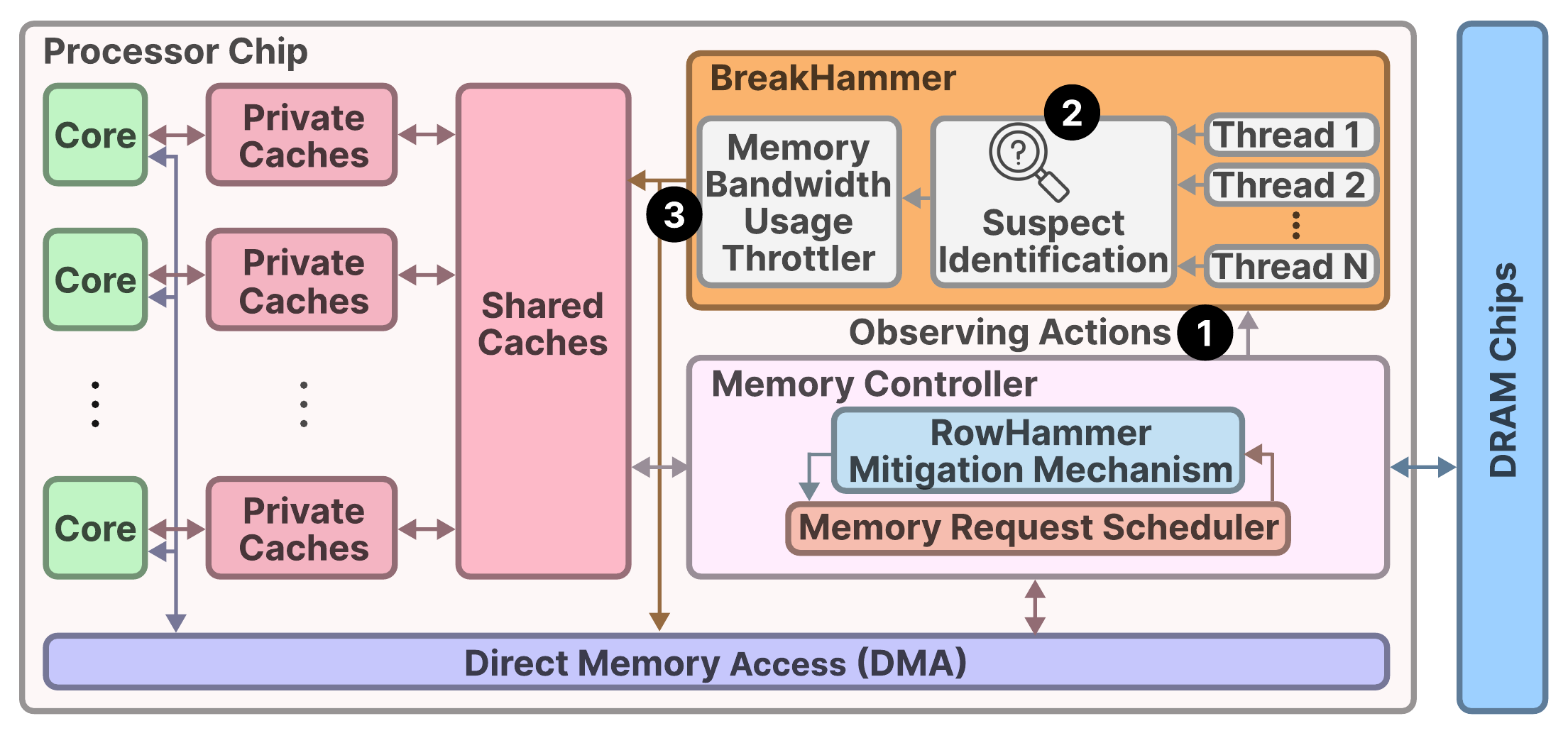}
\caption{\X{} high-level overview}
\label{fig:high_level_mechanism}
\end{figure}

\figref{fig:high_level_mechanism}\ouscomment{4}{Updated labels} depicts a high-level overview of \X{} in a simplified generic system with multiple processors accessing the main memory through the cache hierarchy or via \om{3}{a} direct memory access (DMA) unit~\cite{hennessy2017computer,lee2015decoupled}.
The memory controller implements a memory request scheduler to issue DRAM commands via the DRAM interface (e.g., DDR4~\cite{jedec2017ddr4}).
The memory controller employs a RowHammer mitigation mechanism \agy{3}{or} issues special DRAM commands (e.g., RFM~\cite{jedec2020ddr5} and PRAC~\cite{jedec2024jesd795c}) to provide an on-DRAM-die mitigation mechanism with a time window to perform RowHammer-preventive actions.

\X{} \ous{3}{divides the execution timeline into throttling windows of length $TH_{window}$ (e.g., 64ms) where it} performs \param{three} key operations \ous{3}{when the RowHammer mitigation mechanism triggers a RowHammer-preventive action}.

\head{\circled{1}~Observing RowHammer-Preventive Actions}
When a hardware thread \ous{5}{causes} a RowHammer-preventive action, \X{} increments a per-thread \om{3}{counter called} \emph{RowHammer-preventive score} for the thread.

\head{\circled{2}~\ous{3}{Identifying} Suspect Threads}
\X{} performs outlier analysis on the RowHammer-preventive scores to identify suspect threads.~\X{} identifies a hardware thread as a suspect when the thread's RowHammer-preventive score \ous{3}{1) exceeds a threshold and 2) \om{4}{largely} surpasses the average} RowHammer-preventive score \ous{3}{across} all threads.

\head{\circled{3}~Throttling Memory Bandwidth \om{4}{Usage} of Suspect Threads}
\X{} \ous{2}{limits} the \ous{3}{dynamic request count of a} suspect thread.
To do so, \X{} calculates a memory request resource (i.e., \om{4}{last level} cache-miss buffers~\cite{kroft1981lockup}) quota available to each suspect thread based on \ous{3}{the duration the thread has been a suspect for}.
By doing so, \X{} reduces the performance degradation the RowHammer mitigation mechanism causes by altering the memory access patterns such that the RowHammer mitigation mechanism performs fewer RowHammer-preventive actions.

\head{Optional Feedback to the System Software}
\X{} optionally exposes each hardware thread's RowHammer-preventive score counter to the system \om{4}{software}.
The system software can access these counters similarly to how it accesses \ous{3}{thread-specific special registers (e.g., CR3 in x86)}.
By doing so, \X{} supports associating scores with software threads, address spaces, processes, users, or other identifiers.

\subsection{Observing RowHammer-Preventive Actions}
\label{sec:observed_actions}

\X{} observes a hardware thread's performance overhead due to triggering RowHammer-preventive actions.
To do so, \X{} maintains a RowHammer-preventive score for each hardware thread, which tracks the number of RowHammer-preventive actions \ous{4}{caused by a thread}.
Since one \emph{cannot} directly attribute a RowHammer-preventive action to a thread, we update the RowHammer-preventive score \om{5}{of each thread} proportionally \om{5}{to} the fraction of \om{5}{all activations that thread is responsible for a given} RowHammer-preventive action \ous{5}{(activation count tracking is reset after a given RowHammer-preventive action)}.\omcomment{5}{Write better}
\om{4}{The exact method to update} RowHammer-preventive scores (i.e., score attribution \om{4}{method}) depends on the \om{4}{employed} RowHammer mitigation mechanism.
\om{4}{We} briefly describe the score attribution \om{4}{methods used} for the RowHammer mitigation mechanisms \om{4}{we evaluate}.

\head{PARA~\cite{kim2014flipping}}
PARA probabilistically performs a RowHammer-preventive refresh when the memory controller issues a row activation.
PARA is \emph{stateless}, i.e., does not keep track of any history or statistics.
\emph{Score \ous{3}{Attribution:}} \X{} 1)~tracks the \ous{5}{activations} of \om{3}{each} hardware thread between two consecutive RowHammer-preventive refreshes and 2)~when a RowHammer-preventive refresh is performed, attributes a score to each thread proportional to \ous{3}{its} activation \ous{3}{count}.

\head{Graphene~\cite{park2020graphene}}
Graphene implements the Misra-Gries frequent element finding algorithm~\cite{misra1982finding}, which counts the number of activations for the most frequently accessed DRAM rows.
Graphene issues a RowHammer-preventive refresh when an aggressor row's activation count exceeds Graphene's refresh threshold.
\emph{Score \ous{3}{Attribution:}} \X{} 1)~tracks the \ous{5}{activations} of each hardware thread between two consecutive RowHammer-preventive refreshes and 2)~when a RowHammer-preventive refresh is performed, attributes a score to each thread proportional to its \om{6}{activation count}.

\head{Hydra~\cite{qureshi2022hydra}}\ouscomment{3}{how can we make this better? suggestions?}
Hydra implements a two-part tracking algorithm that 1)~tracks an aggregated group of rows until the collective activation of \ous{3}{the group} reaches a threshold \ous{3}{(i.e., \emph{group threshold})} and 2)~switches to per-row tracking for \ous{3}{that} group.
Hydra stores the per-row tracking \ous{3}{table} in DRAM and keeps a small cache for \ous{3}{the table's entries} in the memory controller.
Hydra performs a RowHammer-preventive refresh when a per-row \ous{3}{table} entry exceeds \agy{3}{a threshold} \ous{3}{(i.e., \emph{refresh threshold})}.
\emph{Score \ous{3}{Attribution:}} \X{} 1)~tracks each hardware thread’s row activation counts between two consecutive RowHammer-preventive actions (i.e., per-row table cache miss/eviction and RowHammer-preventive refresh) and 2)~when a RowHammer-preventive action is performed, attributes a score proportional to its activation count.

\head{TWiCe~\cite{lee2019twice}}
TWiCe implements a counter table of frequently accessed rows.
Table entries have a lifetime, and infrequently accessed rows are periodically pruned.
\emph{Score \ous{3}{Attribution:}} \X{} 1)~tracks the \ous{5}{activations} of each hardware thread between two consecutive RowHammer-preventive refreshes and 2)~when a RowHammer-preventive refresh is performed, attributes a score to each thread proportional to its \om{6}{activation count}.

\head{AQUA~\cite{saxena2022aqua}}
AQUA implements the Misra-Gries frequent element finding algorithm (similar to Graphene) \ous{5}{and} transfers aggressor rows to a quarantine area in DRAM (\om{3}{via} \emph{row migration}) to mitigate attacks.
\emph{Score \ous{3}{Attribution:}} \X{} 1)~tracks the \ous{5}{activations} of each hardware thread between two consecutive row migrations and 2)~when a row migration is performed, attributes a score to each thread proportional to its \om{6}{activation count}.

\head{REGA~\cite{marazzi2023rega}}
REGA modifies the DRAM chip by adding a second row buffer (i.e., \emph{REGA row buffer}) to each subarray.
REGA periodically performs refreshes based on a configurable activation frequency (i.e., $REGA_{T}$) using the \agy{3}{REGA} row buffer while \ous{5}{serving requests} using the first row buffer.
\emph{Score \ous{3}{Attribution:}} \X{} increments a hardware thread's score by one for every $REGA_{T}$ \ous{5}{activations} the thread performs.

\head{Refresh Management (RFM)~\cite{jedec2020ddr5}}
The DDR5 standard \om{3}{introduces} the new \emph{refresh management} (RFM) command.
The memory controller \ous{2}{periodically sends RFM commands} (e.g., once every \param{80} row activations to a bank~\cite{jedec2020ddr5}), \ous{2}{which provides the DRAM chip with a time window to take necessary RowHammer-preventive actions}.
\emph{Score \ous{3}{Attribution:}} \X{} 1)~tracks the \ous{5}{activations} of \om{3}{each} hardware thread between two consecutive RFM commands and 2)~\ous{3}{when the memory controller issues an RFM command}, attributes a score to each thread proportional to \ous{3}{its} \om{6}{activation count}.

\head{Per Row Activation Counting (PRAC)~\cite{jedec2024jesd795c}}
The latest (as of April 2024) JEDEC DDR5 standard introduces a new on-DRAM-die read disturbance mitigation mechanism called \emph{Per Row Activation Counting} (PRAC).
PRAC 1) maintains an activation counter per DRAM row~\cite{kim2014flipping} and 2) proposes a new \emph{back-off} signal to convey the need for RowHammer-preventive refreshes from the DRAM chip to the memory controller, similar to what prior works propose~\cite{bennett2021panopticon, devaux2021method, yaglikci2021security, hassan2022acase, kim2022mithril}.
\emph{Score \ous{3}{Attribution:}} \X{}
1)~tracks the \ous{5}{activations} of \om{3}{each} hardware thread between two consecutive back-off signals and
2)~\ous{3}{when the DRAM chip triggers a back-off}, attributes a score to each thread proportional to \om{3}{its} \om{6}{activation count}.

\subsection{\om{3}{Identifying} Suspect Threads}
\label{sec:anomaly_detection}

\om{3}{\X{} \agy{3}{identifies} \ous{3}{a suspect thread that causes} too much} \ous{3}{performance overhead due to triggering RowHammer-preventive actions.
To do so, it can employ different mechanisms}.
We use \om{3}{a mechanism we call} \emph{thresholded deviation from the mean} with \param{two} tunable configuration parameters:
1)~$TH_{threat}$: the minimum RowHammer-preventive score to consider a thread as a \om{3}{potential} suspect and
2)~$TH_{outlier}$: the maximum allowed divergence from the \ous{6}{average} of all thread \ous{1}{RowHammer-preventive} scores \ous{6}{to mark a thread as suspect}.
\ous{4}{When the RowHammer mitigation mechanism performs a} RowHammer-preventive action, \X{} uses \algref{alg:tmomthresh} to increment RowHammer-preventive scores, detect outliers, and mark suspect threads.\footnote{Our score attribution method \om{6}{described} \om{5}{in \algref{alg:tmomthresh}} (lines 3-7) \om{5}{works for all evaluated RowHammer mitigation mechanisms except for REGA}. We refer \om{6}{the reader} to our open-source repository for the implementation of each mechanism at \url{https://github.com/CMU-SAFARI/BreakHammer}.}

\SetAlFnt{\scriptsize}
\RestyleAlgo{ruled}
\begin{algorithm}
\caption{Identifying suspect threads}
\label{alg:tmomthresh}
\DontPrintSemicolon
\SetKwFunction{TMoM}{incrementThreadScore}
\SetKwProg{Fn}{Function}{:}{}

\tcp{$Threads$: All threads in the system}
\tcp{$Scores$: RowHammer-preventive score of each thread}
\tcp{$Activations$: Row activation \om{6}{count} of each thread}
\tcp{$TH_{threat}$: Threat threshold}
\tcp{$TH_{outlier}$: Outlier threshold}
    \Fn{$updateScores$()}{
        \textbf{// Attribute score to each thread based on its \om{6}{row activation count}}\;
        $totalACTs$ $\gets$ $sum$($Activations$)\;
        \ForEach{$T_{i}$ \textbf{in} $Threads$} {
            $Scores$($T_{i}$) $\gets$ $Activations$($T_{i}$) $/$ $totalACTs$\;
            $Activations$($T_{i}$) $\gets$ $0$\;
        }
        $maxDeviation$ $\gets$ $(1 + TH_{outlier}) * $$mean$($Scores$)\;
        \ForEach{$T_{i}$ \textbf{in} $Threads$} {
            \textbf{// \om{4}{Avoid marking} threads with low scores}\;
            \If{$Scores$($T_{i}$) $<$ $TH_{threat}$} {
                \textbf{continue}\;
            }
            \textbf{// Mark threads that exceed the mean by a factor of $TH_{outlier}$}\;
            \If{$Scores$($T_{i}$) $>$ $maxDeviation$} {
                $markSuspect$($T_{i}$)\;
            }
        }
    }
\end{algorithm}

Our \X{} implementation \ous{3}{attributes score to each thread proportional to its row \om{6}{activation count} since the last RowHammer-preventive action (lines 3-7).
\X{}} performs \param{two} checks \ous{3}{for each thread} to detect outliers each time \X{} increments scores.
First, \X{} compares \ous{3}{a} thread's RowHammer-preventive score against $TH_{threat}$ to determine if the thread \om{3}{has} triggered enough RowHammer-preventive actions to be considered as a \om{3}{potential} suspect thread (line 11).
Second, \X{} checks if \ous{3}{a} thread's RowHammer-preventive score exceeds the mean of all RowHammer-preventive scores by a factor of $TH_{outlier}$ (line 15).
\om{3}{If \ous{3}{a} thread passes both checks, it is} marked as a suspect thread (line 16) for the remainder of the throttling window.

\begin{figure}[h!]
\centering
\includegraphics[width=0.9\linewidth]{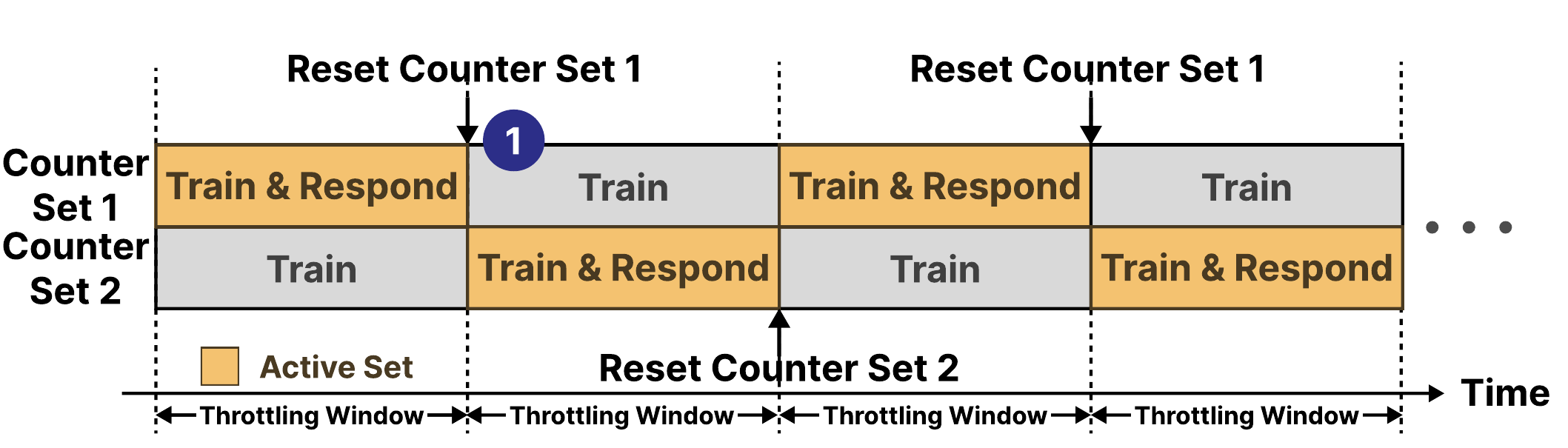}
\vspace{-4pt}\caption{Time interleaving across \X{}'s counters}
\label{fig:time_interleaving}
\end{figure}

\head{Time-interleaving-based continuous monitoring}
Indefinitely incrementing scores \ous{2}{could eventually} cause the counters to saturate, in which case \X{} would mistakenly reduce the memory \om{3}{bandwidth} usage of benign workloads.
To avoid this issue, \X{} periodically resets RowHammer-preventive score counters.
To avoid losing information when resetting counters, \X{} time interleaves across two sets of counters \om{4}{(similar to what prior works do~\cite{yaglikci2021blockhammer, li2012compression})} where \om{4}{each} set contains a RowHammer-preventive score counter for each hardware thread, \om{3}{as shown in \figref{fig:time_interleaving}}.

During a throttling window, \ous{3}{a thread's counters in each set are updated}.
However, only \ous{3}{the counters of one set}, \om{5}{called} the \emph{active set} (\ous{4}{denoted with orange in \figref{fig:time_interleaving}}, e.g., counter set 1 \ous{4}{before \circledblue{1}}), responds to \om{5}{queries needed for suspect identification of a thread}.
At the end of each throttling window, only the counters in the active set are reset (e.g., \ous{4}{counter set 1 is reset after} \circledblue{1}).
\om{4}{After} the reset, \ous{4}{the set with retained counter values becomes the active set} (e.g., counter set 2 after \circledblue{1}) for the next throttling window.\omcomment{4}{Hard to read}
Because counters \ous{3}{in the new active set were already tracking} the RowHammer-preventive actions that took place in the previous throttling window, they start responding to queries with already trained values.
This method ensures continuous monitoring of each thread's activity across \ous{1}{throttling windows}.

\subsection{Throttling Memory Bandwidth \om{4}{Usage} of \\ Suspect Threads}
\label{sec:throttling}

\X{} attributes a hardware thread $i$ with two properties: the dynamic request quota ($Q_{i}$) and the $recent\_suspect_{i}$ flag which indicates whether the thread is identified as a suspect thread (see \algref{alg:tmomthresh}) in the previous throttling window.
In the very first throttling window (i.e., when the system boots), no dynamic request quota is enforced \om{4}{on} any thread (i.e., $Q_i$ \om{4}{of each thread} is \om{4}{equal to} the number of all cache-miss buffers) and \emph{no} hardware thread is identified as a suspect thread (i.e., $recent\_suspect_{i}$ is $false$ for each thread).
\ous{5}{When a hardware thread~$i$ is identified as a suspect (line 16 in \algref{alg:tmomthresh}), \X{} uses \expref{eqn:quota} to reduce the thread's dynamic memory access quota ($Q_{i}$)}.
If the thread is identified as a suspect thread in the previous throttling window (i.e., $recent\_suspect_{i}$ is $true$), the thread's dynamic \om{4}{request} quota is reduced by a constant amount ($P_{oldsuspect}$)
such that $Q_{i}$ becomes $Q_{i}-P_{oldsuspect}$.
If the thread is \emph{not} identified as a suspect thread in the previous throttling window (i.e., $recent\_suspect_{i}$ is $false$), the thread's dynamic \om{4}{request} quota is divided by a constant number ($P_{newsuspect}$) such that $Q_{i}$ becomes $Q_{i}/P_{newsuspect}$.\omcomment{6}{It is hard to believe this is the best mechanism. Reducing by 1 does not seem strong enough. Maybe these thresholds should change dynamically.}\ouscomment{6}{We fiddled with the parameters but we did not try other mechanisms. Extending project document so we can discuss better mechanisms and options for the future.}
\begin{equation}
\label{eqn:quota}
Q_{i} =
\begin{cases} 
      max(Q_{i} - P_{oldsuspect}, 0) & \text{if}~recent\_suspect_{i} \\
      Q_{i} / P_{newsuspect} & \text{else} 
\end{cases}
\end{equation}

\head{\agy{5}{Resetting Reduced Quotas}}
\ous{5}{A suspect thread executes with reduced dynamic request quota as long as it continues to be identified as a suspect (i.e., $recent\_suspect_{i}$ stays $true$ across throttling windows).
If a suspect thread stays benign (i.e., does \emph{not} get marked as a suspect in \algref{alg:tmomthresh}) for the full duration of a throttling window (i.e., $recent\_suspect_{i}$ becomes $false$ at the beginning of a new throttling window), \X{} restores the thread's full dynamic quota (i.e., $Q_{i}$ \om{5}{becomes} equal to the number of all cache-miss buffers \om{5}{again}).}

\head{\ous{5}{Throttling with Cache-Miss Buffers}}
\X{} throttles a suspect thread by limiting the number of cache-miss buffers the thread can allocate in the last level cache (LLC) due to two reasons.
First, by limiting cache-miss buffers, \X{} allows a suspect thread to continue accessing memory locations that result in cache-miss buffer hits.
By doing so, a suspect can access the data 1) that already exists in or 2) being brought to caches.
Second, \X{} is implemented near the memory controller, where it can easily communicate with caches (similar to prior work~\cite{usui2016dash,kim2010thread,subramanian2014bliss,kim2010atlas,nesbit2006fairqueuing,ausavarungnirun2012staged,yaglikci2021blockhammer,mutlu2007stall,mutlu2008parbs}).

\subsection{DMA and Systems without Caches}
\X{} throttles memory \ous{3}{bandwidth usage} by limiting the number of \ous{3}{cache-miss buffers} that track LLC misses each hardware thread can allocate.
However,
1)~the memory request serving unit (e.g., DMA) may lack resources to track memory request status (e.g., \ous{3}{cache-miss buffers}) or
2)~processors in a system may lack caches.
\X{}'s memory throttling support can be implemented in such systems by extending:
1)~the memory request serving unit with a relatively small counter table that tracks the unresolved memory request count of each hardware thread or
2)~the processor's load store unit (LSU) to limit the unresolved memory instruction count of each hardware thread.
We do \emph{not} recommend throttling hardware threads at the memory controller because the unresolved memory requests of a malicious application can potentially reduce performance by blocking the memory request resources of the memory hierarchy (e.g., \ous{3}{cache-miss buffers}) or the processor (e.g., LSU)~\cite{ebrahimi2010fairness}.
\section{Security Analysis}
\label{sec:security_analysis}

We analyze \X{}'s security against 1)~RowHammer attacks, 2)~attacks that exploit RowHammer-preventive actions to hog \ous{3}{DRAM bandwidth availability} (i.e., memory performance attacks~\cite{mutlu2007stall}), and 3)~attacks that try to manipulate score attribution by \ous{4}{causing benign} threads to perform the access that triggers a RowHammer-preventive action.\omcomment{4}{Is there a shorter and better description of this last item?}\ouscomment{4}{i snipped some parts of it}

\subsection{Security Against RowHammer Attacks}
\X{} is \emph{not} a RowHammer defense by itself, but it is a mechanism that reduces the performance overheads incurred by existing RowHammer defenses.
\X{} does \emph{not} interfere with any triggering algorithm and the RowHammer-preventive actions that \om{3}{an} existing RowHammer mitigation mechanism \om{4}{performs}.
Therefore, \om{3}{\X{} preserves the} security guarantees \om{3}{of the RowHammer mitigation mechanism} \om{4}{it is attached to}.
When \X{} observes that the RowHammer mitigation mechanism works aggressively (i.e., performs many RowHammer-preventive actions) as a result of the memory accesses generated by one or few threads, \X{} effectively reduces the memory requests coming from those threads.
Meanwhile, the RowHammer mitigation mechanism \om{3}{continues to execute} its triggering algorithm on the memory accesses that contain fewer requests from the suspect threads.
\om{3}{As such}, the RowHammer mitigation mechanism performs RowHammer-preventive actions less aggressively \emph{without} compromising \om{3}{its} security guarantees.

\subsection{Security Against Memory Performance Attacks}
\label{sec:security_memory_performance}
\X{} increments a hardware thread's RowHammer-preventive score \agy{3}{when} \ous{3}{a} thread causes a RowHammer-preventive action.
Due to time-interleaving-based continuous monitoring across two sets of counters\agy{3}{,} similar to what prior work does~\cite{yaglikci2021blockhammer}, \X{} successfully tracks \agy{3}{\emph{all}} RowHammer-preventive actions.
Therefore, \X{} \ous{3}{1)}~detects a suspect thread responsible for \ous{3}{too many} RowHammer-preventive \ous{3}{actions that reduce DRAM bandwidth availability} and \ous{3}{2)~limits the suspect's dynamic memory access count}.
By doing so, \X{} reduces the interference in the memory subsystem and makes it harder \ous{3}{for an attacker to mount a memory performance attack by exploiting RowHammer-preventive actions}.

\head{\ous{3}{Multi-Threaded Attacks}}
There are two ways that an attacker can try to defeat \X{} using \ous{3}{a group of threads:\omcomment{4}{is (1) a problem? doesn't BH throttle these}\ouscomment{4}{it is a problem, if there are too many attack threads, the attacker can influence the mean and trigger many actions without detection}
1)~increasing the average RowHammer-preventive score across all threads where mounting a memory performance attack becomes the norm (i.e., \emph{rigging suspect identification}) or
2)~intelligently utilizing threads such that \X{} works as intended against each thread, but the coordinated group mounts a memory performance attack (i.e., circumventing suspect identification)}.

\head{\ous{3}{Rigging Suspect Identification}}
An attacker can utilize \ous{3}{many} hardware threads and maliciously increase the \ous{3}{average} RowHammer-preventive score \ous{3}{across all hardware threads} \agy{3}{so that the attacker's behavior becomes the norm instead of an outlier}.
\ous{3}{As such, a thread used by the attacker \agy{3}{(denoted as an \emph{attack thread})} can trigger many RowHammer-preventive actions \emph{without} getting identified as a suspect.
During the attack, an attacker can control the \emph{RowHammer-preventive score of an attack thread $i$} ($\rsatki{}$) but \emph{cannot} control the \emph{RowHammer-preventive score of a benign thread i} ($\rsbeni{}$).
To avoid suspect identification, the \emph{attack thread with maximum RowHammer-preventive score} ($\rsatkmax{}$) should not exceed the average RowHammer-preventive score across all threads by a factor of $TH_{outlier}$.}
\expref{eqn:maxrateattack} calculates $\rsatkmax{}$ before an attack thread is identified as a suspect, \ous{3}{where the number of attack and benign threads in the system are $N_{\text{atk}}$ and $N_{\text{ben}}$, respectively}.
\begin{equation}
\label{eqn:maxrateattack}
\rsatkmax{} < \frac{\sum_{i=1}^{N_{\text{atk}}} \rsatki{} + \sum_{i=1}^{N_{\text{ben}}} \rsbeni{}}{N_{\text{atk}} + N_{\text{ben}}}(1 + TH_{outlier})
\end{equation}

\begin{figure}[h]
    \centering
    \includegraphics[width=1\linewidth]{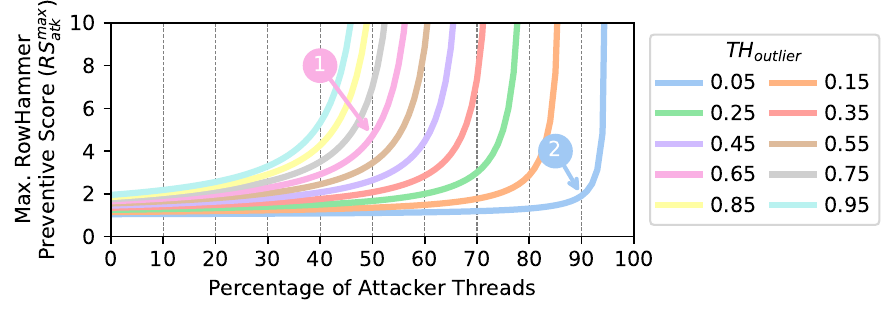}
    \vspace{-16pt}\caption{\ous{3}{$\rsatkmax{}$ normalized to $\rsbenavg{}$ before an attack thread is identified as a suspect}}
    \label{fig:multithreadvisual}
\end{figure}

\ous{3}{\figref{fig:multithreadvisual} presents a visualization of \expref{eqn:maxrateattack}.
The x- and y-axes show the percentage of attack threads in the system ($N_{\text{atk}}/(N_{\text{atk}}+N_{\text{ben}})$) and $\rsatkmax{}$ before an attacker thread is identified as suspect normalized to the \emph{average RowHammer-preventive score of benign threads} ($\rsbenavg{}$), respectively.}
\ous{2}{Different} lines identify $TH_{outlier}$ \ous{3}{configurations}.\omcomment{4}{which THoutlier do we use? how do we determine it?}\ouscomment{4}{we use 0.65, we experimentally found these values to perform well during the ISCA/S\&P submission period (initial submissions of \X{})}

We make two observations from \figref{fig:multithreadvisual}.
First, at a $TH_{outlier}$ value of 0.65 (\circledpastelpink{1}), if the attacker \agy{3}{concurrently} uses \param{50\%} of all \agy{3}{hardware} threads, \agy{3}{an} attack thread can trigger 
$4.71\times$ the number of RowHammer-preventive actions that benign threads trigger on average.
Second, at a $TH_{outlier}$ value of 0.05 (\circledpastelblue{2}), if the attacker \agy{3}{concurrently} uses \param{90\%} of all \agy{3}{hardware} threads, \ous{3}{an} attack thread can trigger 
$1.90\times$ the number of RowHammer-preventive actions that benign threads trigger on average.
We conclude that an attacker needs to \agy{3}{concurrently} use \om{4}{an overwhelming fraction (i.e., $>90$\%)} portion of all \agy{3}{hardware} threads to defeat \X{}.
The system software can determine such an attack by monitoring system-wide resource consumption statistics.
We leave the design of such system-level detection techniques for future research.\footnote{Rigging suspect identification problem can also be solved by developing other suspect identification mechanisms. For example, \om{5}{one can devise a technique that is} sensitive to the fraction of aggressive threads in the system. \om{5}{We leave such techniques to future work.}}

\head{\ous{3}{Circumventing Suspect Identification}}
\ous{3}{An attacker can try to circumvent suspect identification with a large group of attack threads by switching to a new attack thread each time the primary attack thread used to trigger RowHammer-preventive action is identified as a suspect.
Doing so allows the attacker to avoid the limited memory bandwidth and continue triggering RowHammer-preventive actions.
To prevent the attack, the system software can track RowHammer-preventive scores within system software structures (e.g., thread context).
Doing so allows tracking RowHammer-preventive scores at different granularities than a single hardware thread.
For example, the system software can 1) associate a process, address space, or user as the owner of its threads' RowHammer-preventive scores and 2) slow down or stop owners with high cumulative RowHammer-preventive scores.}
We leave such system integration of \X{} to future work.

\subsection{Manipulating Score Attribution}
\ous{3}{Multiple applications can share a DRAM row.
An attacker can try to trick \X{}'s score attribution \om{4}{method} into increasing the RowHammer-preventive score of a benign application by:
1)~activating a row many times \emph{without} causing a RowHammer-preventive action and then
2)~waiting for a benign thread to trigger the RowHammer-preventive action.}
\X{} attributes a score to each hardware thread proportional to its \om{6}{activation count} for a given RowHammer-preventive action.
Therefore, \X{} is secure against score attribution manipulation because it tracks each hardware thread's contribution to a RowHammer-preventive action.
\section{Hardware Complexity}
\label{sec:hardware_complexity}

\X{} is implemented \ous{3}{near} the memory controller and does \ous{0}{\emph{not}} introduce any changes to the DRAM chip \om{3}{or its interfaces}.
We evaluate \X{}'s hardware complexity using CACTI~\cite{cacti} and Synopsys Design Compiler~\cite{synopsys}.
\ous{3}{We implement \X{} in Chisel HDL~\cite{bachrach2012chisel} and synthesize the emitted Verilog HDL design using Synopsys Design Compiler~\cite{synopsys} with a \param{65}nm process technology}.

\head{Area Analysis}
\X{} \om{3}{uses} two 32-bit \ous{5}{RowHammer-preventive score} counters, \ous{5}{one 16-bit activation counter, and two 1-bit suspect flags} per hardware thread, which consume 0.000105$mm^2$ per memory channel \ous{3}{at a 65nm process technology}, consuming an overall area overhead of 0.0002\% of \om{4}{a state-of-the-art} Intel Xeon processor's chip area~\cite{intelxeon} \om{5}{(which is implemented in \om{6}{an} \ous{5}{Intel 10 nm} technology node)}.

\head{Latency Analysis}
Our \X{} implementation is fully \om{3}{pipelined:} it can observe new actions (e.g., memory access or RowHammer-preventive refresh) and make a throttling decision every cycle \om{3}{using an 8-stage} pipeline.\footnote{We note that having a latency in the order of a few nanoseconds is \om{3}{unnecessary} \ous{3}{because} \X{} \ous{3}{1)~slows down \emph{future} memory requests of a thread and 2)~does \emph{not} stop online memory requests. As such, \X{} \emph{can} allow a few more memory requests from a thread before throttling it as long as RowHammer-preventive action tracking is done correctly.}}
According to our Chisel HDL model, \X{} can be clocked at \param{1.5}GHz (\param{$\sim$0.67}ns).\ouscomment{1}{I don't get the comparison against tRC. BreakHammer needs to be able to start processing each request but the latency to respond is not important (because throttling the request a few nanoseconds or one request later is still fine).}
This latency is faster than the latency of regular memory controller operations as it is smaller than \gls{trrd} (e.g., 2.5ns in DDR4~\cite{jedec2017ddr4} and 5ns in DDR5~\cite{jedec2020ddr5}).

\section{Experimental Methodology}
\label{sec:methodology}

We evaluate \X{}'s effect \ous{2}{on system} performance using \ous{2}{cycle-accurate simulations}.
We use Ramulator2~\cite{ramulator2github, luo2023ramulator2} for \ous{2}{our simulations where 1) \X{} is integrated with PARA~\cite{kim2014flipping}, Graphene~\cite{park2020graphene}, Hydra~\cite{qureshi2022hydra}, TWiCe~\cite{lee2019twice}, AQUA~\cite{saxena2022aqua}, REGA~\cite{marazzi2023rega}, RFM~\cite{jedec2024jesd795c}, and PRAC~\cite{jedec2024jesd795c} and 2) BlockHammer~\cite{yaglikci2021blockhammer} is implemented}.
We evaluate system performance using the weighted speedup metric~\cite{eyerman2008systemlevel, snavely2000symbiotic} and unfairness using the maximum slowdown on a benign application~\cite{ebrahimi2010fairness,kim2010atlas,kim2010thread,das2009application}.

\tabref{table:system_configuration} shows our system configuration.
We assume a realistic quad-core system, connected to a dual-rank memory with eight bank groups, each containing two banks (32 banks in total). 
The memory controller employs the \ous{2}{FR-FCFS memory scheduler~\cite{frfcfs, zuravleff1997controller} with a Cap on Column-Over-Row Reordering (FR-FCFS+Cap) of \param{four}~\cite{mutlu2007stall}}.

\vspace{4pt}
\newcolumntype{C}[1]{>{\let\newline\\\arraybackslash\hspace{0pt}}m{#1}}
\begin{table}[ht]
\scriptsize
\centering
\caption{Simulated System Configuration}

\begin{tabular}{l|C{5.8cm}}
    \hline
    \textbf{Processor} & {\SI{4.2}{\giga\hertz}, 4~core, 4-wide issue, {128-entry} instr. window}\\ \hline
    \textbf{Last-Level Cache} & {64-byte} cache line, 8-way {set-associative, \SI{8}{\mega\byte}} \\ \hline
    \textbf{Memory Controller} & {64-entry \ous{2}{read/write request queues}; \ous{2}{FR-FCFS+Cap with \om{6}{Cap=}\param{4}~\cite{mutlu2007stall}}}; Address mapping: MOP~\cite{kaseridis2011minimalistic} \\ \hline
    \textbf{Main Memory} & DDR5, 1 channel, 2 ranks, 8 bank groups, 2 banks/bank group, 64K rows/bank\\ \hline
    \end{tabular}
    \label{table:system_configuration}
\end{table}

\head{Comparison Points}
We pair \X{} with \param{eight} \om{3}{different} state-of-the-art RowHammer mitigation mechanisms that provide RowHammer-safe operation:
one is a probabilistic mechanism~\cite{kim2014flipping},
five are deterministic mechanisms~\cite{park2020graphene,qureshi2022hydra,lee2019twice,saxena2022aqua,marazzi2023rega}, and
two \om{3}{use} the \om{3}{DDR5} RFM command~\cite{jedec2024jesd795c} and PRAC~\cite{jedec2024jesd795c}.
We configure all RowHammer mitigation mechanisms except RFM and PRAC according to the methodology described by their studies and scale them to the \gls{nrh} values \ous{2}{used in} our evaluation.
For RFM and PRAC, we assume that the DRAM chip maintains an activation counter for each DRAM row~\cite{bennett2021panopticon, kim20231perfecttrack, jedec2024jesd795c} and use mathematically-proven RowHammer-secure \ous{2}{configurations} from prior work~\cite{canpolat2024understanding}.

We compare \X{}-paired versions of these \param{eight} RowHammer mitigation mechanisms to their baseline implementations \ous{2}{without \X{}}.
We also compare these eight \X{}-paired mechanisms to BlockHammer, the state-of-the-art \om{3}{throttling-based} RowHammer mitigation mechanism.\omcomment{4}{static threshold might be bad}\ouscomment{4}{added to doc for future discussion}\omcomment{4}{how does a thread get back its quota?}\ouscomment{4}{it is now explicitly clarified in \secref{sec:throttling}}
\ous{3}{We configure the duration of a throttling window as 64ms to match the refresh window~\cite{jedec2017ddr4}, similar to counter reset periods of prior work~\cite{park2020graphene,lee2018twice,yaglikci2021blockhammer}.
We empirically configure $TH_{threat}$ and $TH_{outlier}$ on a smaller set of workloads separate from our evaluations.}
Table~\ref{tab:tsars_conf}\ouscomment{6}{Removed counter size.} summarizes \X{}'s configuration parameters for our evaluation.

\vspace{4pt}
\begin{table}[ht]
  \centering
  \footnotesize
  \caption{\X{} Configuration}
    \begin{tabular}{|l||l|}
        \hline
        {{\bf Component}} & \textbf{Parameters} \\
        \hline
        \hline
                                        & $TH_{window}:$ $64$ms \\ 
        \ous{3}{Suspect Identification} & $TH_{threat}:$ $32$ \\
                                        & $TH_{outlier}:$ $0.65$ \\
        \hline
        \ous{2}{Memory Throttling}      & \makecell[l]{$P_{oldsuspect}: 1$ \\ $P_{newsuspect}: 10$}  \\
        \hline
    \end{tabular}
    \label{tab:tsars_conf}
\end{table}

\head{Workloads}
We evaluate \X{} with applications from five benchmark suites: SPEC CPU2006~\cite{spec2006}, SPEC CPU2017~\cite{spec2017}, TPC~\cite{tpc}, MediaBench~\cite{fritts2009media}, and YCSB~\cite{ycsb}.
We group applications into three memory intensity groups based on their \ous{3}{row buffer misses-per-kilo-instructions (RBMPKIs)}.
These groups are High (H), Medium (M), and Low (L) for the lowest \ous{3}{RBMPKI} values of \param{\ous{3}{20, 10, and 0}}, respectively.
\ous{3}{We create \param{six} four-core workload mixes (HHHH, HHMM, MMMM, HHLL, MMLL, and LLLL)}\footnote{\ous{3}{Each letter in a mix denotes the memory intensity of an application.}} \ous{3}{where each mix contains 15 four-core workloads (90 in total)}.
We simulate these workloads until each \ous{4}{benign} core \om{4}{completes} 100M instructions.\footnote{We do \emph{not} wait for attacker cores to complete 100M instructions because \X{} significantly slows down the attacker's progression \om{5}{and attacker performance is not important (nor evaluated)}.}
Table~\ref{tab:mpki_act_table} summarizes RBMPKI\ouscomment{4}{TODO: get for 1K (i don't have the data at the moment)} and the average number of rows with more than \om{4}{512, 128, and 64} activations per 64ms time window for the \param{8} most memory-intensive workloads.
From this table, we observe that \ous{3}{as \gls{nrh} decreases}, many benign \ous{3}{applications} \om{4}{become} capable of triggering RowHammer-preventive actions.

\vspace{4pt}
\begin{table}[h!]
  \centering
  \footnotesize
  \captionsetup{justification=centering, singlelinecheck=false, labelsep=colon}
  \caption{Workload Characteristics: RBMPKI and \ous{6}{Average Number of Rows with More Than} 512+, 128+, and 64+ \ous{6}{Activations} per 64ms Time Window}
    \begin{tabular}{|c||c|ccc|}
        \hline
        {{\bf Workload}} & \textbf{RBMPKI} & {{\bf ACT-512+}}  & {{\bf ACT-128+}} & {{\bf ACT-64+}}\\
        \hline
        \hline
        429.mcf         &  68.27  & 2564 & 2564 &  2564 \\
        470.lbm         &  28.09  &  664 & 6596 &  7089 \\
        462.libquantum  &  25.95  &    0 &    0 &     1 \\
        549.fotonik3d   &  25.28  &    0 &   88 & 10065 \\
        459.GemsFDTD    &  24.93  &    0 &  218 & 10572 \\
        519.lbm         &  24.37  & 2482 & 5455 &  5824 \\
        434.zeusmp      &  22.24  &  292 & 4825 & 11085 \\
        510.parest      &  17.79  &   94 &  185 &   803 \\
        \hline
        \hline
        Average         &  29.615 & 762 & 2491 & 6000 \\
        \hline
    \end{tabular}
  \label{tab:mpki_act_table}
\end{table}
\section{Performance Evaluation}
\label{sec:perf_evaluation}

\addtocounter{figure}{2}
\begin{figure*}[t]
    \centering
    \includegraphics[width=1\linewidth]{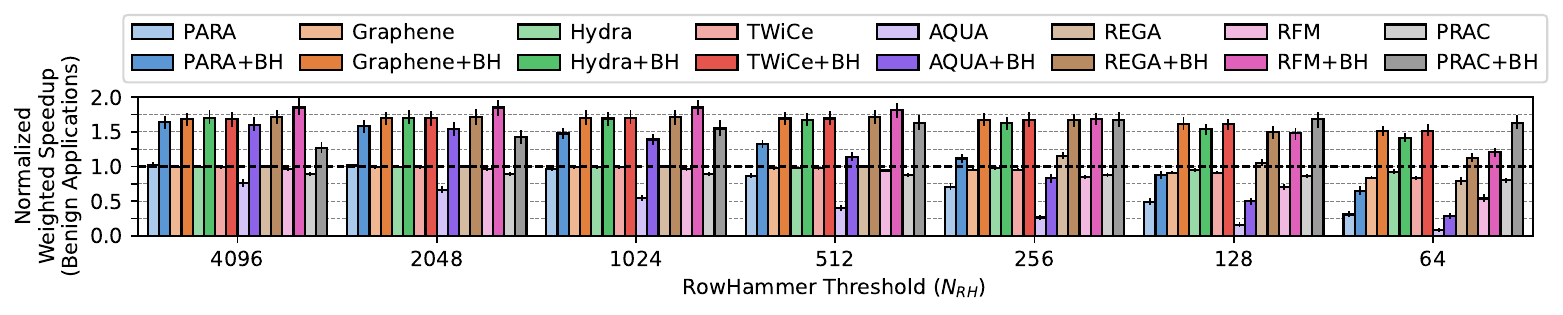}
    \vspace{-16pt}\caption{\X{}'s performance scaling for existing RowHammer mitigation mechanisms \ous{3}{with an attacker present}}
    \label{fig:a_fourcore_underattack_scaling}
\end{figure*}
\addtocounter{figure}{-3}

We evaluate \X{}'s performance \ous{2}{and energy} impact in four parts.
First \om{3}{(\secref{subsec:perfunderattack})}, we experimentally demonstrate that
1)~a malicious hardware thread can incur significant performance and energy overheads on concurrently running multi-programmed workloads, and
2)~\X{} prevents such performance and energy overheads by significantly reducing the number of RowHammer-preventive actions and increasing DRAM bandwidth availability.
Second, we show that \X{} does \emph{not} incur significant performance and DRAM energy overheads when there is \emph{no} malicious thread in the system \om{3}{(\secref{subsec:perfnoattack})}.
Third, we compare \X{} against the state-of-the-art memory access throttling-based RowHammer mitigation mechanism, BlockHammer~\cite{yaglikci2021blockhammer} \om{3}{(\secref{subsec:compareblockhammer})}.
Our analysis shows that \X{}
1)~outperforms BlockHammer across \emph{all} evaluated \gls{nrh} values and
2)~maintains a minimal area overhead with \om{3}{only} two counters per hardware thread whereas BlockHammer requires a significantly growing history buffer \om{4}{as \gls{nrh} decreases}.
Fourth, we analyze \X{}'s sensitivity to configuration parameters \om{3}{(\secref{sec:sensitivity_to_configuration_parameters})}.

\subsection{Under RowHammer Attack}
\label{subsec:perfunderattack}
\begin{figure}[H]
    \centering
    \includegraphics[width=1\linewidth]{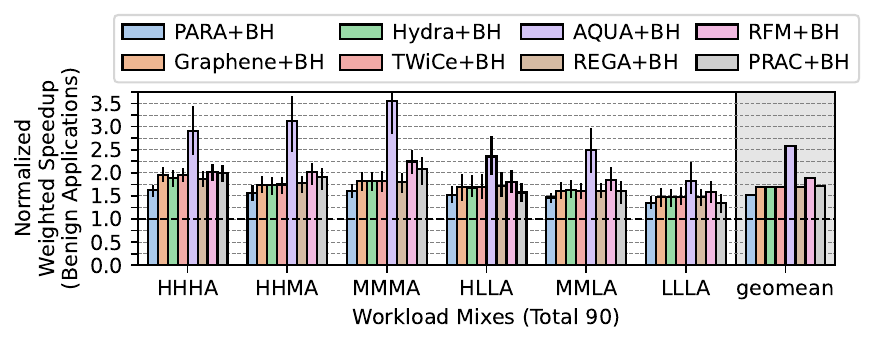}
    \vspace{-16pt}\caption{\X{}'s impact on performance for existing RowHammer mitigation mechanisms \ous{3}{with an attacker present}}
    \label{fig:a_fourcore_underattack}
\end{figure}

\head{System Performance}
\figref{fig:a_fourcore_underattack} presents the average performance impact of \X{} when paired with \param{eight} state-of-the-art RowHammer mitigation mechanisms on four-core workloads, with an attacker present, at an \gls{nrh} of $1K$.
The x-axis shows the memory intensity of applications in the mix \ous{3}{(15 workloads per mix)} where H, M, L, and A denote High, Medium, Low, and Attacker, respectively, alongside the geometric-mean (geomean) of performance across \emph{all} 90 workloads.
\om{4}{Different bars identify RowHammer mitigation mechanisms paired with \X{} (e.g., PARA+\Xshort{})}.
The y-axis shows the performance of benign workloads normalized to \om{3}{each} baseline RowHammer mitigation mechanism \emph{without} \X{}.

From \figref{fig:a_fourcore_underattack}, we observe that \X{} increases the system performance of benign workloads by \om{5}{an} \om{4}{average (maximum) of} \param{84.6}\% (\param{170.0}\%) \om{3}{when the system is under RowHammer attack from one malicious thread}.
We note that \X{} detects and throttles the attacker in \ous{3}{\emph{all} 90} workloads.

\begin{figure}[H]
    \centering
    \includegraphics[width=1\linewidth]{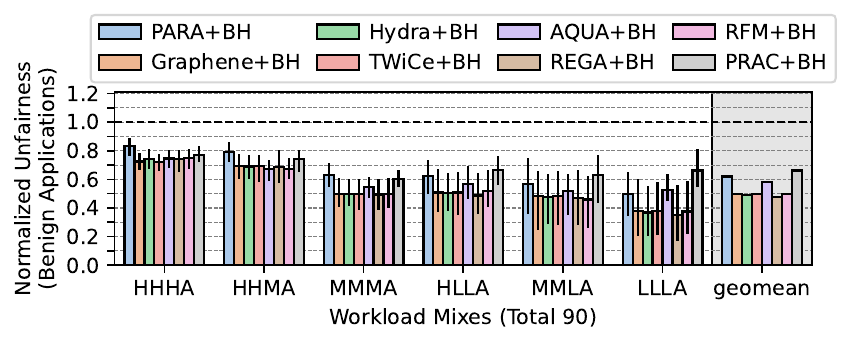}
    \vspace{-16pt}\caption{\X{}'s impact on unfairness for existing RowHammer mitigation mechanisms with an attacker present}
    \label{fig:fourcore_underattack_unfairness}
\end{figure}

\head{Unfairness}
\figref{fig:fourcore_underattack_unfairness} presents the average unfairness impact of \X{} when paired with \param{eight} state-of-the-art RowHammer mitigation mechanisms on four-core workloads, with an attacker present, at an \gls{nrh} of $1K$.
The x- and y-axes show the memory intensity of the applications in the mix and unfairness on benign applications normalized to each baseline RowHammer mitigation mechanism \emph{without} \X{}.\omcomment{4}{repeating different bars is unnecessary but ok to have}\ouscomment{4}{removed them after this point (unless we introduce a different type for the first time, e.g., lines)}

\addtocounter{figure}{1}

We make \param{two} observations from \figref{fig:fourcore_underattack_unfairness}.
First, \X{} reduces unfairness on benign workloads by \om{4}{an average (maximum) of \param{45.8}\% (\param{90.6}\%)}.
Second, \X{} reduces unfairness the most (\param{55.9}\% on average) for LLLA and the least (\param{24.7}\% on average) for HHHA mixes.
We attribute this trend to high memory intensity applications causing more memory interference against the attacker and reducing the attacker's \om{4}{effectiveness at mounting a memory performance attack}.

\head{System Performance with Future DRAM Chips}
\figref{fig:a_fourcore_underattack_scaling} presents the \om{3}{average} performance \ous{2}{impact} of \X{} when paired with \param{eight} state-of-the-art RowHammer mitigation mechanisms on \ous{2}{four-core} workloads, with an attacker present, \ous{2}{as \gls{nrh} decreases from \param{$4K$} to \param{$64$}}.\omcomment{4}{No 32 and 16?}\ouscomment{4}{we don't have the data and they will take time to obtain, i will try after sensitivity}
\ous{2}{The x- and y-axes show the \gls{nrh} values and system performance \om{3}{of benign workloads} normalized to a baseline with \emph{no} RowHammer mitigation mechanism, respectively}.

We make \param{three} observations from \figref{fig:a_fourcore_underattack_scaling}.
First, across \om{3}{\emph{all}} evaluated \gls{nrh} values, \X{} increases the system performance of benign workloads by \om{4}{an average (maximum) of} \param{90.1}\% (\param{162.4}\%).
Second, as \gls{nrh} decreases, \X{} maintains a speedup above the no defense baseline for \ous{3}{\emph{all}} mechanisms except for PARA at $\nrh{} < 256$ and AQUA at $\nrh{} < 512$.
\om{3}{At low \gls{nrh} values (e.g., $< 1K$), PARA and AQUA significantly degrade system performance even after the attacker is throttled}.\omcomment{4}{Maybe we do not throttle the attacker enough?}\ouscomment{4}{Because these mechanisms perform poorly at low \gls{nrh}, the attacker takes time to get throttled. We should be throttling the attacker enough already (further aggressiveness will hurt benign fairness more)}
PARA \ous{3}{is \emph{stateless}, where it probabilistically performs a RowHammer-preventive refresh \emph{even} for a benign application that does \emph{not} activate a row many times}.
On the other hand, AQUA uses row migration, which is a costly RowHammer-preventive action in itself.
Third, RFM and PRAC \ous{3}{induce} high system performance overheads \emph{even} with \emph{per row activation counters} in DRAM.
However, \X{} still reduces the performance overheads of RFM and PRAC \emph{without} any online information about the on-DRAM-die mechanisms.

\begin{figure}[h]
    \centering
    \includegraphics[width=1\linewidth]{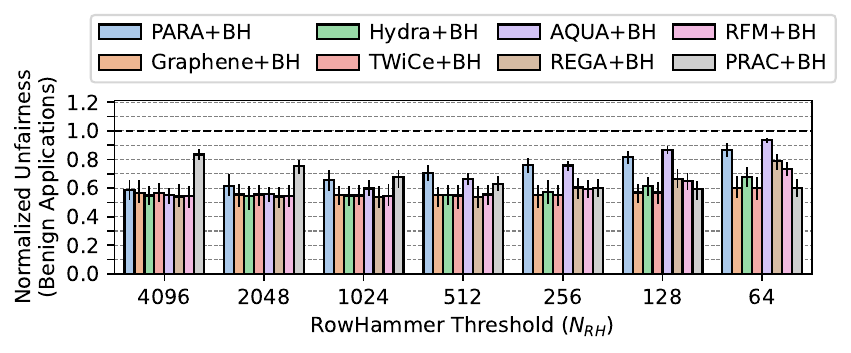}
    \vspace{-16pt}\caption{\ous{3}{\X{}'s impact on unfairness for existing RowHammer mitigation mechanisms with an attacker present}}
    \label{fig:fourcore_underattack_unfairness_scaling}
\end{figure}

\head{Unfairness with Future DRAM Chips}\omcomment{4}{add unfairness with future DRAM chips}
\figref{fig:fourcore_underattack_unfairness_scaling} presents the unfairness impact of \X{} when paired with \param{eight} state-of-the-art RowHammer mitigation mechanisms on four-core workloads, with an attacker present, as \gls{nrh} decreases from \param{$4K$} to \param{$64$}.
The x- and y-axes show the \gls{nrh} values and unfairness \om{3}{on benign workloads} normalized to a baseline with \emph{no} RowHammer mitigation mechanism, respectively.

We make \param{two} observations from \figref{fig:fourcore_underattack_unfairness_scaling}.
First, across \emph{all} evaluated \gls{nrh} values, \X{} reduces the unfairness on benign workloads by an average (maximum) of \param{31.5}\% (\param{99.1}\%).
Second, as \gls{nrh} decreases, \X{}'s unfairness benefits for 1) PRAC increases while 2) PARA and AQUA decreases.
We attribute PRAC's unfairness benefit improvements to it accurately tracking each row's activation count, where PRAC does \emph{not} perform many RowHammer-preventive refreshes at relatively high \gls{nrh} values (i.e., $>1K$).
As such, when \gls{nrh} decreases, the attacker triggers RowHammer-preventive refreshes frequently and deviates from the average RowHammer-preventive score of benign threads faster.
On the other hand, PARA and AQUA's unfairness benefits decrease.
This is because PARA and AQUA increase memory interference for benign applications and make it harder to detect the attacker as
1)~PARA's starts performing many RowHammer-preventive refreshes for the benign applications and
2)~AQUA uses time-consuming row migration operations.

\addtocounter{figure}{2}
\begin{figure*}[b]
    \centering
    \includegraphics[width=1\linewidth]{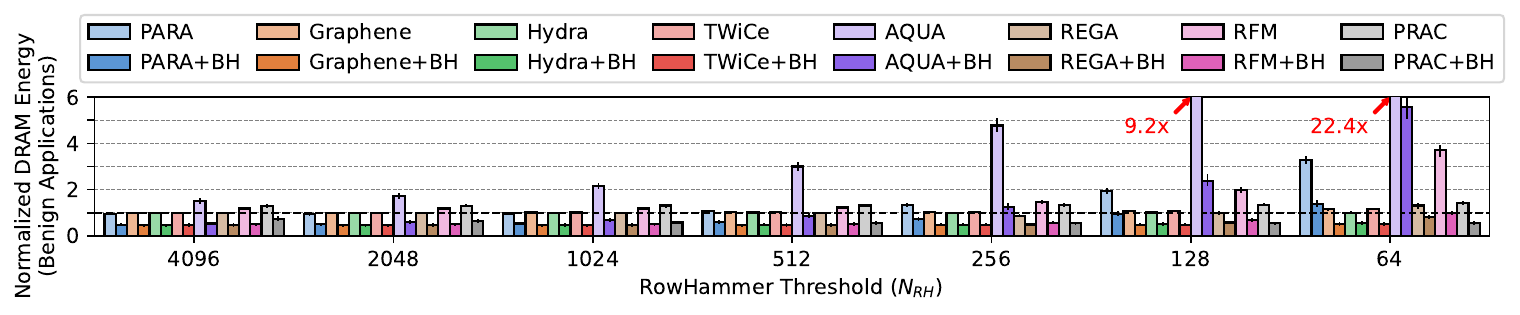}
    \caption{\X{}'s energy scaling for existing RowHammer mitigation mechanisms with an attacker present}
    \label{fig:a_fourcore_underattack_energy}
\end{figure*}
\addtocounter{figure}{-3}

\begin{figure}[H]
    \centering
    \includegraphics[width=1\linewidth]{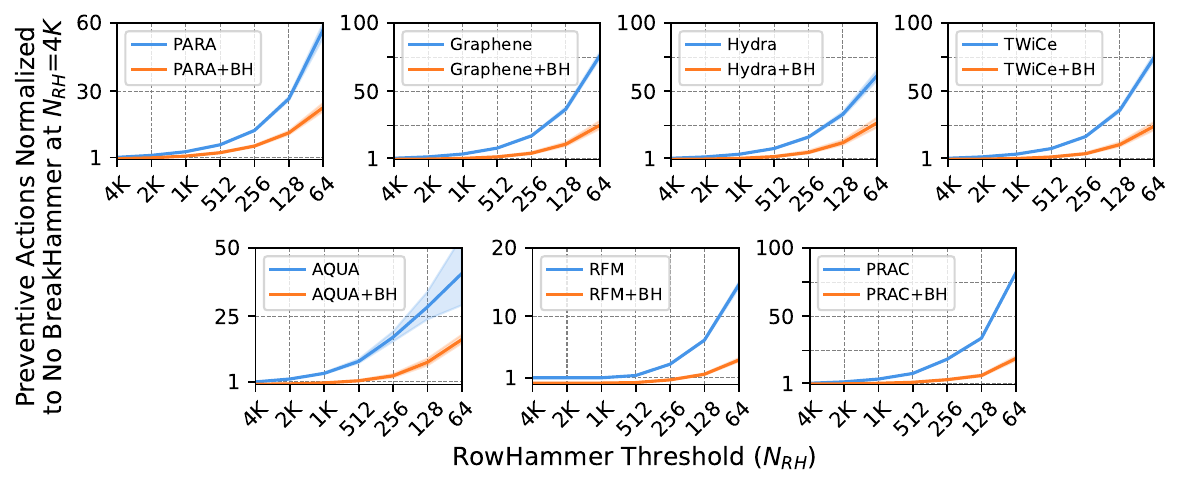}
    \vspace{-16pt}\caption{\X{}'s impact on \om{3}{the number of} RowHammer-preventive actions performed \om{3}{by} existing RowHammer mitigation mechanisms \ous{3}{with an attacker present}}
    \label{fig:a_fourcore_underattack_mitigative_actions}
\end{figure}

\head{Effect on Performed RowHammer-Preventive Actions}\ouscomment{4}{numbers seem correct, maybe Ramulator2's PARA configuration isn't confidently secure? might need probability analysis (added to gdoc)}\ouscomment{4}{REGA just has a flat-ish line because it is equal to the number of activations at this threshold (email discussion for brainstorming, REGA throttles attacker and reduces ACTs)}
\figref{fig:a_fourcore_underattack_mitigative_actions} presents the performed RowHammer-preventive action counts of \ous{3}{PARA, Graphene, Hydra, TWiCe, AQUA, RFM, and PRAC} 1)~by themselves and 2)~when paired with \X{}, \ous{2}{as \gls{nrh} decreases from \param{$4K$} to \param{$64$}}.\footnote{\ous{3}{We do \emph{not} include REGA in this analysis because it performs refreshes in parallel to row activations. However, we still evaluate REGA's performance and energy overheads based on its impact on DRAM timing constraints.}}
Each subplot displays a different mechanism where the x-axis shows the different \gls{nrh} values and the y-axis shows the \om{3}{number of} RowHammer-preventive actions \om{3}{taken with \X{}} normalized to without \X{} at \ous{2}{an \gls{nrh} of \param{$4K$}}.
The \param{orange} and \param{blue} lines respectively depict the 1)~\X{}-paired and 2)~baseline mechanisms.
The band shade around each line marks the 100\% confidence interval across \om{4}{\emph{all}} \param{90} workloads.

We make \param{two} observations from \figref{fig:a_fourcore_underattack_mitigative_actions}.
First, as \gls{nrh} decreases, RowHammer-preventive \ous{5}{actions} of \ous{3}{\emph{all}} \om{3}{mitigation} mechanisms increase.
Second, across all evaluated \gls{nrh} values, \X{} decreases the RowHammer-preventive \ous{5}{actions} of \emph{all} mechanisms by \om{4}{an average (maximum) of} \param{71.6}\% (\param{91.8}\%).

\begin{figure}[h]
    \centering
    \includegraphics[width=1\linewidth]{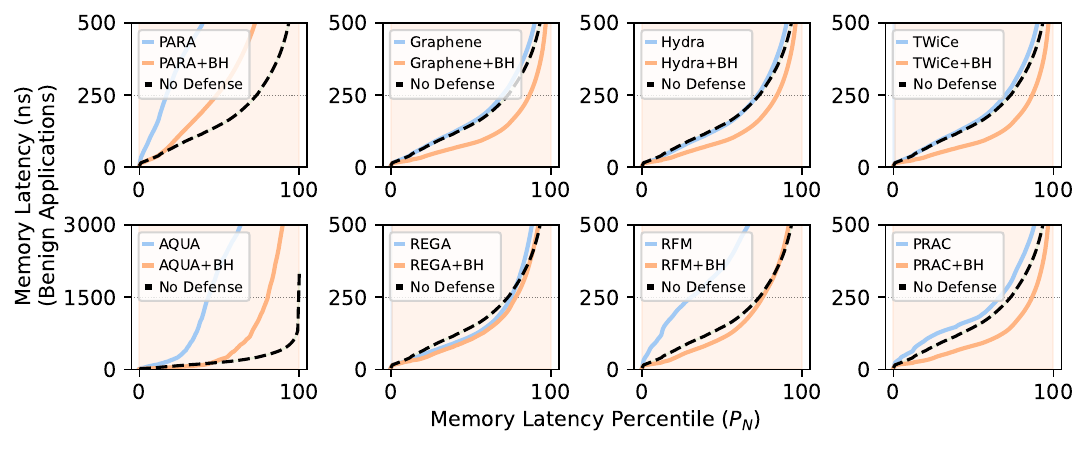}
    \vspace{-16pt}\caption{\ous{3}{\X{}'s \ous{3}{impact on memory latency} for existing RowHammer mitigations with an attacker present}}
    \label{fig:fourcore_underattack_memlat}
\end{figure}

\head{Memory Latency}
\figref{fig:fourcore_underattack_memlat} presents the memory latency percentiles ($P_{N}$) of \X{} when paired with \param{eight} state-of-the-art RowHammer mitigation mechanisms on four-core workloads, with an attacker present, at an \gls{nrh} of \param{$64$}.
Each subplot depicts the memory latency results of a different RowHammer mitigation mechanism.
In subplots, the x- and y-axes show the percentile values (e.g., 90th percentile shows that 90\% of memory latencies are below this value) and the memory latency in nanoseconds, respectively.
Different curves identify the baseline with no RowHammer mitigation and a RowHammer mitigation mechanism 1)~by itself and 2)~paired with BreakHammer.
The background color indicates the mechanism with lower memory latency for a given percentile of accesses, e.g., orange means that the tested mechanism results in a lower memory latency when used with \X{}, compared to without \X{}.\footnote{\ous{5}{AQUA's subplot uses a different scale because AQUA incurs relatively large memory latencies due to costly row migration operations.}}

\addtocounter{figure}{1}

From \figref{fig:fourcore_underattack_memlat}, we observe that at an \gls{nrh} of \param{$64$}, \X{} reduces the memory latency \om{4}{benign applications experience} for \emph{all} RowHammer mitigation mechanisms, sometimes to values \emph{even} lower than the no defense baseline.
This is because \X{} \emph{significantly} reduces the memory interference caused by the attacker at multiple levels of the memory hierarchy, e.g., memory request \om{4}{scheduler} and caches.

\head{DRAM Energy}
\figref{fig:a_fourcore_underattack_energy} presents the \ous{2}{DRAM} energy \ous{2}{impact} of \X{} when paired with \param{eight} state-of-the-art RowHammer mitigation mechanisms on \ous{2}{four-core} workloads, with an attacker present, \ous{2}{as \gls{nrh} decreases from \param{$4K$} to \param{$64$}}.
\ous{3}{The x- and y-axes} shows the \gls{nrh} values and \ous{4}{DRAM energy of benign workloads normalized to a baseline with \emph{no} RowHammer mitigation mechanism}, \ous{3}{respectively}.

\ous{3}{We make \param{three} observations from \figref{fig:a_fourcore_underattack_energy}.}
\ous{3}{First}, across \emph{all} evaluated \gls{nrh} values, \X{} significantly reduces \om{4}{energy by an average (maximum) of} \param{55.4}\% (\param{67.4}\%).
\ous{3}{Second}, as \gls{nrh} decreases from $4K$ to $64$, \emph{all} baseline RowHammer mitigation mechanisms consume more energy (\param{4.4}x on average).
\ous{3}{Third}, at an \gls{nrh} of $64$, AQUA and RFM incur the highest energy overhead by 22.3x and 3.7x on avearge, respectively.

We conclude that
1)~RowHammer mitigation mechanisms incur increasing performance and energy overheads due to performing many RowHammer-preventive actions as \gls{nrh} decreases and
2)~\X{} significantly improves performance, unfairness, memory latency, and energy when an attacker is present, \om{4}{at \emph{all} evaluated \gls{nrh} values}.
\subsection{No RowHammer Attack}
\label{subsec:perfnoattack}

\begin{figure}[h]
    \centering
    \includegraphics[width=1\linewidth]{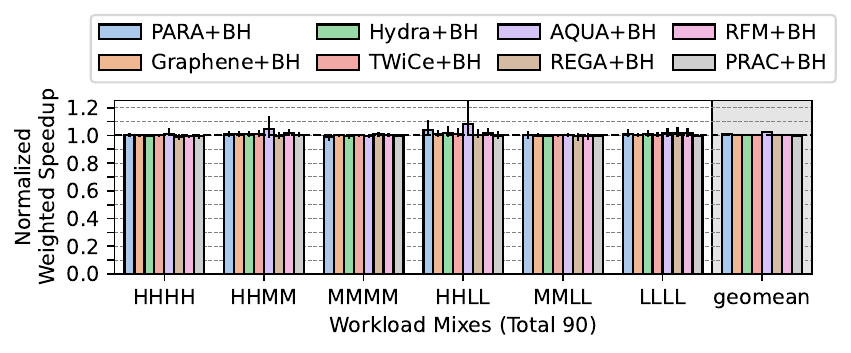}
    \vspace{-16pt}\caption{\X{}'s impact on performance for existing RowHammer mitigation mechanisms with no attacker}
    \label{fig:b_fourcore_noattack}
\end{figure}

\head{System Performance}
\figref{fig:b_fourcore_noattack} presents the performance impact of \X{} when paired with \param{eight} state-of-the-art RowHammer mitigation mechanisms on \ous{2}{four-core} workloads, where all applications are benign, \ous{2}{at an \gls{nrh} of $64$}.
The x- and y-axes show the \ous{5}{workload mixes and} the system performance normalized to \ous{5}{each} baseline RowHammer mitigation mechanism \emph{without} \X{}.

From \figref{fig:b_fourcore_noattack}, we observe that \X{} increases the system performance of benign \om{4}{application} mixes by \om{4}{an average (maximum) of} \param{0.7}\% (\param{2.4}\%).

\begin{figure}[h]
    \centering
    \includegraphics[width=1\linewidth]{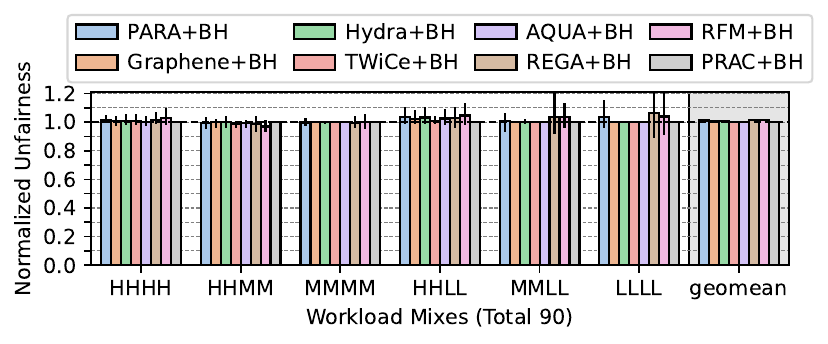}
    \vspace{-16pt}\caption{\X{}'s impact on unfairness for existing RowHammer mitigation mechanisms with no attacker}
    \label{fig:fourcore_benign_unfairness}
\end{figure}

\head{Unfairness}
\figref{fig:fourcore_benign_unfairness} presents the unfairness impact of \X{} when paired with \param{eight} state-of-the-art RowHammer mitigation mechanisms on four-core workloads, where all applications are benign, at an \gls{nrh} of $1K$.
The x- and y-axes show the memory intensity of the applications in the mix and unfairness normalized to each baseline RowHammer mitigation mechanism \emph{without} \X{}.

\ous{4}{From \figref{fig:fourcore_benign_unfairness}, we observe that} \X{} \ous{4}{\emph{slightly} increases} unfairness by \param{0.9}\% on average when compared to the baseline mechanisms \emph{without} \X{}.
Our results \ous{4}{at an \gls{nrh} of \param{1K}, across} eight mitigation mechanisms, and 90 workloads show that \X{} identifies a benign application as suspect in \param{2.2}\% of the simulations (not shown in \figref{fig:fourcore_benign_unfairness}).

\begin{figure}[h]
    \centering
    \includegraphics[width=1\linewidth]{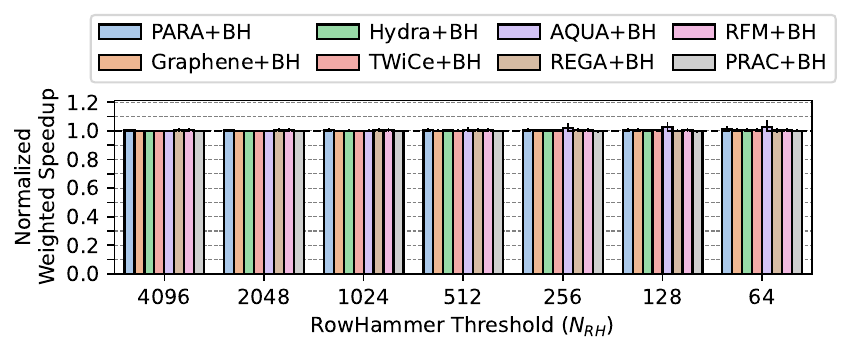}
    \vspace{-16pt}\caption{\X{}'s impact on performance for existing RowHammer mitigation mechanisms at various \gls{nrh} values}
    \label{fig:b_fourcore_noattack_scaling}
\end{figure}

\head{System Performance with Future DRAM Chips}
\figref{fig:b_fourcore_noattack_scaling} presents the performance \ous{2}{impact} of \X{} when paired with \param{eight} state-of-the-art RowHammer mitigation mechanisms on \ous{2}{four-core} workloads, where all applications are benign, \ous{2}{as \gls{nrh} decreases from \param{$4K$} to \param{$64$}}.
\ous{2}{The x- and y-axes show the \gls{nrh} values and system performance normalized to the baseline RowHammer mitigation mechanism \emph{without} \X{}, respectively}.

From \figref{fig:b_fourcore_noattack_scaling}, we observe that below an $\nrh{}$ of 1024, \X{} slightly improves the average (maximum) performance of \om{3}{\emph{all}} tested RowHammer mitigation mechanisms.

\head{Unfairness with Future DRAM Chips}\omcomment{4}{add unfairness with future DRAM chips}
\figref{fig:fourcore_underattack_unfairness_scaling} presents the average unfairness impact of \X{} when paired with \param{eight} state-of-the-art RowHammer mitigation mechanisms on four-core workloads, where all applications are benign, as \gls{nrh} decreases from \param{$4K$} to \param{$64$}.
The x- and y-axes show the \gls{nrh} values and unfairness normalized to a baseline with \emph{no} RowHammer mitigation mechanism, respectively.

\begin{figure}[h]
    \centering
    \includegraphics[width=1\linewidth]{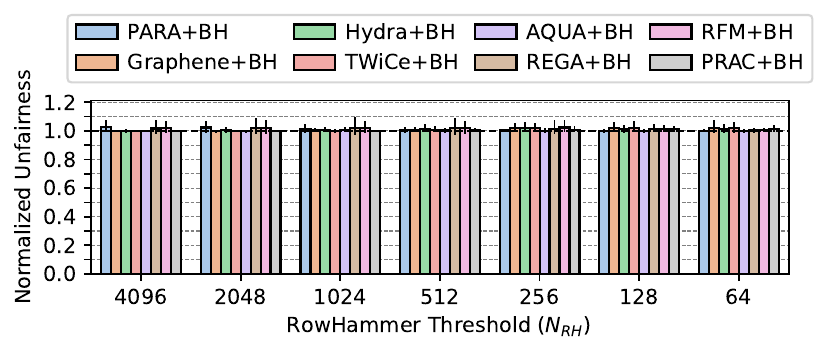}
    \vspace{-16pt}\caption{\X{}'s impact on unfairness for existing RowHammer mitigation mechanisms at various \gls{nrh} values}
    \label{fig:fourcore_benign_unfairness_scaling}
\end{figure}

We make \param{two} observations from \figref{fig:fourcore_benign_unfairness_scaling}.
First, as \gls{nrh} decreases, \X{} \emph{slightly} increases unfairness by \param{0.9}\% on average compared to the baseline mechanisms \emph{without} \X{}.
Second, \X{} has a best-case unfairness reduction of \param{29.1}\% and a worst-case unfairness increase of \param{36.4}\%.
Our results across \emph{all} \gls{nrh} values, eight mitigation mechanisms, and 90 workloads show that \X{} marks a benign application as suspect in \param{18.7}\% of the simulations.

\begin{figure}[h]
    \centering
    \includegraphics[width=1\linewidth]{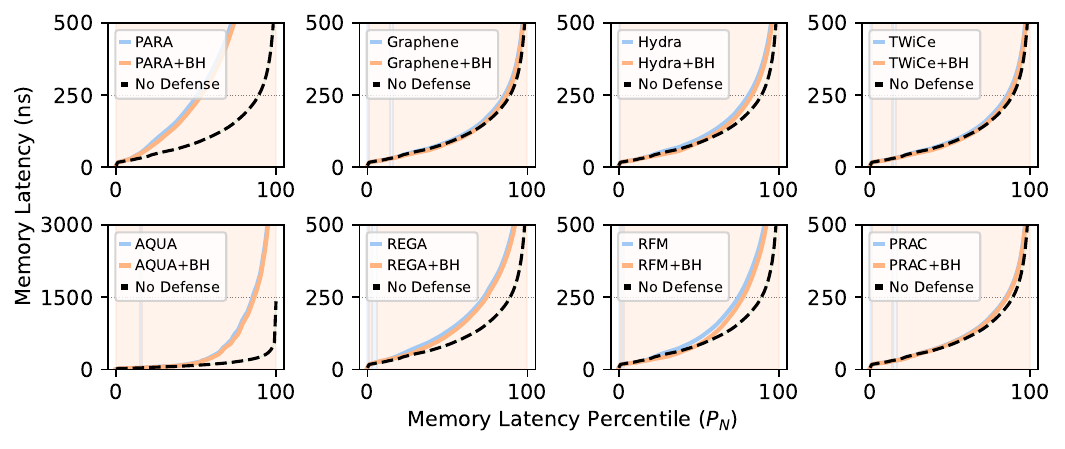}
    \caption{\X{}'s impact on memory latency for existing RowHammer mitigations with no attacker}
    \label{fig:fourcore_ben_memlat}
\end{figure}

\head{Memory Latency}
\figref{fig:fourcore_ben_memlat} presents the memory latency percentiles \ous{2}{($P_{N}$)} of \X{} when paired with \param{eight} state-of-the-art RowHammer mitigation mechanisms on \ous{2}{four-core} workloads, where all applications are benign, \ous{2}{at an \gls{nrh} of \param{$64$}}.
\ous{2}{Each subplot depicts the memory latency results of a different RowHammer mitigation mechanism}.
\ous{2}{In subplots, the x- and y-axes show the percentile values \ous{3}{(e.g., 90th percentile shows that 90\% of memory latencies are below this value)} and the memory latency in nanoseconds, respectively.}

From \figref{fig:fourcore_ben_memlat}, we observe that \ous{2}{at an \gls{nrh} of \param{$64$}}, \X{} \ous{2}{induces no} overhead and \ous{2}{improves} the memory latency \ous{2}{at all $P_{N}$ values across all RowHammer mitigation mechanisms}.

We conclude that \X{} \om{5}{either \emph{slightly} improves or does \emph{not} degrade} average performance and energy for benign applications for \emph{all} evaluated RowHammer mitigation mechanisms.
\subsection{Comparison to BlockHammer}
\label{subsec:compareblockhammer}

BlockHammer~\cite{yaglikci2021blockhammer,blockhammergithub} is the state-of-the-art throttling based RowHammer mitigation mechanism.
It works by blacklisting rows \om{3}{that are frequently activated}.
To understand \X{}'s performance benefits compared to BlockHammer we study \ous{2}{four-core} workloads with an attacker present at \param{seven} \gls{nrh} values \om{3}{(from \param{$4K$} to \param{$64$})}.

\begin{figure}[h]
    \centering
    \includegraphics[width=1\linewidth]{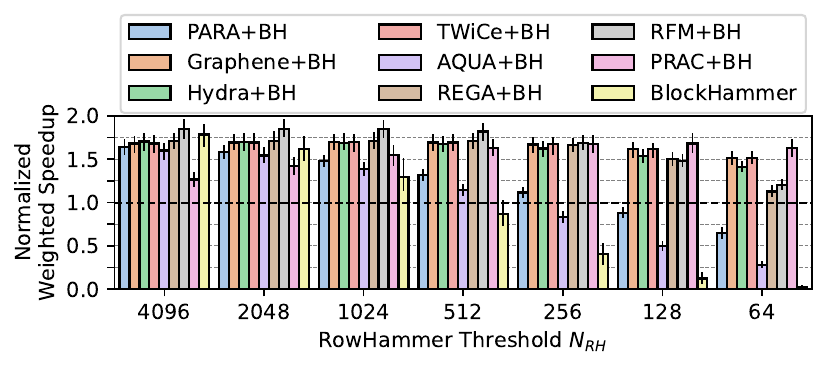}
    \caption{\X{}'s \ous{3}{impact on performance} compared to BlockHammer~\cite{yaglikci2021blockhammer}, the state-of-the-art throttling-based RowHammer mitigation mechanism}
    \label{fig:c_fourcore_blockhammer}
\end{figure}

\figref{fig:c_fourcore_blockhammer} presents the performance \ous{2}{impact of 1)} \X{} \ous{3}{when paired with \param{eight} RowHammer mitigation mechanisms} \ous{2}{and 2) BlockHammer on four-core} workloads, with \ous{5}{an attacker} present, as \gls{nrh} decreases from \param{$4K$} to \param{$64$}.
\ous{2}{The x- and y-axes show \gls{nrh} values and system performance normalized to a baseline with no RowHammer mitigation, respectively}.

We make \param{two} observations from \figref{fig:c_fourcore_blockhammer}.
First, across \ous{3}{\emph{all}} evaluated \gls{nrh} values, \om{3}{a mechanism paired with \X{} outperforms} BlockHammer.
Second, as \gls{nrh} decreases from $4K$ to $64$, BlockHammer's average system performance impact drastically drops from \param{78.6\%} improvement to \param{98.0\%} \emph{degradation}.
This is because BlockHammer \om{4}{blocks accesses} to a row that is close to causing RowHammer bitflips until the row's victims are periodically refreshed.
As \gls{nrh} decreases, even a benign application activates DRAM rows many times before they are periodically refreshed (see Table~\ref{tab:mpki_act_table}).\ouscomment{4}{added BlockHammer + refresh discussion item to gdoc}
Therefore, at low \gls{nrh} values, BlockHammer blocks access to increasingly many rows and significantly degrades performance.
\X{} does \emph{not} suffer from such a weakness because it allows refreshing victim rows when necessary and only throttles suspect applications.
Based on these observations, we conclude that \X{} outperforms BlockHammer, the previous state-of-the-art throttling based mechanism at \emph{all} \gls{nrh} values even when combined with the \emph{worst}-scaling RowHammer mitigations, i.e., AQUA and PARA.

\subsection{Sensitivity to Configuration Parameters}
\label{sec:sensitivity_to_configuration_parameters}

\X{} has \param{three} tunable configuration parameters:
1)~$TH_{window}$: \ous{3}{the length of a throttling window},
2)~$TH_{threat}$: \ous{3}{the minimum RowHammer-preventive score to consider a thread as a potential suspect}, and
3)~$TH_{outlier}$: \ous{3}{the maximum allowed divergence from the average of all thread RowHammer-preventive scores \om{6}{to mark a thread as suspect}}.
We set $TH_{window}$ as 64ms \ous{3}{to match with the refresh interval~\cite{jedec2017ddr4}}, similar to prior work~\cite{yaglikci2021blockhammer, lee2019twice}.
We set $TH_{outlier}$ as $0.65$ to limit the RowHammer-preventive score of attackers (e.g., $\approx5$x the total score of benign threads) when the fraction of attack threads in a system is near 50\% (see \secref{sec:security_memory_performance}).\ouscomment{5}{I cannot obtain a sensible TH\_outlier sweep. I would like to argue in security direction for this choice.}

\begin{figure}[h]
    \centering
    \includegraphics[width=1\linewidth]{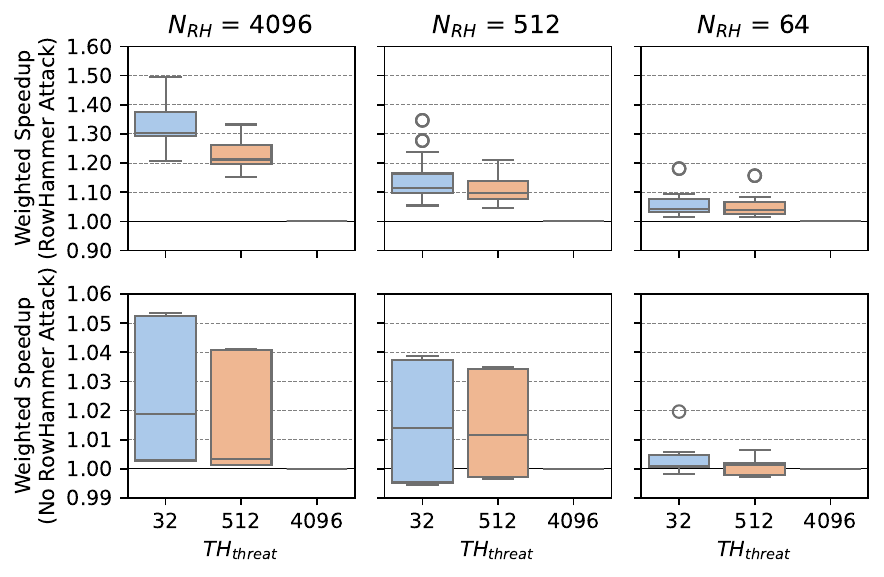}
    \caption{\X{}'s sensitivity to $TH_{threat}$}
    \label{fig:sensitivity_to_configuration_parameters}
\end{figure}

To determine $TH_{threat}$ we sweep our parameters from $4K$ to $32$.
\figref{fig:sensitivity_to_configuration_parameters} shows the performance impact of \X{} for varying $TH_{threat}$ configurations, as \gls{nrh} decreases from $4K$ to $64$, in a box-and-whisker plot.\footnote{The box is lower-bounded by the first quartile (i.e., the median of the first half of the ordered set of data points) and upper-bounded by the third quartile (i.e., the median of the second half of the ordered set of data points). The interquartile range ($IQR$) is the distance between the first and third quartiles (i.e., box size). Whiskers mark the central 1.5$IQR$ range.}
The rows of subplots show the workloads with an attacker present (top) and where all applications are benign (bottom).
The columns of subplots show \gls{nrh} values and for each subplot, the x-axis shows the $TH_{threat}$ values and the y-axis shows the weighted speedup normalized to a baseline system with $TH_{threat}$ of \param{$4K$}.

We make \param{two} observations from \figref{fig:sensitivity_to_configuration_parameters}.\omcomment{6}{Are the weighted speedup calculations done correctly for all these figures?}\ouscomment{6}{Yes, they should be. However, I will triple check while going over all of our numbers.}
First, as $TH_{outlier}$ decreases, \X{}'s average system performance benefit increases for both RowHammer attack (top) and no RowHammer attack (bottom) workloads.
Second, as \gls{nrh} decreases, \ous{6}{more} \om{6}{all-benign workload mixes} \ous{6}{observe reduced} system performance \ous{6}{compared to a baseline system with $TH_{threat}$ of \param{$4K$}}.
This is because as \gls{nrh} decreases, many benign applications become capable of triggering RowHammer-preventive actions (see Table~\ref{tab:mpki_act_table}).
Therefore, more applications are identified as suspects when $TH_{outlier}$ and \gls{nrh} are low enough.

We choose a $TH_{outlier}$ value of 32 because such configuration 1) provides high performance benefits with an attacker present and 2) induces either little or \emph{no} performance overheads when all applications are benign.

\section{Related Work}
\label{sec:related_work}

To our knowledge, this is the first work that provides throttling support to reduce the performance \om{3}{and energy overheads of RowHammer mitigation mechanisms}.
\secref{sec:perf_evaluation} qualitatively and quantitatively compares \X{} with the most relevant prior mechanisms. This section discusses other related works.

\head{RowHammer Mitigation Mechanisms}
There have been \om{4}{various} mitigation \om{4}{mechanisms}~\mitigatingRowHammerAllCitations{} proposed (both by academia and industry) against RowHammer~\cite{kim2014flipping}.~DRAM manufacturers implement RowHammer mitigation \agy{5}{mechanisms}, also known as \emph{Target Row Refresh} (TRR)~\cite{jedec2020ddr5,jedec2017ddr4,hassan2021utrr,frigo2020trrespass}, in commercial DRAM chips, but they do \agy{5}{\emph{not}} openly disclose specific designs.
Recent research shows that custom attack patterns can bypass these mechanisms~\cite{frigo2020trrespass, hassan2021utrr, jattke2022blacksmith, deridder2021smash}.
With worsening RowHammer vulnerability, TRR mechanisms \agy{5}{need to} perform more preventive refresh operations.
To provide \agy{5}{TRR mechanisms with} the necessary time window \agy{5}{to perform preventive refreshes}, the JEDEC DDR5 standard\om{6}{\cite{jedec2020jesd795}} introduces the \emph{Refresh Management} (RFM) command and \ous{4}{(as of April 2024)} \om{6}{the} \emph{Per Row Activation Counting} (PRAC) mechanism~\agy{5}{\cite{jedec2024jesd795c}}.
RFM and PRAC are secure for future DRAM chips with worse RowHammer vulnerabilities, at the cost of high performance overheads~\cite{canpolat2024understanding}.
Our goal in this paper is \emph{not} to propose a new RowHammer mitigation mechanism, but rather to reduce the performance \om{5}{and energy overheads} of RowHammer mitigation mechanisms by carefully reducing the number of RowHammer-preventive actions \om{5}{they perform}, \emph{without} compromising system robustness.\omcomment{5}{Is eight mechanisms correct? shouldn't include BlockHammer}\ouscomment{5}{yes, we evaluate eight mechanisms excluding BlockHammer}
As we show in \secref{sec:perf_evaluation}, \om{5}{\X{} \emph{significantly}} reduces the performance and energy overheads of \om{5}{\param{eight} state-of-the-art RowHammer mitigation mechanisms}

\head{Memory Performance Attacks}\omcomment{4}{Distinguish attack and defense papers}
Several prior works propose techniques for \ous{4}{mounting}~\cite{moscibroda2007memory, mutlu2007stall} and preventing~\cite{mutlu2008parbs, subramanian2014bliss, kim2010atlas} memory performance attacks.
These works tackle memory performance attacks from the memory request scheduling algorithms' point of view.
However, \ous{4}{they}
1)~do \emph{not} consider RowHammer-preventive actions and
2)~are \emph{unaware} of the DRAM bandwidth \om{4}{usage} caused by RowHammer mitigation mechanisms.
In contrast, \X{} tackles the performance and energy overheads induced by RowHammer mitigation mechanisms.
Therefore, \ous{4}{memory performance attack mitigation proposals (e.g., \cite{mutlu2008parbs, subramanian2014bliss, kim2010atlas})} can be \om{5}{combined} with \X{} to achieve better security against \ous{3}{memory performance attacks}.

\head{Other Uses of Throttling}
\omcomment{4}{\X{} is not only about RH attacks}
Prior works on quality-of-service and fairness-oriented architectures propose selectively throttling main memory accesses to provide latency guarantees and/or improve fairness across applications (e.g.,~\cite{ausavarungnirun2012staged, chang2012hat, ebrahimi2010fairness, ebrahimi2011prefetch, ebrahimi2011parallel, ebrahimi2012fairness, kim2010atlas, kim2010thread, lee2008prefetch, moscibroda2007memory, mutlu2007stall, mutlu2008parbs, nychis2010next}).
These mechanisms are
1) \emph{unaware} of RowHammer-preventive actions,
2) \emph{not} designed to reduce the overheads of RowHammer mitigation mechanisms, and
3) \emph{not} designed to prevent RowHammer attacks.
Thus, these mechanisms do \emph{not} interfere with memory performance attacks \om{5}{that} \ous{5}{arise from the overheads of the RowHammer mitigation mechanisms}.\omcomment{5}{Write this sentence better}
\ous{4}{As such, \X{} is complementary to these mechanisms and can work together with them}.

\section{Conclusion}

\om{5}{We introduced \X{}, the first mechanism to reduce performance and energy overheads of existing RowHammer solutions.
The key idea of \X{} is \ous{5}{to limit the dynamic memory request count of a hardware thread based on how frequently the thread triggers RowHammer-preventive actions.}}
We show that \X{} is compatible with both memory controller-based and on-DRAM-die RowHammer mitigation mechanisms by implementing \X{} with \param{eight} state-of-the-art RowHammer mitigation mechanisms.
Our rigorous experimental evaluations show that \X{} significantly 1) reduces the number of performed RowHammer-preventive actions and \om{3}{thereby} 2) improves system performance and DRAM energy.
\om{3}{\X{}'s benefits increase as DRAM chips become more vulnerable to RowHammer}.
We hope and expect that \X{} will \om{3}{help} researchers and engineers to build \om{3}{higher-performance and lower energy read disturbance solutions that will be increasingly needed as DRAM technology scaling continues}.

\section*{Acknowledgments} {
We thank the anonymous reviewers of \ous{0}{S\&P 2024 and MICRO 2024 (\om{4}{both} main submission and artifact evaluation)} for the encouraging feedback.
We thank the SAFARI Research Group members for valuable feedback and the stimulating scientific and intellectual environment.
We acknowledge the generous gift funding provided by our industrial partners (especially Google, Huawei, Intel, Microsoft, VMware), which has been instrumental in enabling the research we have been conducting on read disturbance in DRAM since 2011~\cite{mutlu2023retrospective}.
This work was also in part supported by the Google Security and Privacy Research Award, the Microsoft Swiss Joint Research Center, and the ETH Future Computing Laboratory (EFCL).}\ouscomment{5}{MICRO CR guidelines advise the order: paper, appendix, and references. I also think the appendix here looks better (because check-list table doesn't fit nicely due to spacing when we put it after references)}

\newpage
\balance
\bibliographystyle{unsrt}
\bibliography{newrefs}

\begin{thebibliography}{100}

\bibitem{dennard1968dram}
Robert~H. Dennard.
\newblock {Field-Effect Transistor Memory}, 1968.
\newblock {U.S.}\ Patent 3,387,286.

\bibitem{kim2014flipping}
Y.~{Kim}, R.~{Daly}, J.~{Kim}, C.~{Fallin}, J.~H. {Lee}, D.~{Lee}, C.~{Wilkerson}, K.~{Lai}, and O.~{Mutlu}.
\newblock {Flipping Bits in Memory Without Accessing Them: An Experimental Study of DRAM Disturbance Errors}.
\newblock In {\em ISCA}, 2014.

\bibitem{mutlu2017rowhammer}
Onur Mutlu.
\newblock {The RowHammer Problem and Other Issues We May Face as Memory Becomes Denser}.
\newblock In {\em DATE}, 2017.

\bibitem{mutlu2019rowhammer}
Onur Mutlu and Jeremie~S Kim.
\newblock {RowHammer: A Retrospective}.
\newblock {\em TCAD}, 2019.

\bibitem{mutlu2023fundamentally}
Onur Mutlu, Ataberk Olgun, and A.~Giray Yaglikci.
\newblock {Fundamentally Understanding and Solving RowHammer}.
\newblock In {\em ASP-DAC}, 2023.

\bibitem{fournaris2017exploiting}
Apostolos~P Fournaris, Lidia Pocero~Fraile, and Odysseas Koufopavlou.
\newblock {Exploiting Hardware Vulnerabilities to Attack Embedded System Devices: A Survey of Potent Microarchitectural Attacks}.
\newblock {\em Electronics}, 2017.

\bibitem{poddebniak2018attacking}
Damian Poddebniak, Juraj Somorovsky, Sebastian Schinzel, Manfred Lochter, and Paul R{\"o}sler.
\newblock {Attacking Deterministic Signature Schemes using Fault Attacks}.
\newblock In {\em EuroS\&P}, 2018.

\bibitem{tatar2018throwhammer}
Andrei Tatar, Radhesh~Krishnan Konoth, Elias Athanasopoulos, Cristiano Giuffrida, Herbert Bos, and Kaveh Razavi.
\newblock {Throwhammer: {Rowhammer} {Attacks} Over the {Network} and {Defenses}}.
\newblock In {\em {USENIX} {ATC}}, 2018.

\bibitem{carre2018openssl}
Sebastien Carre, Matthieu Desjardins, Adrien Facon, and Sylvain Guilley.
\newblock {OpenSSL Bellcore's Protection Helps Fault Attack}.
\newblock In {\em DSD}, 2018.

\bibitem{barenghi2018software}
Alessandro Barenghi, Luca Breveglieri, Niccol{\`o} Izzo, and Gerardo Pelosi.
\newblock {Software-Only Reverse Engineering of Physical DRAM Mappings for Rowhammer Attacks}.
\newblock In {\em IVSW}, 2018.

\bibitem{zhang2018triggering}
Zhenkai Zhang, Zihao Zhan, Daniel Balasubramanian, Xenofon Koutsoukos, and Gabor Karsai.
\newblock {Triggering Rowhammer Hardware Faults on ARM: A Revisit}.
\newblock In {\em ASHES}, 2018.

\bibitem{bhattacharya2018advanced}
Sarani Bhattacharya and Debdeep Mukhopadhyay.
\newblock {Advanced Fault Attacks in Software: Exploiting the Rowhammer Bug}.
\newblock In {\em Fault Tolerant Architectures for Cryptography and Hardware Security}, pages 111--135. Springer, 2018.

\bibitem{google-project-zero}
Mark Seaborn and Thomas Dullien.
\newblock {Exploiting the DRAM Rowhammer Bug to Gain Kernel Privileges}.
\newblock \url{http://googleprojectzero.blogspot.com.tr/2015/03/exploiting-dram-rowhammer-bug-to-gain.html}, 2015.

\bibitem{rowhammergithub}
{SAFARI Research Group}.
\newblock {RowHammer --- GitHub Repository}.
\newblock \url{https://github.com/CMU-SAFARI/rowhammer}, 2014.

\bibitem{seaborn2015exploiting}
Mark Seaborn and Thomas Dullien.
\newblock {Exploiting the DRAM Rowhammer Bug to Gain Kernel Privileges}.
\newblock {\em Black Hat}, 2015.

\bibitem{van2016drammer}
Victor van~der Veen, Yanick Fratantonio, Martina Lindorfer, Daniel Gruss, Clementine Maurice, Giovanni Vigna, Herbert Bos, Kaveh Razavi, and Cristiano Giuffrida.
\newblock {Drammer: Deterministic Rowhammer Attacks on Mobile Platforms}.
\newblock In {\em CCS}, 2016.

\bibitem{gruss2016rowhammer}
Daniel Gruss, Cl{\'e}mentine Maurice, and Stefan Mangard.
\newblock {Rowhammer.js: A Remote Software-Induced Fault Attack in Javascript}.
\newblock arXiv:1507.06955 [cs.CR], 2016.

\bibitem{razavi2016flip}
Kaveh Razavi, Ben Gras, Erik Bosman, Bart Preneel, Cristiano Giuffrida, and Herbert Bos.
\newblock {Flip Feng Shui: Hammering a Needle in the Software Stack}.
\newblock In {\em USENIX Security}, 2016.

\bibitem{pessl2016drama}
Peter Pessl, Daniel Gruss, Cl{\'e}mentine Maurice, Michael Schwarz, and Stefan Mangard.
\newblock {DRAMA: Exploiting DRAM Addressing for Cross-CPU Attacks}.
\newblock In {\em USENIX Security}, 2016.

\bibitem{xiao2016one}
Yuan Xiao, Xiaokuan Zhang, Yinqian Zhang, and Radu Teodorescu.
\newblock {One Bit Flips, One Cloud Flops: Cross-VM Row Hammer Attacks and Privilege Escalation}.
\newblock In {\em USENIX Security}, 2016.

\bibitem{bosman2016dedup}
Erik Bosman, Kaveh Razavi, Herbert Bos, and Cristiano Giuffrida.
\newblock {Dedup Est Machina: Memory Deduplication as An Advanced Exploitation Vector}.
\newblock In {\em S\&P}, 2016.

\bibitem{bhattacharya2016curious}
Sarani Bhattacharya and Debdeep Mukhopadhyay.
\newblock {Curious Case of Rowhammer: Flipping Secret Exponent Bits Using Timing Analysis}.
\newblock In {\em CHES}, 2016.

\bibitem{burleson2016invited}
Wayne Burleson, Onur Mutlu, and Mohit Tiwari.
\newblock {Invited: Who is the Major Threat to Tomorrow's Security? You, the Hardware Designer}.
\newblock In {\em DAC}, 2016.

\bibitem{qiao2016new}
Rui Qiao and Mark Seaborn.
\newblock {A New Approach for RowHammer Attacks}.
\newblock In {\em HOST}, 2016.

\bibitem{brasser2017can}
Ferdinand Brasser, Lucas Davi, David Gens, Christopher Liebchen, and Ahmad-Reza Sadeghi.
\newblock {Can't Touch This: Software-Only Mitigation Against Rowhammer Attacks Targeting Kernel Memory}.
\newblock In {\em USENIX Security}, 2017.

\bibitem{jang2017sgx}
Yeongjin Jang, Jaehyuk Lee, Sangho Lee, and Taesoo Kim.
\newblock {SGX-Bomb: Locking Down the Processor via Rowhammer Attack}.
\newblock In {\em SOSP}, 2017.

\bibitem{aga2017good}
Misiker~Tadesse Aga, Zelalem~Birhanu Aweke, and Todd Austin.
\newblock {When Good Protections Go Bad: Exploiting Anti-DoS Measures to Accelerate Rowhammer Attacks}.
\newblock In {\em HOST}, 2017.

\bibitem{tatar2018defeating}
Andrei Tatar, Cristiano Giuffrida, Herbert Bos, and Kaveh Razavi.
\newblock {Defeating Software Mitigations Against Rowhammer: A Surgical Precision Hammer}.
\newblock In {\em RAID}, 2018.

\bibitem{gruss2018another}
Daniel Gruss, Moritz Lipp, Michael Schwarz, Daniel Genkin, Jonas Juffinger, Sioli O'Connell, Wolfgang Schoechl, and Yuval Yarom.
\newblock {Another Flip in the Wall of Rowhammer Defenses}.
\newblock In {\em S\&P}, 2018.

\bibitem{lipp2018nethammer}
Moritz Lipp, Misiker~Tadesse Aga, Michael Schwarz, Daniel Gruss, Cl{\'e}mentine Maurice, Lukas Raab, and Lukas Lamster.
\newblock {Nethammer: Inducing Rowhammer Faults Through Network Requests}.
\newblock arXiv:1805.04956 [cs.CR], 2018.

\bibitem{van2018guardion}
Victor van~der Veen, Martina Lindorfer, Yanick Fratantonio, Harikrishnan~Padmanabha Pillai, Giovanni Vigna, Christopher Kruegel, Herbert Bos, and Kaveh Razavi.
\newblock {GuardION: Practical Mitigation of DMA-Based Rowhammer Attacks on ARM}.
\newblock In {\em {DIMVA}}, 2018.

\bibitem{frigo2018grand}
Pietro Frigo, Cristiano Giuffrida, Herbert Bos, and Kaveh Razavi.
\newblock {Grand Pwning Unit: Accelerating Microarchitectural Attacks with the GPU}.
\newblock In {\em S\&P}, 2018.

\bibitem{cojocar2019eccploit}
Lucian Cojocar, Kaveh Razavi, Cristiano Giuffrida, and Herbert Bos.
\newblock {Exploiting Correcting Codes: On the Effectiveness of ECC Memory Against Rowhammer Attacks}.
\newblock In {\em S\&P}, 2019.

\bibitem{ji2019pinpoint}
Sangwoo Ji, Youngjoo Ko, Saeyoung Oh, and Jong Kim.
\newblock {Pinpoint Rowhammer: Suppressing Unwanted Bit Flips on Rowhammer Attacks}.
\newblock In {\em ASIACCS}, 2019.

\bibitem{hong2019terminal}
Sanghyun Hong, Pietro Frigo, Yi\u{g}itcan Kaya, Cristiano Giuffrida, and Tudor Dumitra\c{s}.
\newblock {Terminal Brain Damage: Exposing the Graceless Degradation in Deep Neural Networks Under Hardware Fault Attacks}.
\newblock In {\em USENIX Security}, 2019.

\bibitem{kwong2020rambleed}
Andrew Kwong, Daniel Genkin, Daniel Gruss, and Yuval Yarom.
\newblock {RAMBleed: Reading Bits in Memory Without Accessing Them}.
\newblock In {\em S\&P}, 2020.

\bibitem{frigo2020trrespass}
Pietro Frigo, Emanuele Vannacci, Hasan Hassan, Victor van~der Veen, Onur Mutlu, Cristiano Giuffrida, Herbert Bos, and Kaveh Razavi.
\newblock {TRRespass: Exploiting the Many Sides of Target Row Refresh}.
\newblock In {\em {S\&P}}, 2020.

\bibitem{cojocar2020rowhammer}
Lucian Cojocar, Jeremie Kim, Minesh Patel, Lillian Tsai, Stefan Saroiu, Alec Wolman, and Onur Mutlu.
\newblock {Are We Susceptible to Rowhammer? An End-to-End Methodology for Cloud Providers}.
\newblock In {\em S\&P}, 2020.

\bibitem{weissman2020jackhammer}
Zane Weissman, Thore Tiemann, Daniel Moghimi, Evan Custodio, Thomas Eisenbarth, and Berk Sunar.
\newblock {JackHammer: Efficient Rowhammer on Heterogeneous FPGA--CPU Platforms}.
\newblock arXiv:1912.11523 [cs.CR], 2020.

\bibitem{zhang2020pthammer}
Zhi Zhang, Yueqiang Cheng, Dongxi Liu, Surya Nepal, Zhi Wang, and Yuval Yarom.
\newblock {PThammer: Cross-User-Kernel-Boundary Rowhammer through Implicit Accesses}.
\newblock In {\em MICRO}, 2020.

\bibitem{yao2020deephammer}
Fan Yao, Adnan~Siraj Rakin, and Deliang Fan.
\newblock {Deephammer: Depleting the Intelligence of Deep Neural Networks Through Targeted Chain of Bit Flips}.
\newblock In {\em USENIX Security}, 2020.

\bibitem{deridder2021smash}
Finn de~Ridder, Pietro Frigo, Emanuele Vannacci, Herbert Bos, Cristiano Giuffrida, and Kaveh Razavi.
\newblock {SMASH}: {Synchronized} {Many-Sided} {Rowhammer} {Attacks} from {JavaScript}.
\newblock In {\em {USENIX Security}}, 2021.

\bibitem{hassan2021utrr}
Hasan Hassan, Yahya~Can Tugrul, Jeremie~S. Kim, Victor van~der Veen, Kaveh Razavi, and Onur Mutlu.
\newblock {Uncovering in-DRAM RowHammer Protection Mechanisms: A New Methodology, Custom RowHammer Patterns, and Implications}.
\newblock In {\em MICRO}, 2021.

\bibitem{jattke2022blacksmith}
Patrick Jattke, Victor van~der Veen, Pietro Frigo, Stijn Gunter, and Kaveh Razavi.
\newblock {Blacksmith: Scalable Rowhammering in the Frequency Domain}.
\newblock In {\em S\&P}, 2022.

\bibitem{tol2022toward}
M~Caner Tol, Saad Islam, Berk Sunar, and Ziming Zhang.
\newblock {Toward Realistic Backdoor Injection Attacks on DNNs using RowHammer}.
\newblock {arXiv:2110.07683}, 2022.

\bibitem{kogler2022half}
Andreas Kogler, Jonas Juffinger, Salman Qazi, Yoongu Kim, Moritz Lipp, Nicolas Boichat, Eric Shiu, Mattias Nissler, and Daniel Gruss.
\newblock {Half-Double: Hammering From the Next Row Over}.
\newblock In {\em USENIX Security}, 2022.

\bibitem{orosa2022spyhammer}
Lois Orosa, Ulrich R{\"u}hrmair, A~Giray Yaglikci, Haocong Luo, Ataberk Olgun, Patrick Jattke, Minesh Patel, Jeremie Kim, Kaveh Razavi, and Onur Mutlu.
\newblock {SpyHammer: Using RowHammer to Remotely Spy on Temperature}.
\newblock arXiv:2210.04084, 2022.

\bibitem{zhang2022implicit}
Zhi Zhang, Wei He, Yueqiang Cheng, Wenhao Wang, Yansong Gao, Dongxi Liu, Kang Li, Surya Nepal, Anmin Fu, and Yi~Zou.
\newblock {Implicit Hammer: Cross-Privilege-Boundary Rowhammer through Implicit Accesses}.
\newblock {\em IEEE TDSC}, 2022.

\bibitem{liu2022generating}
Liang Liu, Yanan Guo, Yueqiang Cheng, Youtao Zhang, and Jun Yang.
\newblock {Generating Robust DNN with Resistance to Bit-Flip based Adversarial Weight Attack}.
\newblock {\em IEEE TC}, 2022.

\bibitem{cohen2022hammerscope}
Yaakov Cohen, Kevin~Sam Tharayil, Arie Haenel, Daniel Genkin, Angelos~D Keromytis, Yossi Oren, and Yuval Yarom.
\newblock {HammerScope: Observing DRAM Power Consumption Using Rowhammer}.
\newblock In {\em CCS}, 2022.

\bibitem{zheng2022trojvit}
Mengxin Zheng, Qian Lou, and Lei Jiang.
\newblock {TrojViT: Trojan Insertion in Vision Transformers}.
\newblock arXiv:2208.13049, 2022.

\bibitem{fahr2022frodo}
Michael Fahr~Jr, Hunter Kippen, Andrew Kwong, Thinh Dang, Jacob Lichtinger, Dana Dachman-Soled, Daniel Genkin, Alexander Nelson, Ray Perlner, Arkady Yerukhimovich, et~al.
\newblock {When Frodo Flips: End-to-End Key Recovery on FrodoKEM via Rowhammer}.
\newblock {\em CCS}, 2022.

\bibitem{tobah2022spechammer}
Youssef Tobah, Andrew Kwong, Ingab Kang, Daniel Genkin, and Kang~G. Shin.
\newblock {SpecHammer: Combining Spectre and Rowhammer for New Speculative Attacks}.
\newblock In {\em S\&P}, 2022.

\bibitem{rakin2022deepsteal}
Adnan~Siraj Rakin, Md~Hafizul~Islam Chowdhuryy, Fan Yao, and Deliang Fan.
\newblock {DeepSteal: Advanced Model Extractions Leveraging Efficient Weight Stealing in Memories}.
\newblock In {\em S\&P}, 2022.

\bibitem{park2016statistical}
Kyungbae Park, Donghyuk Yun, and Sanghyeon Baeg.
\newblock {Statistical Distributions of Row-Hammering Induced Failures in DDR3 Components}.
\newblock {\em Microelectronics Reliability}, 2016.

\bibitem{park2016experiments}
Kyungbae Park, Chulseung Lim, Donghyuk Yun, and Sanghyeon Baeg.
\newblock {Experiments and Root Cause Analysis for Active-Precharge Hammering Fault in DDR3 SDRAM under 3xnm Technology}.
\newblock {\em Microelectronics Reliability}, 2016.

\bibitem{lim2017active}
Chulseung Lim, Kyungbae Park, and Sanghyeon Baeg.
\newblock {Active Precharge Hammering to Monitor Displacement Damage Using High-Energy Protons in 3x-nm SDRAM}.
\newblock {\em TNS}, 2017.

\bibitem{ryu2017overcoming}
Seong-Wan Ryu, Kyungkyu Min, Jungho Shin, Heimi Kwon, Donghoon Nam, Taekyung Oh, Tae-Su Jang, Minsoo Yoo, Yongtaik Kim, and Sungjoo Hong.
\newblock {Overcoming the Reliability Limitation in the Ultimately Scaled DRAM using Silicon Migration Technique by Hydrogen Annealing}.
\newblock In {\em IEDM}, 2017.

\bibitem{yun2018study}
Donghyuk Yun, Myungsang Park, Chulseung Lim, and Sanghyeon Baeg.
\newblock {Study of TID Effects on One Row Hammering using Gamma in DDR4 SDRAMs}.
\newblock In {\em IRPS}, 2018.

\bibitem{yang2019trap}
Thomas Yang and Xi-Wei Lin.
\newblock {Trap-Assisted DRAM Row Hammer Effect}.
\newblock {\em EDL}, 2019.

\bibitem{walker2021ondramrowhammer}
Andrew~J. Walker, Sungkwon Lee, and Dafna Beery.
\newblock {On DRAM RowHammer and the Physics on Insecurity}.
\newblock {\em IEEE TED}, 2021.

\bibitem{kim2020revisiting}
Jeremie~S. Kim, Minesh Patel, Abdullah~Giray Ya\u{g}l{\i}k\c{c}{\i}, Hasan Hassan, Roknoddin Azizi, Lois Orosa, and Onur Mutlu.
\newblock {Revisiting RowHammer: An Experimental Analysis of Modern Devices and Mitigation Techniques}.
\newblock In {\em ISCA}, 2020.

\bibitem{orosa2021deeper}
Lois Orosa, A~Giray Ya{\u{g}}l{\i}k{\c{c}}{\i}, Haocong Luo, Ataberk Olgun, Jisung Park, Hasan Hassan, Minesh Patel, Jeremie~S. Kim, and Onur Mutlu.
\newblock {A Deeper Look into RowHammer's Sensitivities: Experimental Analysis of Real DRAM Chips and Implications on Future Attacks and Defenses}.
\newblock In {\em MICRO}, 2021.

\bibitem{yaglikci2022understanding}
A.~Giray Ya{\u{g}}l{\i}k{c}{\i}, Haocong Luo, Geraldo~F De~Oliviera, Ataberk Olgun, Minesh Patel, Jisung Park, Hasan Hassan, Jeremie~S Kim, Lois Orosa, and Onur Mutlu.
\newblock {Understanding RowHammer Under Reduced Wordline Voltage: An Experimental Study Using Real DRAM Devices}.
\newblock In {\em DSN}, 2022.

\bibitem{khan2018analysis}
Mohammad Nasim~Imtiaz Khan and Swaroop Ghosh.
\newblock {Analysis of Row Hammer Attack on STTRAM}.
\newblock In {\em ICCD}, 2018.

\bibitem{agarwal2018rowhammer}
S.~Agarwal, H.~Dixit, D.~Datta, M.~Tran, D.~Houssameddine, D.~Shum, and F.~Benistant.
\newblock {Rowhammer for Spin Torque based Memory: Problem or not?}
\newblock In {\em INTERMAG}, 2018.

\bibitem{li2014write}
Haitong Li, Hong-Yu Chen, Zhe Chen, Bing Chen, Rui Liu, Gang Qiu, Peng Huang, Feifei Zhang, Zizhen Jiang, Bin Gao, Lifeng Liu, Xiaoyan Liu, Shimeng Yu, H.-S.~Philip Wong, and Jinfeng Kang.
\newblock {Write Disturb Analyses on Half-Selected Cells of Cross-Point RRAM Arrays}.
\newblock In {\em IRPS}, 2014.

\bibitem{ni2018write}
Kai Ni, Xueqing Li, Jeffrey~A. Smith, Matthew Jerry, and Suman Datta.
\newblock {Write Disturb in Ferroelectric FETs and Its Implication for 1T-FeFET AND Memory Arrays}.
\newblock {\em IEEE EDL}, 2018.

\bibitem{genssler2022reliability}
Paul~R. Genssler, Victor~M. van Santen, Jörg Henkel, and Hussam Amrouch.
\newblock {On the Reliability of FeFET On-Chip Memory}.
\newblock {\em TC}, 2022.

\bibitem{tol2023dont}
M~Caner Tol, Saad Islam, Andrew~J Adiletta, Berk Sunar, and Ziming Zhang.
\newblock {Don't Knock! Rowhammer at the Backdoor of DNN Models}.
\newblock In {\em DSN}, 2023.

\bibitem{aydin2022cyber}
Hakan Aydin and Ahmet Sertba{\c{s}}.
\newblock {Cyber Security in Industrial Control Systems (ICS): A Survey of RowHammer Vulnerability}.
\newblock {\em Applied Computer Science}, 2022.

\bibitem{mus2022jolt}
Koksal Mus, Yark{\i}n Dor{\"o}z, M~Caner Tol, Kristi Rahman, and Berk Sunar.
\newblock {Jolt: Recovering TLS Signing Keys via Rowhammer Faults}.
\newblock {\em Cryptology ePrint Archive}, 2022.

\bibitem{wang2022research}
Jianxin Wang, Hongke Xu, Chaoen Xiao, Lei Zhang, and Yuzheng Zheng.
\newblock {Research and Implementation of Rowhammer Attack Method based on Domestic NeoKylin Operating System}.
\newblock In {\em ICFTIC}, 2022.

\bibitem{lefforge2023reverse}
Sam Lefforge.
\newblock {Reverse Engineering Post-Quantum Cryptography Schemes to Find Rowhammer Exploits}.
\newblock Master's thesis, {University of Arkansas}, 2023.

\bibitem{fahr2022effects}
Michael~J Fahr.
\newblock {\em {The Effects of Side-Channel Attacks on Post-Quantum Cryptography: Influencing FrodoKEM Key Generation Using the Rowhammer Exploit}}.
\newblock PhD thesis, University of Arkansas, 2022.

\bibitem{kaur2022work}
Anandpreet Kaur, Pravin Srivastav, and Bibhas Ghoshal.
\newblock {Work-in-Progress: DRAM-MaUT: DRAM Address Mapping Unveiling Tool for ARM Devices}.
\newblock In {\em CASES}, 2022.

\bibitem{cai2022feasibility}
Kunbei Cai, Zhenkai Zhang, and Fan Yao.
\newblock {On the Feasibility of Training-time Trojan Attacks through Hardware-based Faults in Memory}.
\newblock In {\em HOST}, 2022.

\bibitem{li2022cyberradar}
Dawei Li, Di~Liu, Yangkun Ren, Ziyi Wang, Yu~Sun, Zhenyu Guan, Qianhong Wu, and Jianwei Liu.
\newblock {CyberRadar: A PUF-based Detecting and Mapping Framework for Physical Devices}.
\newblock arXiv:2201.07597, 2022.

\bibitem{roohi2022efficient}
Arman Roohi and Shaahin Angizi.
\newblock {Efficient Targeted Bit-Flip Attack Against the Local Binary Pattern Network}.
\newblock In {\em HOST}, 2022.

\bibitem{staudigl2022neurohammer}
Felix Staudigl, Hazem Al~Indari, Daniel Sch{\"o}n, Dominik Sisejkovic, Farhad Merchant, Jan~Moritz Joseph, Vikas Rana, Stephan Menzel, and Rainer Leupers.
\newblock {NeuroHammer: Inducing Bit-Flips in Memristive Crossbar Memories}.
\newblock In {\em DATE}, 2022.

\bibitem{yang2022socially}
Li-Hsing Yang, Shin-Shan Huang, Tsai-Ling Cheng, Yi-Ching Kuo, and Jian-Jhih Kuo.
\newblock {Socially-Aware Collaborative Defense System against Bit-Flip Attack in Social Internet of Things and Its Online Assignment Optimization}.
\newblock In {\em ICCCN}, 2022.

\bibitem{islam2022signature}
Saad Islam, Koksal Mus, Richa Singh, Patrick Schaumont, and Berk Sunar.
\newblock {Signature Correction Attack on Dilithium Signature Scheme}.
\newblock In {\em Euro S\&P}, 2022.

\bibitem{france2022modeling}
Lo{\"\i}c France, Florent Bruguier, Maria Mushtaq, David Novo, and Pascal Benoit.
\newblock {Modeling Rowhammer in the gem5 Simulator}.
\newblock In {\em CHES 2022-Conference on Cryptographic Hardware and Embedded Systems}, 2022.

\bibitem{kurmus2017from}
Anil Kurmus, Nikolas Ioannou, Matthias Neugschwandtner, Nikolaos Papandreou, and Thomas Parnell.
\newblock {From Random Block Corruption to Privilege Escalation: A Filesystem Attack Vector for RowHammer-like Attacks}.
\newblock In {\em USENIX WOOT}, 2017.

\bibitem{li2023fphammer}
Dawei Li, Di~Liu, Yangkun Ren, Ziyi Wang, Yu~Sun, Zhenyu Guan, Qianhong Wu, and Jianwei Liu.
\newblock {FPHammer: A Device Identification Framework based on DRAM Fingerprinting}.
\newblock In {\em TrustCom}, 2023.

\bibitem{tomita2022extracting}
Chihiro Tomita, Makoto Takita, Kazuhide Fukushima, Yuto Nakano, Yoshiaki Shiraishi, and Masakatu Morii.
\newblock {Extracting the Secrets of OpenSSL with RAMBleed}.
\newblock {\em Sensors}, 2022.

\bibitem{luo2023rowpress}
Haocong Luo, Ataberk Olgun, Abdullah~Giray Ya{\u{g}}l{\i}k{\c{c}}{\i}, Yahya~Can Tu{\u{g}}rul, Steve Rhyner, Meryem~Banu Cavlak, Jo{\"e}l Lindegger, Mohammad Sadrosadati, and Onur Mutlu.
\newblock {RowPress: Amplifying Read Disturbance in Modern DRAM Chips}.
\newblock In {\em ISCA}, 2023.

\bibitem{yaglikci2021blockhammer}
A~Giray Ya{\u{g}}l{\i}k{\c{c}}{\i}, Minesh Patel, Jeremie~S. Kim, Roknoddin Azizibarzoki, Ataberk Olgun, Lois Orosa, Hasan Hassan, Jisung Park, Konstantinos Kanellopoullos, Taha Shahroodi, Saugata Ghose, and Onur Mutlu.
\newblock {BlockHammer: Preventing RowHammer at Low Cost by Blacklisting Rapidly-Accessed DRAM Rows}.
\newblock In {\em HPCA}, 2021.

\bibitem{park2020graphene}
Yeonhong Park, Woosuk Kwon, Eojin Lee, Tae~Jun Ham, Jung~Ho Ahn, and Jae~W Lee.
\newblock {Graphene: Strong yet Lightweight Row Hammer Protection}.
\newblock In {\em MICRO}, 2020.

\bibitem{yaglikci2022hira}
A~Giray Ya{\u{g}}lik{c}i, Ataberk Olgun, Minesh Patel, Haocong Luo, Hasan Hassan, Lois Orosa, O{\u{g}}uz Ergin, and Onur Mutlu.
\newblock {HiRA: Hidden Row Activation for Reducing Refresh Latency of Off-the-Shelf DRAM Chips}.
\newblock In {\em MICRO}, 2022.

\bibitem{de2022ancient}
Arthur De~Graauw.
\newblock {Ancient Port Structures. Parallels between the Ancient and the Modern}.
\newblock {\em M{\'e}diterran{\'e}e. Revue g{\'e}ographique des pays m{\'e}diterran{\'e}ens/Journal of Mediterranean geography}, 2022.

\bibitem{saxena2022aqua}
Anish Saxena, Gururaj Saileshwar, Prashant~J. Nair, and Moinuddin Qureshi.
\newblock {AQUA: Scalable Rowhammer Mitigation by Quarantining Aggressor Rows at Runtime}.
\newblock In {\em MICRO}, 2022.

\bibitem{qureshi2022hydra}
Moinuddin Qureshi, Aditya Rohan, Gururaj Saileshwar, and Prashant~J Nair.
\newblock {Hydra: Enabling Low-Overhead Mitigation of Row-Hammer at Ultra-Low Thresholds via Hybrid Tracking}.
\newblock In {\em ISCA}, 2022.

\bibitem{lee2018twice}
Eojin Lee, Sukhan Lee, G~Edward Suh, and Jung~Ho Ahn.
\newblock {TWiCe: Time Window Counter Based Row Refresh to Prevent Row-Hammering}.
\newblock {\em IEEE CAL}, 2018.

\bibitem{marazzi2023rega}
Michele Marazzi, Flavien Solt, Patrick Jattke, Kubo Takashi, and Kaveh Razavi.
\newblock {{REGA}: Scalable Rowhammer Mitigation with Refresh-Generating Activations}.
\newblock In {\em {S\&P}}, 2023.

\bibitem{jedec2020ddr5}
{JEDEC}.
\newblock {\em {JESD79-5: DDR5 SDRAM Standard}}, 2020.

\bibitem{jedec2024jesd795c}
{JEDEC}.
\newblock {\em {JESD79-5c: DDR5 SDRAM Standard}}, 2024.

\bibitem{lee2019twice}
Eojin Lee, Ingab Kang, Sukhan Lee, G~{Edward Suh}, and Jung {Ho Ahn}.
\newblock {TWiCe: Preventing Row-Hammering by Exploiting Time Window Counters}.
\newblock In {\em ISCA}, 2019.

\bibitem{jedec2020jesd795}
{JEDEC}.
\newblock {\em {JESD79-5: DDR5 SDRAM Standard}}, 2020.

\bibitem{kroft1981lockup}
David Kroft.
\newblock {Lockup-Free Instruction Fetch/Prefetch Cache Organization}.
\newblock In {\em ISCA}, 1981.

\bibitem{bachrach2012chisel}
Jonathan Bachrach, Huy Vo, Brian Richards, Yunsup Lee, Andrew Waterman, Rimas Avi{\v{z}}ienis, John Wawrzynek, and Krste Asanovi{\'c}.
\newblock Chisel: constructing hardware in a scala embedded language.
\newblock In {\em DAC}, pages 1216--1225, 2012.

\bibitem{synopsys}
{Synopsys, Inc.}
\newblock {Synopsys Design Compiler}.
\newblock \url{https://www.synopsys.com/support/training/rtl-synthesis/design-compiler-rtl-synthesis.html}.

\bibitem{cacti}
Rajeev Balasubramonian, Andrew~B. Kahng, Naveen Muralimanohar, Ali Shafiee, and Vaishnav Srinivas.
\newblock {CACTI: An integrated cache and memory access time, cycle time, area, leakage, and dynamic power model}.
\newblock \url{https://www.hpl.hp.com/research/cacti/}, 2017.

\bibitem{jedec2017ddr4}
{JEDEC}.
\newblock {\em {JESD79-4C: DDR4 SDRAM Standard}}, 2020.

\bibitem{luo2023ramulator2}
Haocong Luo, Yahya~Can Tu\u{g}rul, F.~Nisa Bostancı, Ataberk Olgun, A.~Giray Ya\u{g}l{\i}k\c{c}{\i}, , and Onur Mutlu.
\newblock {Ramulator 2.0: A Modern, Modular, and Extensible DRAM Simulator}, 2023.

\bibitem{ramulator2github}
SAFARI~Research Group.
\newblock {Ramulator V2.0}.
\newblock \url{https://github.com/CMU-SAFARI/ramulator2}.

\bibitem{spec2006}
{Standard Performance Evaluation Corp.}
\newblock {SPEC CPU 2006}.
\newblock \url{http://www.spec.org/cpu2006/}.

\bibitem{spec2017}
{Standard Performance Evaluation Corp.}
\newblock {SPEC CPU2017 Benchmarks}.
\newblock \url{http://www.spec.org/cpu2017/}.

\bibitem{tpc}
{\em {Transaction Processing Performance Council}}.

\bibitem{fritts2009media}
Jason~E. Fritts, Frederick~W. Steiling, Joseph~A. Tucek, and Wayne Wolf.
\newblock {MediaBench II Video: Expediting the next Generation of Video Systems Research}.
\newblock {\em Microprocess. Microsyst.}, 2009.

\bibitem{ycsb}
Brian Cooper, Adam Silberstein, Erwin Tam, Raghu Ramakrishnan, and Russell Sears.
\newblock {Benchmarking Cloud Serving Systems with {YCSB}}.
\newblock In {\em SoCC}, 2010.

\bibitem{mutlu2019retrospective}
Onur Mutlu and Jeremie Kim.
\newblock {RowHammer: A Retrospective}.
\newblock {\em IEEE TCAD Special Issue on Top Picks in Hardware and Embedded Security}, 2019.

\bibitem{luo2024rowpress}
Haocong Luo, Ataberk Olgun, A~Giray Ya{\u{g}}l{\i}k{\c{c}}{\i}, Yahya~Can Tu{\u{g}}rul, Steve Rhyner, Meryem~Banu Cavlak, Jo{\"e}l Lindegger, Mohammad Sadrosadati, and Onur Mutlu.
\newblock {RowPress Vulnerability in Modern DRAM Chips}.
\newblock {\em IEEE Micro}, 2024.

\bibitem{redeker2002investigation}
Michael Redeker, Bruce~F Cockburn, and Duncan~G Elliott.
\newblock {An Investigation into Crosstalk Noise in DRAM Structures}.
\newblock In {\em MTDT}, 2002.

\bibitem{park2014active}
Kyungbae Park, Sanghyeon Baeg, ShiJie Wen, and Richard Wong.
\newblock {Active-Precharge Hammering on a Row-Induced Failure in DDR3 SDRAMs Under 3x nm Technology}.
\newblock In {\em IIRW}, 2014.

\bibitem{yang2016suppression}
Chia Yang, Chen~Kang Wei, Yu~Jing Chang, Tieh~Chiang Wu, Hsiu~Pin Chen, and Chao~Sung Lai.
\newblock {Suppression of RowHammer Effect by Doping Profile Modification in Saddle-Fin Array Devices for Sub-30-nm DRAM Technology}.
\newblock {\em TDMR}, 2016.

\bibitem{yang2017scanning}
Chia-Ming Yang, Chen-Kang Wei, Hsiu-Pin Chen, Jian-Shing Luo, Yu~Jing Chang, Tieh-Chiang Wu, and Chao-Sung Lai.
\newblock {Scanning Spreading Resistance Microscopy for Doping Profile in Saddle-Fin Devices}.
\newblock {\em IEEE Transactions on Nanotechnology}, 2017.

\bibitem{lim2018study}
Chulseung Lim, Kyungbae Park, Geunyong Bak, Donghyuk Yun, Myungsang Park, Sanghyeon Baeg, Shi-Jie Wen, and Richard Wong.
\newblock {Study of Proton Radiation Effect to Row Hammer Fault in DDR4 SDRAMs}.
\newblock {\em Microelectronics Reliability}, 2018.

\bibitem{gautam2019rowhammering}
S.~K. Gautam, S.~K. Manhas, Arvind Kumar, Mahendra Pakala, and Yiehm Ellie.
\newblock {Row Hammering Mitigation Using Metal Nanowire in Saddle Fin DRAM}.
\newblock {\em IEEE TED}, 2019.

\bibitem{jiang2021quantifying}
Yichen Jiang, Huifeng Zhu, Dean Sullivan, Xiaolong Guo, Xuan Zhang, and Yier Jin.
\newblock {Quantifying RowHammer Vulnerability for DRAM Security}.
\newblock In {\em DAC}, 2021.

\bibitem{he2023whistleblower}
Wei He, Zhi Zhang, Yueqiang Cheng, Wenhao Wang, Wei Song, Yansong Gao, Qifei Zhang, Kang Li, Dongxi Liu, and Surya Nepal.
\newblock {WhistleBlower: A System-level Empirical Study on RowHammer}.
\newblock {\em IEEE Transactions on Computers}, 2023.

\bibitem{baeg2022estimation}
Sanghyeon Baeg, Donghyuk Yun, Myungsun Chun, and Shi-Jie Wen.
\newblock {Estimation of the Trap Energy Characteristics of Row Hammer-Affected Cells in Gamma-Irradiated DDR4 DRAM}.
\newblock {\em IEEE Transactions on Nuclear Science}, 2022.

\bibitem{mutlu2018rowhammer}
Onur Mutlu.
\newblock {RowHammer}.
\newblock \url{https://people.inf.ethz.ch/omutlu/pub/onur-Rowhammer-TopPicksinHardwareEmbeddedSecurity-November-8-2018.pdf}, 2018.

\bibitem{olgun2023hbm}
Ataberk Olgun, Majd Osseiran, Abdullah~Giray Yaglikci, Yahya~Can Tugrul, Haocong Luo, Steve Rhyner, Behzad Salami, Juan Gomez~Luna, and Onur Mutlu.
\newblock {An Experimental Analysis of RowHammer in HBM2 DRAM Chips}.
\newblock In {\em DSN Disrupt}, 2023.

\bibitem{olgun2023drambender}
Ataberk Olgun, Hasan Hassan, A.~Giray Ya{\u{g}}l{\i}k{c}{\i}, Yahya~Can Tu{\u{g}}rul, Lois Orosa, Haocong Luo, Minesh Patel, Ergin O{\u{g}}uz, and Onur Mutlu.
\newblock {DRAM Bender: An Extensible and Versatile FPGA-based Infrastructure to Easily Test State-of-the-art DRAM Chips}.
\newblock {\em {TCAD}}, 2023.

\bibitem{zhou2023double}
Longda Zhou, Jie Li, Zheng Qiao, Pengpeng Ren, Zixuan Sun, Jianping Wang, Blacksmith Wu, Zhigang Ji, Runsheng Wang, Kanyu Cao, and Ru~Huang.
\newblock {Double-sided Row Hammer Effect in Sub-20 nm DRAM: Physical Mechanism, Key Features and Mitigation}.
\newblock In {\em IRPS}, 2023.

\bibitem{lang2023blaster}
Zhenrong Lang, Patrick Jattke, Michele Marazzi, and Kaveh Razavi.
\newblock Blaster: Characterizing the blast radius of rowhammer.
\newblock In {\em 3rd Workshop on DRAM Security (DRAMSec) co-located with ISCA 2023}. ETH Zurich, 2023.

\bibitem{li2024understanding}
Jie Li, Longda Zhou, Sheng Ye, Zheng Qiao, and Zhigang Ji.
\newblock {Understanding the Competitive Interaction in Leakage Mechanisms for Effective Row Hammer Mitigation in Sub-20nm DRAM}.
\newblock {\em IEEE EDL}, 2024.

\bibitem{zhou2024understanding}
Longda Zhou, Jie Li, Pengpeng Ren, Sheng Ye, Da~Wang, Zheng Qiao, and Zhigang Ji.
\newblock {Understanding the Physical Mechanism of RowPress at the Device-Level in Sub-20 nm DRAM}.
\newblock In {\em IRPS}, 2024.

\bibitem{zhou2024unveiling}
Longda Zhou, Sheng Ye, Runsheng Wang, and Zhigang Ji.
\newblock {Unveiling RowPress in Sub-20 nm DRAM Through Comparative Analysis With Row Hammer: From Leakage Mechanisms to Key Features}.
\newblock {\em IEEE TED}, 2024.

\bibitem{luo2024experimental}
Haocong Luo, Ismail~Emir Y{\"u}ksel, Ataberk Olgun, A~Giray Ya{\u{g}}l{\i}k{\c{c}}{\i}, Mohammad Sadrosadati, and Onur Mutlu.
\newblock {An Experimental Characterization of Combined RowHammer and RowPress Read Disturbance in Modern DRAM Chips}.
\newblock {\em DSN Disrupt}, 2024.

\bibitem{apple2015about}
{Apple Inc.}
\newblock {About the Security Content of Mac EFI Security Update 2015-001}.
\newblock \url{https://support.apple.com/en-us/HT204934}, 2015.

\bibitem{hp2015rowhammer}
{Hewlett-Packard Enterprise}.
\newblock {HP Moonshot Component Pack Version 2015.05.0}.
\newblock \url{http://h17007.www1.hp.com/us/en/enterprise/servers/products/moonshot/component-pack/index.aspx}, 2015.

\bibitem{lenovo2015rowhammer}
{Lenovo}.
\newblock {Row Hammer Privilege Escalation}.
\newblock \url{https://support.lenovo.com/us/en/product_security/row_hammer}, 2015.

\bibitem{greenfield2012throttling}
Zvika Greenfield and Tomer Levy.
\newblock {Throttling Support for Row-Hammer Counters}, 2016.

\bibitem{kim2014architectural}
Dae-Hyun Kim, Prashant~J Nair, and Moinuddin~K Qureshi.
\newblock {Architectural Support for Mitigating Row Hammering in DRAM Memories}.
\newblock {\em CAL}, 2014.

\bibitem{aichinger2015ddr}
Barbara Aichinger.
\newblock {DDR Memory Errors Caused by Row Hammer}.
\newblock In {\em HPEC}, 2015.

\bibitem{aweke2016anvil}
Zelalem~Birhanu Aweke, Salessawi~Ferede Yitbarek, Rui Qiao, Reetuparna Das, Matthew Hicks, Yossi Oren, and Todd Austin.
\newblock {ANVIL: Software-Based Protection Against Next-Generation Rowhammer Attacks}.
\newblock In {\em ASPLOS}, 2016.

\bibitem{bains-merged}
K.~Bains et~al.
\newblock {Row Hammer Refresh Command}.
\newblock US Patents: 9,117,544 9,236,110 10,210,925, 2015.

\bibitem{bains2015row}
Kuljit Bains, John Halbert, Christopher Mozak, Theodore Schoenborn, and Zvika Greenfield.
\newblock {Row Hammer Refresh Command}, 2015.

\bibitem{bains2016distributed}
Kuljit~S Bains and John~B Halbert.
\newblock {Distributed Row Hammer Tracking}, 2016.

\bibitem{bains2016row}
Kuljit~S Bains and John~B Halbert.
\newblock {Row Hammer Monitoring Based on Stored Row Hammer Threshold Value}.
\newblock US Patent: 10,083,737, 2016.

\bibitem{gomez2016dummy}
H.~{Gomez}, A.~{Amaya}, and E.~{Roa}.
\newblock {DRAM} row-hammer attack reduction using dummy cells.
\newblock In {\em NORCAS}, 2016.

\bibitem{son2017making}
Mungyu Son, Hyunsun Park, Junwhan Ahn, and Sungjoo Yoo.
\newblock {Making DRAM Stronger Against Row Hammering}.
\newblock In {\em DAC}, 2017.

\bibitem{seyedzadeh2018mitigating}
S.~M. {Seyedzadeh}, A.~K. {Jones}, and R.~{Melhem}.
\newblock {Mitigating Wordline Crosstalk Using Adaptive Trees of Counters}.
\newblock In {\em ISCA}, 2018.

\bibitem{irazoqui2016mascat}
Gorka Irazoqui, Thomas Eisenbarth, and Berk Sunar.
\newblock {MASCAT: Stopping Microarchitectural Attacks Before Execution}.
\newblock {\em IACR Cryptology}, 2016.

\bibitem{you2019mrloc}
Jung~Min You and Joon-Sung Yang.
\newblock {MRLoc: Mitigating Row-Hammering Based on Memory Locality}.
\newblock In {\em DAC}, 2019.

\bibitem{yaglikci2021security}
A.~Giray Ya{\u{g}}l{\i}k{\c{c}}{\i}, Jeremie~S. Kim, Fabrice Devaux, and Onur Mutlu.
\newblock {Security Analysis of the Silver Bullet Technique for RowHammer Prevention}.
\newblock arXiv:2106.07084, 2021.

\bibitem{kang2020cattwo}
Ingab Kang, Eojin Lee, and Jung~Ho Ahn.
\newblock {CAT-TWO: Counter-Based Adaptive Tree, Time Window Optimized for {DRAM} Row-Hammer Prevention}.
\newblock {\em {IEEE} Access}, 2020.

\bibitem{saileshwar2022randomized}
Gururaj Saileshwar, Bolin Wang, Moinuddin Qureshi, and Prashant~J Nair.
\newblock {Randomized Row-Swap: Mitigating Row Hammer by Breaking Spatial Correlation Between Aggressor and Victim Rows}.
\newblock In {\em ASPLOS}, 2022.

\bibitem{konoth2018zebram}
Radhesh~Krishnan Konoth, Marco Oliverio, Andrei Tatar, Dennis Andriesse, Herbert Bos, Cristiano Giuffrida, and Kaveh Razavi.
\newblock {ZebRAM: Comprehensive and Compatible Software Protection Against Rowhammer Attacks}.
\newblock In {\em OSDI}, 2018.

\bibitem{vig2018rapid}
Saru Vig, Sarani Bhattacharya, Debdeep Mukhopadhyay, and Siew-Kei Lam.
\newblock {Rapid Detection of Rowhammer Attacks Using Dynamic Skewed Hash Tree}.
\newblock In {\em HASP}, 2018.

\bibitem{hassan2019crow}
H.~{Hassan}, M.~{Patel}, J.~S. {Kim}, A.~G. {Ya\u{g}l{\i}k\c{c}{\i}}, N.~{Vijaykumar}, N.~{Mansouri Ghiasi}, S.~{Ghose}, and O.~{Mutlu}.
\newblock {CROW: A Low-Cost Substrate for Improving DRAM Performance, Energy Efficiency, and Reliability}.
\newblock In {\em ISCA}, 2019.

\bibitem{kim2022mithril}
Michael~Jaemin Kim, Jaehyun Park, Yeonhong Park, Wanju Doh, Namhoon Kim, Tae~Jun Ham, Jae~W Lee, and Jung~Ho Ahn.
\newblock {Mithril: Cooperative Row Hammer Protection on Commodity DRAM Leveraging Managed Refresh}.
\newblock In {\em HPCA}, 2022.

\bibitem{lee2021cryoguard}
Gyu-Hyeon Lee, Seongmin Na, Ilkwon Byun, Dongmoon Min, and Jangwoo Kim.
\newblock {CryoGuard: A Near Refresh-Free Robust DRAM Design for Cryogenic Computing}.
\newblock In {\em ISCA}, 2021.

\bibitem{marazzi2022protrr}
Michele Marazzi, Patrick Jattke, Flavien Solt, and Kaveh Razavi.
\newblock {ProTRR}: {Principled} yet {Optimal} {In-DRAM} {Target Row Refresh}.
\newblock In {\em {S\&P}}, 2022.

\bibitem{zhang2022softtrr}
Zhi Zhang, Yueqiang Cheng, Minghua Wang, Wei He, Wenhao Wang, Surya Nepal, Yansong Gao, Kang Li, Zhe Wang, and Chenggang Wu.
\newblock {SoftTRR: Protect Page Tables against Rowhammer Attacks using Software-only Target Row Refresh}.
\newblock In {\em USENIX ATC}, 2022.

\bibitem{joardar2022learning}
Biresh~Kumar Joardar, Tyler~K Bletsch, and Krishnendu Chakrabarty.
\newblock {Learning to Mitigate RowHammer Attacks}.
\newblock In {\em DATE}, 2022.

\bibitem{juffinger2023csi}
Jonas Juffinger, Lukas Lamster, Andreas Kogler, Maria Eichlseder, Moritz Lipp, and Daniel Gruss.
\newblock {CSI: Rowhammer--Cryptographic Security and Integrity against Rowhammer (to appear)}.
\newblock In {\em S\&P}, 2023.

\bibitem{manzhosov2022revisiting}
Evgeny Manzhosov, Adam Hastings, Meghna Pancholi, Ryan Piersma, Mohamed Tarek~Ibn Ziad, and Simha Sethumadhavan.
\newblock {Revisiting Residue Codes for Modern Memories}.
\newblock In {\em MICRO}, 2022.

\bibitem{ajorpaz2022evax}
Samira~Mirbagher Ajorpaz, Daniel Moghimi, Jeffrey~Neal Collins, Gilles Pokam, Nael Abu-Ghazaleh, and Dean Tullsen.
\newblock {EVAX: Towards a Practical, Pro-active \& Adaptive Architecture for High Performance \& Security}.
\newblock In {\em MICRO}, 2022.

\bibitem{naseredini2022alarm}
Amir Naseredini, Martin Berger, Matteo Sammartino, and Shale Xiong.
\newblock {ALARM: Active LeArning of Rowhammer Mitigations}.
\newblock \url{https://users.sussex.ac.uk/~mfb21/rh-draft.pdf}, 2022.

\bibitem{joardar2022machine}
Biresh~Kumar Joardar, Tyler~K. Bletsch, and Krishnendu Chakrabarty.
\newblock {Machine Learning-based Rowhammer Mitigation}.
\newblock {\em TCAD}, 2022.

\bibitem{hassan2022case}
Hasan Hassan, Ataberk Olgun, A~Giray Yaglikci, Haocong Luo, and Onur Mutlu.
\newblock {A Case for Self-Managing DRAM Chips: Improving Performance, Efficiency, Reliability, and Security via Autonomous in-DRAM Maintenance Operations}.
\newblock arXiv:2207.13358, 2022.

\bibitem{zhang2020leveraging}
Zhenkai Zhang, Zihao Zhan, Daniel Balasubramanian, Bo~Li, Peter Volgyesi, and Xenofon Koutsoukos.
\newblock {Leveraging EM Side-Channel Information to Detect Rowhammer Attacks}.
\newblock In {\em S\&P}, 2020.

\bibitem{loughlin2021stop}
Kevin Loughlin, Stefan Saroiu, Alec Wolman, and Baris Kasikci.
\newblock {Stop! Hammer Time: Rethinking Our Approach to Rowhammer Mitigations}.
\newblock In {\em HotOS}, 2021.

\bibitem{devaux2021method}
Fabrice Devaux and Renaud Ayrignac.
\newblock {Method and Circuit for Protecting a DRAM Memory Device from the Row Hammer Effect}.
\newblock US Patent: 10,885,966, 2021.

\bibitem{han2021surround}
Jin Han, Jungsik Kim, Dafna Beery, K~Deniz Bozdag, Peter Cuevas, Amitay Levi, Irwin Tain, Khai Tran, Andrew~J Walker, Senthil~Vadakupudhu Palayam, et~al.
\newblock {Surround Gate Transistor With Epitaxially Grown Si Pillar and Simulation Study on Soft Error and Rowhammer Tolerance for DRAM}.
\newblock {\em TED}, 2021.

\bibitem{fakhrzadehgan2022safeguard}
Ali Fakhrzadehgan, Yale~N. Patt, Prashant~J. Nair, and Moinuddin~K. Qureshi.
\newblock {SafeGuard: Reducing the Security Risk from Row-Hammer via Low-Cost Integrity Protection}.
\newblock In {\em HPCA}, 2022.

\bibitem{saroiu2022price}
Stefan Saroiu, Alec Wolman, and Lucian Cojocar.
\newblock {The Price of Secrecy: How Hiding Internal DRAM Topologies Hurts Rowhammer Defenses}.
\newblock In {\em IRPS}, 2022.

\bibitem{saroiu2022configure}
Stefan Saroiu and Alec Wolman.
\newblock {How to Configure Row-Sampling-Based Rowhammer Defenses}.
\newblock {\em DRAMSec}, 2022.

\bibitem{loughlin2022moesiprime}
Kevin Loughlin, Stefan Saroiu, Alec Wolman, Yatin~A. Manerkar, and Baris Kasikci.
\newblock {MOESI-Prime: Preventing Coherence-Induced Hammering in Commodity Workloads}.
\newblock In {\em ISCA}, 2022.

\bibitem{zhou2022lt}
Ranyang Zhou, Sepehr Tabrizchi, Arman Roohi, and Shaahin Angizi.
\newblock {LT-PIM: An LUT-Based Processing-in-DRAM Architecture With RowHammer Self-Tracking}.
\newblock {\em IEEE CAL}, 2022.

\bibitem{hong2023dsac}
Seungki Hong, Dongha Kim, Jaehyung Lee, Reum Oh, Changsik Yoo, Sangjoon Hwang, and Jooyoung Lee.
\newblock {DSAC: Low-Cost Rowhammer Mitigation Using In-DRAM Stochastic and Approximate Counting Algorithm}.
\newblock arXiv:2302.03591, 2023.

\bibitem{di2023copy}
Andrea Di~Dio, Koen Koning, Herbert Bos, and Cristiano Giuffrida.
\newblock {Copy-on-Flip: Hardening ECC Memory Against Rowhammer Attacks}.
\newblock In {\em NDSS}, 2023.

\bibitem{sharma2022review}
Sonia Sharma, Debdeep Sanyal, Arpit Mukhopadhyay, and Ramij~Hasan Shaik.
\newblock {A Review on Study of Defects of DRAM-RowHammer and Its Mitigation}.
\newblock {\em Journal For Basic Sciences}, 2022.

\bibitem{woo2023scalable}
Jeonghyun Woo, Gururaj Saileshwar, and Prashant~J Nair.
\newblock {Scalable and Secure Row-Swap: Efficient and Safe Row Hammer Mitigation in Memory Systems}.
\newblock In {\em HPCA}, 2023.

\bibitem{park2022row}
Jin~Hyo Park, Su~Yeon Kim, Dong~Young Kim, Geon Kim, Je~Won Park, Sunyong Yoo, Young-Woo Lee, and Myoung~Jin Lee.
\newblock {RowHammer Reduction Using a Buried Insulator in a Buried Channel Array Transistor}.
\newblock {\em IEEE Transactions on Electron Devices}, 2022.

\bibitem{wi2023shadow}
Minbok Wi, Jaehyun Park, Seoyoung Ko, Michael~Jaemin Kim, Nam~Sung Kim, Eojin Lee, and Jung~Ho Ahn.
\newblock {SHADOW: Preventing Row Hammer in DRAM with Intra-Subarray Row Shuffling}.
\newblock In {\em HPCA}, 2023.

\bibitem{kim2023a11v}
Woongrae Kim, Chulmoon Jung, Seongnyuh Yoo, Duckhwa Hong, Jeongjin Hwang, Jungmin Yoon, Ohyong Jung, Joonwoo Choi, Sanga Hyun, Mankeun Kang, Sangho Lee, Dohong Kim, Sanghyun Ku, Donhyun Choi, Nogeun Joo, Sangwoo Yoon, Junseok Noh, Byeongyong Go, Cheolhoe Kim, Sunil Hwang, Mihyun Hwang, Min Seol-Yi, Hyungmin Kim, Sanghyuk Heo, Yeonsu Jang, Kyoungchul Jang, Shinho Chu, Yoonna Oh, Kwidong Kim, Junghyun Kim, Soohwan Kim, Jeongtae Hwang, Sangil Park, Junphyo Lee, Inchul Jeong, Joohwan Cho, and Jonghwan Kim.
\newblock {A 1.1 V 16Gb DDR5 DRAM with Probabilistic-Aggressor Tracking, Refresh-Management Functionality, Per-Row Hammer Tracking, a Multi-Step Precharge, and Core-Bias Modulation for Security and Reliability Enhancement}.
\newblock In {\em ISSCC}, 2023.

\bibitem{gude2023defending}
C~Gude~Ramarao, K~Tejesh Kumar, G~Ujjinappa, and B~Vasu~Deva Naidu.
\newblock {Defending SoCs with FPGAs from Rowhammer Attacks}.
\newblock {\em Material Science}, 2023.

\bibitem{guha2022criticality}
Krishnendu Guha and Amlan Chakrabarti.
\newblock {Criticality based Reliability from Rowhammer Attacks in Multi-User-Multi-FPGA Platform}.
\newblock In {\em VLSID}, 2022.

\bibitem{france2022reducing}
Lo{\"\i}c France, Florent Bruguier, David Novo, Maria Mushtaq, and Pascal Benoit.
\newblock {Reducing the Silicon Area Overhead of Counter-Based Rowhammer Mitigations}.
\newblock In {\em 18th CryptArchi Workshop}, 2022.

\bibitem{bennett2021panopticon}
Tanj Bennett, Stefan Saroiu, Alec Wolman, and Lucian Cojocar.
\newblock {Panopticon: A Complete In-DRAM Rowhammer Mitigation}.
\newblock In {\em DRAMSec}, 2021.

\bibitem{enomoto2022efficient}
Shuhei Enomoto, Hiroki Kuzuno, and Hiroshi Yamada.
\newblock {Efficient Protection Mechanism for CPU Cache Flush Instruction Based Attacks}.
\newblock {\em IEICE Transactions on Information and Systems}, 2022.

\bibitem{arikan2022processor}
Kerem Ar{\i}kan, Alessandro Palumbo, Luca Cassano, Pedro Reviriego, Salvatore Pontarelli, Giuseppe Bianchi, O{\u{g}}uz Ergin, and Marco Ottavi.
\newblock {Processor security: Detecting microarchitectural attacks via count-min sketches}.
\newblock {\em VLSI}, 2022.

\bibitem{saxena2023pt}
Anish Saxena, Gururaj Saileshwar, Jonas Juffinger, Andreas Kogler, Daniel Gruss, and Moinuddin Qureshi.
\newblock {PT-Guard: Integrity-Protected Page Tables to Defend Against Breakthrough Rowhammer Attacks}.
\newblock In {\em DSN}, 2023.

\bibitem{zhou2023dnndefender}
Ranyang Zhou, Sabbir Ahmed, Adnan~Siraj Rakin, and Shaahin Angizi.
\newblock {DNN-Defender: An in-DRAM Deep Neural Network Defense Mechanism for Adversarial Weight Attack}.
\newblock arXiv:2305.08034, 2023.

\bibitem{bostanci2024comet}
F.~Nisa Bostancı, Ismail~Emir Yüksel, Ataberk Olgun, Konstantinos Kanellopoulos, Yahya~Can Tugrul, A.~Giray Yaglıkçı, Mohammad Sadrosadati, and Onur Mutlu.
\newblock {CoMeT: Count-Min-Sketch-Based Row Tracking to Mitigate RowHammer at Low Cost}.
\newblock In {\em HPCA}, 2024.

\bibitem{didio2023copyonflip}
Andrea Di~Dio, Koen Koning, Herbert Bos, and Cristiano Giuffrida.
\newblock {Copy-on-Flip: Hardening ECC Memory Against Rowhammer Attacks}.
\newblock In {\em NDSS}, 2023.

\bibitem{seyedzadeh2017counterbased}
Seyed~Mohammad Seyedzadeh, Alex~K. Jones, and Rami Melhem.
\newblock {Counter-Based Tree Structure for Row Hammering Mitigation in DRAM}.
\newblock {\em IEEE CAL}, 2017.

\bibitem{seyedzadeh2017mitigating}
Seyed~Mohammad Seyedzadeh, Donald Kline~Jr, Alex~K Jones, and Rami Melhem.
\newblock {Mitigating Bitline Crosstalk Noise in DRAM Memories}.
\newblock In {\em MEMSYS}, 2017.

\bibitem{van2024dram}
Victor van~der Veen, Pankaj Deshmukh, Behnam Dashtipour, David Hartley, and Mosaddiq Saifuddin.
\newblock {Dynamic Random Access Memory (DRAM) Row Hammering Mitigation}.
\newblock US Patent App. 17/890,022, 2024.

\bibitem{van2024dynamic}
Victor van~der Veen, Pankaj Deshmukh, Behnam Dashtipour, and David Hartley.
\newblock {Dynamic Rowhammer Management}.
\newblock US Patent App. 17/940,430, 2024.

\bibitem{verma2023defense}
Akash Verma, Victor van~der Veen, Joona Kannisto, and Marcel Selhorst.
\newblock {Defense Against Row Hammer Attacks}.
\newblock US Patent App. 17/842,606, 2023.

\bibitem{olgun2024abacus}
Ataberk Olgun, Yahya~Can Tugrul, F.~Nisa Bostancı, Ismail~Emir Yüksel, Haocong Luo, Steve Rhyner, A.~Giray Yaglıkçı, Geraldo~F. Oliveira, and Onur Mutlu.
\newblock {ABACuS: All-Bank Activation Counters for Scalable and Low Overhead RowHammer Mitigation}.
\newblock In {\em USENIX Security}, 2024.

\bibitem{seyedzadeh2017cbt}
Seyed~Mohammad Seyedzadeh, Alex~K. Jones, and Rami Melhem.
\newblock {Counter-Based Tree Structure for Row Hammering Mitigation in DRAM}.
\newblock {\em CAL}, 2017.

\bibitem{eyerman2008systemlevel}
Stijn Eyerman and Lieven Eeckhout.
\newblock {System-Level Performance Metrics for Multiprogram Workloads}.
\newblock {\em IEEE Micro}, 2008.

\bibitem{snavely2000symbiotic}
Allan Snavely and Dean~M Tullsen.
\newblock {Symbiotic Jobscheduling for A Simultaneous Multithreaded Processor}.
\newblock In {\em {ASPLOS}}, 2000.

\bibitem{hennessy2017computer}
John~L Hennessy and David~A Patterson.
\newblock {\em {Computer Architecture: A Quantitative Approach}}.
\newblock Morgan kaufmann, 2017.

\bibitem{lee2015decoupled}
Donghyuk Lee, Lavanya Subramanian, Rachata Ausavarungnirun, Jongmoo Choi, and Onur Mutlu.
\newblock {Decoupled Direct Memory Access: Isolating CPU and IO Traffic by Leveraging a Dual-Data-Port DRAM}.
\newblock In {\em PACT}, 2015.

\bibitem{misra1982finding}
Jayadev Misra and David Gries.
\newblock {Finding Repeated Elements}.
\newblock {\em {Science of Computer Programming}}, 1982.

\bibitem{hassan2022acase}
Hasan Hassan, Ataberk Olgun, A~Giray Yaglikci, Haocong Luo, and Onur Mutlu.
\newblock {A Case for Self-Managing DRAM Chips: Improving Performance, Efficiency, Reliability, and Security via Autonomous in-DRAM Maintenance Operations}.
\newblock arXiv:2207.13358 [cs.AR], 2022.

\bibitem{li2012compression}
Zhaogeng Li, Jun Bi, Sen Wang, and Xiaoke Jiang.
\newblock {Compression of Pending Interest Table with Bloom Filter in Content Centric Network}.
\newblock In {\em CFI}, 2012.

\bibitem{usui2016dash}
Hiroyuki Usui, Lavanya Subramanian, Kevin Kai-Wei Chang, and Onur Mutlu.
\newblock {DASH: Deadline-Aware High-Performance Memory Scheduler for Heterogeneous Systems with Hardware Accelerators}.
\newblock {\em TACO}, 2016.

\bibitem{kim2010thread}
Yoongu Kim, Michael Papamichael, Onur Mutlu, and Mor Harchol-Balter.
\newblock {Thread Cluster Memory Scheduling: Exploiting Differences in Memory Access Behavior}.
\newblock In {\em MICRO}, 2010.

\bibitem{subramanian2014bliss}
Lavanya Subramanian, Donghyuk Lee, Vivek Seshadri, Harsha Rastogi, and Onur Mutlu.
\newblock {The Blacklisting Memory Scheduler: Achieving High Performance and Fairness at Low Cost}.
\newblock In {\em ICCD}, 2014.

\bibitem{kim2010atlas}
Yoongu Kim, Dongsu Han, Onur Mutlu, and Mor Harchol-Balter.
\newblock {ATLAS: A Scalable and High-Performance Scheduling Algorithm for Multiple Memory Controllers}.
\newblock In {\em HPCA}, 2010.

\bibitem{nesbit2006fairqueuing}
K.~J. Nesbit, N.~Aggarwal, J.~Laudon, and J.~E. Smith.
\newblock {Fair Queuing Memory Systems}.
\newblock In {\em International Symposium on Microarchitecture (MICRO-39)}, 2006.

\bibitem{ausavarungnirun2012staged}
Rachata Ausavarungnirun, Kevin Kai-Wei Chang, Lavanya Subramanian, Gabriel~H Loh, and Onur Mutlu.
\newblock {Staged Memory Scheduling: Achieving High Performance and Scalability in Heterogeneous Systems}.
\newblock In {\em ISCA}, 2012.

\bibitem{mutlu2007stall}
Onur Mutlu and Thomas Moscibroda.
\newblock {Stall-Time Fair Memory Access Scheduling for Chip Multiprocessors}.
\newblock In {\em MICRO}, 2007.

\bibitem{mutlu2008parbs}
Onur Mutlu and Thomas Moscibroda.
\newblock {Parallelism-Aware Batch Scheduling: Enhancing Both Performance and Fairness of Shared DRAM Systems}.
\newblock In {\em ISCA}, 2008.

\bibitem{ebrahimi2010fairness}
Eiman Ebrahimi, Chang~Joo Lee, Onur Mutlu, and Yale~N Patt.
\newblock {Fairness via Source Throttling: A Configurable and High-Performance Fairness Substrate for Multi-Core Memory Systems}.
\newblock In {\em ASPLOS}, 2010.

\bibitem{intelxeon}
{Intel Inc.}
\newblock {3rd Gen Intel Xeon Scalable Processors}.
\newblock \url{https://www.intel.com/content/dam/www/public/us/en/documents/a1171486-icelake-productbrief-updates-r1v2.pdf}.

\bibitem{das2009application}
Reetuparna Das, Onur Mutlu, Thomas Moscibroda, and Chita~R Das.
\newblock {Application-Aware Prioritization Mechanisms for On-Chip Networks}.
\newblock In {\em MICRO}, 2009.

\bibitem{frfcfs}
Scott Rixner et~al.
\newblock {Memory Access Scheduling}.
\newblock In {\em ISCA}, 2000.

\bibitem{zuravleff1997controller}
William~K Zuravleff and Timothy Robinson.
\newblock {Controller for a Synchronous DRAM That Maximizes Throughput by Allowing Memory Requests and Commands to Be Issued Out of Order}, 1997.

\bibitem{kaseridis2011minimalistic}
Dimitris Kaseridis, Jeffrey Stuecheli, and Lizy~Kurian John.
\newblock {Minimalist Open-Page: A DRAM Page-Mode Scheduling Policy for the Many-Core Era}.
\newblock In {\em MICRO}, 2011.

\bibitem{kim20231perfecttrack}
Woongrae Kim, Chulmoon Jung, Seongnyuh Yoo, Duckhwa Hong, Jeongjin Hwang, Jungmin Yoon, Ohyong Jung, Joonwoo Choi, Sanga Hyun, Mankeun Kang, et~al.
\newblock A 1.1 v 16gb ddr5 dram with probabilistic-aggressor tracking, refresh-management functionality, per-row hammer tracking, a multi-step precharge, and core-bias modulation for security and reliability enhancement.
\newblock In {\em 2023 IEEE International Solid-State Circuits Conference (ISSCC)}, pages 1--3. IEEE, 2023.

\bibitem{canpolat2024understanding}
O{\u{g}}uzhan Canpolat, A~Giray Ya{\u{g}}l{\i}k{\c{c}}{\i}, Geraldo~F Oliveira, Ataberk Olgun, O{\u{g}}uz Ergin, and Onur Mutlu.
\newblock {Understanding the Security Benefits and Overheads of Emerging Industry Solutions to DRAM Read Disturbance}.
\newblock {\em DRAMSec}, 2024.

\bibitem{blockhammergithub}
{SAFARI Research Group}.
\newblock {BlockHammer --- GitHub Repository}.
\newblock \url{https://github.com/CMU-SAFARI/blockhammer}, 2021.

\bibitem{moscibroda2007memory}
Thomas Moscibroda and Onur Mutlu.
\newblock {Memory Performance Attacks: Denial of Memory Service in Multi-Core Systems}.
\newblock In {\em USENIX Security}, 2007.

\bibitem{chang2012hat}
Kevin~KaiWei Chang, Rachata Ausavarungnirun, Chris Fallin, and Onur Mutlu.
\newblock {HAT: Heterogeneous Adaptive Throttling for On-Chip Networks}.
\newblock In {\em SBAC-PAD}, 2012.

\bibitem{ebrahimi2011prefetch}
Eiman Ebrahimi, Chang~Joo Lee, Onur Mutlu, and Yale~N. Patt.
\newblock {Prefetch-Aware Shared Resource Management for Multi-Core Systems}.
\newblock In {\em ISCA}, 2011.

\bibitem{ebrahimi2011parallel}
Eiman Ebrahimi, Rustam Miftakhutdinov, Chris Fallin, Chang~Joo Lee, Jos{\'e}~A Joao, Onur Mutlu, and Yale~N Patt.
\newblock {Parallel Application Memory Scheduling}.
\newblock In {\em MICRO}, 2011.

\bibitem{ebrahimi2012fairness}
Eiman Ebrahimi, Chang~Joo Lee, Onur Mutlu, and Yale~N. Patt.
\newblock {Fairness via Source Throttling: A Configurable and High Performance Fairness Substrate for Multi Core Memory Systems}.
\newblock In {\em ASPLOS}, 2010.

\bibitem{lee2008prefetch}
Chang~Joo Lee, Onur Mutlu, Veynu Narasiman, and Yale~N Patt.
\newblock {Prefetch-Aware DRAM Controllers}.
\newblock In {\em MICRO}, 2008.

\bibitem{nychis2010next}
George Nychis, Chris Fallin, Thomas Moscibroda, and Onur Mutlu.
\newblock {Next Generation On-Chip Networks: What Kind of Congestion Control Do We Need?}
\newblock In {\em HOTNETS}, 2010.

\bibitem{mutlu2023retrospective}
Onur Mutlu.
\newblock Retrospective: Flipping bits in memory without accessing them: An experimental study of dram disturbance errors.
\newblock {\em Retrospective Issue for ISCA-50}, 2023.

\end{thebibliography}
\newpage
%
%
%
%
%


\appendix
\section{Artifact Appendix}

\subsection{Abstract}

Our artifact contains the data, source code, and scripts needed to reproduce our results.
We provide: 1) the source code of our simulation infrastructure based on Ramulator2 and 2) all evaluated memory access traces and all major evaluation results.
We provide Bash and Python scripts to analyze and plot the results automatically.

\subsection{Artifact Check-list (meta-information)}

\urlstyle{sf}
\begin{table}[H]
  \centering
  \scriptsize
  \setlength{\tabcolsep}{0.7\tabcolsep}
  \captionsetup{justification=centering, singlelinecheck=false, labelsep=colon}
    \begin{tabular}{ll}
        {{\bf Parameter}} & \textbf{Value} \\
        \hline
                        &  C++ program \\
        Program         &  Python3 scripts \\
                        &  Shell scripts \\
        \hline
        Compilation     &  C++ compiler with c++20 features \\
        \hline
                             &  Ubuntu 20.04 (or similar) Linux \\
                             &  C++20 build toolchain (tested with GCC 10) \\
        Run-time environment &  Python 3.10+ \\
                             &  Podman 4.5+ \\
                             &  Git \\
        \hline
                &  Weighted speedup \\
        Metrics &  Maximum slowdown \\
                &  DRAM energy \\
        \hline
        Experiment workflow & \makecell[l]{Perform simulations, aggregate results, and \\ run analysis scripts on the result} \\
        \hline
        Experiment customization & Possible. See \secref{sec:expcustom} \\
        \hline
        Disk space requirement & $\approx$ 30GiB \\
        \hline
        Workflow preparation time & $\approx$ 30 minutes \\
        \hline
        Experiment completion time & $\approx$ 2 days (on a compute cluster with 250 cores) \\
        \hline
                            &  Benchmarks (\url{https://zenodo.org/records/13293692})  \\
        Publicly available? &  Zenodo (\url{https://zenodo.org/record/13638017}) \\
                            &  GitHub (\url{https://github.com/CMU-SAFARI/BreakHammer})  \\
        \hline
        Code licences & MIT \\
        \hline
    \end{tabular}
    \label{tab:artifact_table}
\end{table}
\urlstyle{tt}

\subsection{Description}
 
\noindent\emph{We highly recommend using Slurm with a cluster that can run experiments in bulk.}

\subsubsection{How To Access}

Source code and scripts are available at \url{https://github.com/CMU-SAFARI/BreakHammer}.

\subsubsection{Hardware Dependencies}

We recommend using a PC with 32 GiB of main memory.
Approximately 30 GiB of disk space is needed to store intermediate and final evaluation results.

\subsubsection{Software Dependencies}

\begin{itemize}
    \item GNU Make, CMake 3.20+
    \item C++20 build toolchain (tested with GCC 10)
    \item Python 3.9+
    \item pip packages: matplotlib, pandas, seaborn, pyyaml, wget, and scipy
    \item Ubuntu 22.04
    \item (Optional) Slurm 20+
    \item (Optional) Podman 4.5+
\end{itemize}

\subsubsection{Benchmarks}

We use workload memory traces collected from SPEC2006, SPEC2017, TPC, MediaBench, and YCSB benchmark suites.
These traces are available at \url{https://zenodo.org/records/13293692}.
Install scripts will download and extract the traces.

\subsection{Installation}

\lstset{
    backgroundcolor=\color{gray!20}, 
    basicstyle=\ttfamily\bfseries\footnotesize,
    columns=fullflexible,
    frame=single,
    breaklines=true,
    postbreak=\mbox{\textcolor{red}{$\hookrightarrow$}\space},
    showstringspaces=false,
    numbersep=5pt,
    xleftmargin=6pt,
    xrightmargin=4pt,
    numbers=none,
    keywordstyle=\color{black},  
    identifierstyle=\color{black},  
    commentstyle=\color{black},  
    stringstyle=\color{black}  
}

The following steps will download and prepare the repository for the main experiments:
\begin{enumerate}
    \item Clone the git repository.
    \begin{lstlisting}[language=bash]
$ git clone git@github.com:CMU-SAFARI/BreakHammer.git
    \end{lstlisting}
    \item (Optional) Build the Podman container to run the scripts.
    \begin{lstlisting}[language=bash]
$ podman build . -t breakhammer_artifact
    \end{lstlisting}
    The following command runs a script using the container:
    \begin{lstlisting}[language=bash]
$ podman run --rm -v $PWD:/app \
    breakhammer_artifact <script>
    \end{lstlisting}
    \item Install Python dependencies, compile Ramulator2, download workload traces, and run a small sanity check.
    \begin{lstlisting}[language=bash]
$ ./run_simple_test.sh
    \end{lstlisting}
\end{enumerate}

\subsection{Evaluation and Expected Results}

\head{Claim 1 (C1)}
RowHammer mitigation mechanisms induce increasing performance overheads as the RowHammer threshold decreases.
This property is proven by evaluating the performance of the state-of-the-art RowHammer mitigation mechanisms on benign workloads (E1) as described in \secref{sec:motivation} whose results are illustrated in \figref{fig:performance_motivation}.

\head{Claim 2 (C2)}
\X{} improves system performance and DRAM energy overheads of state-of-the-art RowHammer mitigation mechanisms by detecting and stopping attackers that trigger many preventive actions.
This property is proven by evaluating the performance and DRAM energy of the state-of-the-art RowHammer mitigation mechanisms on workloads that include an attacker 1) when paired with \X{} and 2) without \X{} (E2) as described in \secref{subsec:perfunderattack} whose results are illustrated in \figsref{fig:a_fourcore_underattack}, \ref{fig:a_fourcore_underattack_scaling}, \ref{fig:a_fourcore_underattack_mitigative_actions}, \ref{fig:a_fourcore_underattack_energy}, and \ref{fig:c_fourcore_blockhammer}.

\head{Claim 3 (C3)}
\X{} does \emph{not} degrade system performance or DRAM energy when all applications are benign.
This property is proven by evaluating the performance and DRAM energy of the state-of-the-art RowHammer mitigation mechanisms on benign workloads 1) when paired with \X{} and 2) without \X{} (E3) as described in \secref{subsec:perfnoattack} whose results are illustrated in \figsref{fig:b_fourcore_noattack}, \ref{fig:b_fourcore_noattack_scaling}, and \ref{fig:fourcore_ben_memlat}.

\head{Experiments (E1, E2, and E3)}
[Ramulator2 simulations]
[10 human-minutes + 40 compute-hours (assuming $\sim$600 Ramulator2 simulations run in parallel) + 30GiB disk]

We prove our claims in two steps:
1) Execute Ramulator2 simulations to generate data supporting C1, C2, and C3.
2) Plot all figures that prove C1, C2 and C3.

\begin{enumerate}
    \item Launch all Ramulator2 simulation jobs.
    \begin{lstlisting}[language=bash]
$ ./run_with_personalcomputer.sh
(or ./run_with_slurm.sh if Slurm is available)
    \end{lstlisting}
    \item Wait for the simulations to end. The following displays the status and generates scripts to restart failed runs:
    \begin{lstlisting}[language=bash]
$ ./check_run_status.sh
    \end{lstlisting}
    \item Parse simulation results and collect statistics.
    \begin{lstlisting}[language=bash]
$ ./parse_results.sh
    \end{lstlisting}
    \item Generate all figures that support C1, C2, and C3.
    \begin{lstlisting}[language=bash]
$ ./plot_all_figures.sh
    \end{lstlisting}
\end{enumerate}

\subsection{Experiment Customization}
\label{sec:expcustom}
Our scripts provide easy configuration of the 1) evaluated RowHammer mitigation mechanisms, 2) tested RowHammer thresholds, 3) simulation duration, and 4) simulated workload combinations. The run parameters are configurable in \texttt{scripts/run\_config.py} with 1) \textit{mitigation\_list}, 2) \textit{tRH\_list}, and 3) \textit{NUM\_EXPECTED\_INSTS} or \textit{NUM\_MAX\_CYCLES}, respectively. Simulated attacker and benign workload combinations can be updated in \texttt{mixes/microattack.mix} and \texttt{mixes/microbenign.mix}, respectively.

\subsection{Methodology}

Submission, reviewing and badging methodology:

\begin{itemize}
  \item \url{https://www.acm.org/publications/policies/artifact-review-and-badging-current}
  \item \url{https://cTuning.org/ae}
\end{itemize}


\end{document}